\documentclass[aps,prb,superscriptaddress,amsmath,amssymb,floatfix,twocolumn,showpacs,amsfonts,longbibliography]{revtex4-2}
\usepackage{times}
\usepackage[varg]{txfonts}
\usepackage{textcomp}
\usepackage{graphicx,makecell}
\usepackage{subfigure}
\usepackage{color}
\usepackage[colorlinks=true,citecolor=blue,urlcolor=blue,linkcolor=blue,hyperindex]{hyperref}
\usepackage{braket}
\usepackage{tikz}
\usepackage{overpic}
\usepackage{bm}

\begin{document}
	
	\title{Topological superconducting states and quasiparticle transport on kagome lattice}
	
	\author{Zi-Qian Zhou}
	\email[These authors contributed equally to this work.]{} 
	\affiliation{School of Physics, Sun Yat-Sen University, Guangzhou 510275, China}
	
	\author{Weimin Wang}
	\email[These authors contributed equally to this work.]{} 
	\affiliation{School of Physics, Sun Yat-Sen University, Guangzhou 510275, China}
	
	\author{Zhi Wang}
	\affiliation{School of Physics, Sun Yat-Sen University, Guangzhou 510275, China}
	
	\author{Dao-Xin Yao}
	\email[Corresponding author:]{yaodaox@mail.sysu.edu.cn}
	\affiliation{School of Physics, Sun Yat-Sen University, Guangzhou 510275, China}
	\affiliation{State Key Laboratory of Optoelectronic Materials and Technologies, Center for Neutron Science and Technology, Guangdong Provincial Key Laboratory of Magnetoelectric Physics and Devices, Guangzhou 510275, China}
	
	\date{\today}
	\begin{abstract}
		The pairing symmetry of superconducting state is a critical topic in the realm of topological superconductivity. However, the pairing symmetry of the $A\mathrm{V_3Sb_5}$ family, wherein $A=\mathrm{K,Rb,Cs}$, remains indeterminate. To address this issue, we formulate an effective model on the kagome lattice to describe topological superconducting states featuring chiral charge density wave. Through this model, we explore the topological phase diagrams and thermal Hall conductivity under various parameters, with and without spin-orbit coupling. Our analysis reveals that the disparities in thermal Hall conductivity curves among different pairing symmetry states are safeguarded by the topology resulting from the interplay of spin-orbital coupling and superconductivity. Remarkably, this theoretical prediction can potentially enable the differentiation of various superconducting pairing symmetry states in materials via experimental measurements of thermal Hall conductivity curves.
	\end{abstract}

	\maketitle
	\section{Introduction}\label{sec:intro}
	
	Revealing specific physical phenomena through minimal lattice models is a fundamental undertaking within the realm of contemporary condensed matter physics. The kagome lattice has garnered considerable attention due to its distinctive and exotic electronic properties, along with its unique topo- logical features. These encompass both theoretically predicted topological bands and flat bands \cite{ghimire2020topology}. When the rotational and spin symmetries are perturbed within the kagome lattice struc- ture, intriguing outcomes emerge, including the emergence of nontrivial $\mathrm{Z}_2$ invariants and the presence of gapless edge states \cite{guo2009topological}. Furthermore, the kagome lattice has been instru- mental in realizing higher-order topological insulators \cite{xue2019acoustic}. More recently, the family of $A\mathrm{V_3Sb_5}$ (where $ A $ can be K, Rb, or Cs) has come into focus as the first example of quasi-two- dimensional kagome superconductors. The crystal structure of these materials is depicted in Fig. \ref{fig.kagome lattice 1}. This discovery of- fers an exceptional platform for investigating superconducting properties within the kagome lattice framework. Addition- ally, these materials have been demonstrated to exhibit charge density wave (CDW) order and superconducting (SC) properties \cite{ortiz2020cs, Ortiz2021Superconductivity, Yin2021Superconductivity}, sparking considerable interest in understanding the characteristics of each and their coexistence \cite{Jiang2021Unconventional, Zhao2021Cascade, Liang2021Three, Wang2021Unconventional, li2021spatial, ptok2021dynamical, zhao2021nodal, Chen2021Roton, Duan2021Nodeless, Xu2021Multiband, Chen2021Double, Du2021Pressure, song2021competing, Song2021Competition,Springer2021Unusual, oey2021fermi, yang2021doping}. Additionally, anomalous Hall effect (AHE), magneto-Seebeck effect, and Nernst effect have been observed in these materials, fur- ther expanding the range of intriguing phenomena associated with them \cite{Yu2021Concurrence, zheng2021gatecontrollable, Gan2021Magneto, li2021unconventional, chen2021anomalous, zhou2021anomalous}.

	The determination of the pairing symmetry underlying superconductivity within the $A\mathrm{V_3Sb_5}$ family remains a highly contentious issue that has yet to find a definitive resolution. Although a majority of findings point towards conventional $ s $-wave pairing, certain experimental observations have suggested unconventional behavior. The presence of U-shaped differential conductivity and the absence of in-gap states lend support to the notion of s-wave pairing \cite{Xu2021Multiband, Liang2021Three, Chen2021Roton}. In materials exhibiting $ s $-wave pairing, in-gap states are only induced by magnetic impurities, not by states with sign-changing properties. Notably, a Cr cluster only triggers the formation of an in-gap bound state, which strongly implies $ s $-wave pairing in $A\mathrm{V_3Sb_5}$ \cite{Xu2021Multiband}. Moreover, the clear Hebel Slichter coherence peak observed in nuclear magnetic resonance serves as additional robust evidence for $ s $-wave pairing \cite{Mu2021S, Mu2022Tri}. Additionally, two-gap $ s $-wave pairing models have provided the most comprehensive explanations for measurements related to resistance, penetration depth, and superfluid density \cite{Duan2021Nodeless, Ni2021Anisotropic}. Nevertheless, two independent experiments involving differential conductivity have identified an unsplit zero bias conductivity peak, suggesting the potential existence of $ p $-wave pairing \cite{Liang2021Three, Wang2021Electronic}. The thermal conductivity exhibits similarities with the $ d $-wave superconductor Tl-2201 \cite{zhao2021nodal}, featuring residual conductivity at 0 K. Furthermore, certain theoretical frameworks have predicted the presence of nodal $ s $-wave, $ p $-wave, and $ d $-wave pairings within the $\mathrm{AV_3Sb_5}$ system \cite{tazai2021mechanism, lin2021kagome}. Benefiting from the intense competition between chiral charge density wave and superconducting phases \cite{zhao2021nodal, Chen2021Double, Du2021Pressure}, we have developed an effective model aimed at capturing the critical physical properties.
	
	The paper is structured as follows. Our objective is to dis- tinguish various superconducting pairing symmetry states via the observable quasiparticle transport properties, specifically, the thermal Hall effect. After this introductory section, Sec. \ref{sec:model} expounds upon an effective model applied to a kagome lattice characterized by the presence of chiral charge density wave and superconductivity. In Sec. \ref{Sec.Method}, we elucidate the methodology employed for the computation of Berry curvature, Chern number, and thermal Hall conductivity. Subsequently, in Sec. \ref{Sec.noSOC}, we outline the phase diagrams and thermal Hall conductivity profiles for the model in the absence of spin- orbit coupling (SOC), as previously mentioned. Section \ref{Sec.SOC} then presents the outcomes concerning phase diagrams and thermal Hall conductivity for the model incorporating SOC, considering different chemical potential values. Section \ref{Sec.Conclusion} engages in a comprehensive discussion on the means by which we can discern distinct pairing symmetries by scrutinizing the thermal Hall conductivity profiles, ultimately concluding the paper’s presentation. Supplementary materials are included in the Appendices.
	
	\section{MODEL}\label{sec:model}
	
	We develop an effective model on a two-dimensional kagome lattice that incorporates chiral charge density wave (CDW), spin-orbit coupling (SOC), and superconductivity (SC). The primary goal of this model is to examine the topo- logical properties of the system’s superconducting states and the transport properties of quasiparticles in the presence of time-reversal symmetry breaking. Furthermore, we intend to propose a method for discriminating between different pairing symmetries based on our results.
	
	Taking inspiration from the properties exhibited by $ A\mathrm{V_3Sb_5} $, we have constructed our model on a two-dimensional kagome lattice, consisting of three atoms in each unit cell of the basic lattice. When we consider the $ 2\times 2 $ chiral CDW, the unit cell expands by a factor of four, while the Brillouin region shrinks to one quarter of its original size. This results in the number of sublattice atoms becoming 12. Upon the introduction of a nonzero SOC, the spin symmetry is broken, thereby doubling the number of bands to 24.
	
	We decompose the Hamiltonian into four parts: the nearest- neighbor tight-binding part, the CDW part, the SOC part, and the SC part, as expressed by Eq. (\ref{formula.Hamiltonian}). The first three terms
	ˆ combined are referred to as $\hat{H}_0$.
	
	\begin{equation}\label{formula.Hamiltonian}
		\hat{H} = \hat{H}_{TB}+\hat{H}_{CDW}+\hat{H}_{SOC}+\hat{H}_{SC} ,
	\end{equation}
	
	The nearest-neighbor tight-binding model for the kagome lattice is given by
	\begin{equation}\label{formula.tb}
		\begin{aligned}
			\hat{H}_{TB} &= \sum_{k,\sigma} \sum_{\alpha,\alpha'} (\mathcal{H}_{TB})_{\alpha,\alpha'}c^{\dagger}_{k\sigma,\alpha}c_{k\sigma,\alpha'}\\
			&= \sum_k \sum_{\alpha,\alpha'}\left[-\mu\delta_{\alpha,\alpha'}-2t\cos\left(\frac{k_l}{2}|\epsilon_{\alpha \alpha' l}|\right)\right]c^{\dagger}_{k\sigma,\alpha}c_{k\sigma,\alpha'}\quad ,
		\end{aligned}
	\end{equation}
	where the sublattice indexes $\alpha,\alpha'=A,B,C$ are extended to $\alpha,\alpha'=1,2,\cdots ,11,12$ with the inclusion of CDW. 
	 {$ \epsilon_{\alpha\alpha'l} $ is the Levi-Civita symbol.} We only consider the case where $\alpha$ and $\alpha'$ represent the nearest neighbor sublattices. Spin indices are denoted by $\sigma=\uparrow,\ \downarrow$, and the hopping $t$ is chosen to be isotropic, implying that we are studying the low-energy state (Appendix \ref{App.A}). For the sake of simplicity, we choose $t=1$ as energy unit throughout the paper, and $\mu$ represents the chemical potential.
	
	
	\begin{figure}[t]
		\centering
		\subfigure[]
		{
			\includegraphics[width=1.65in]{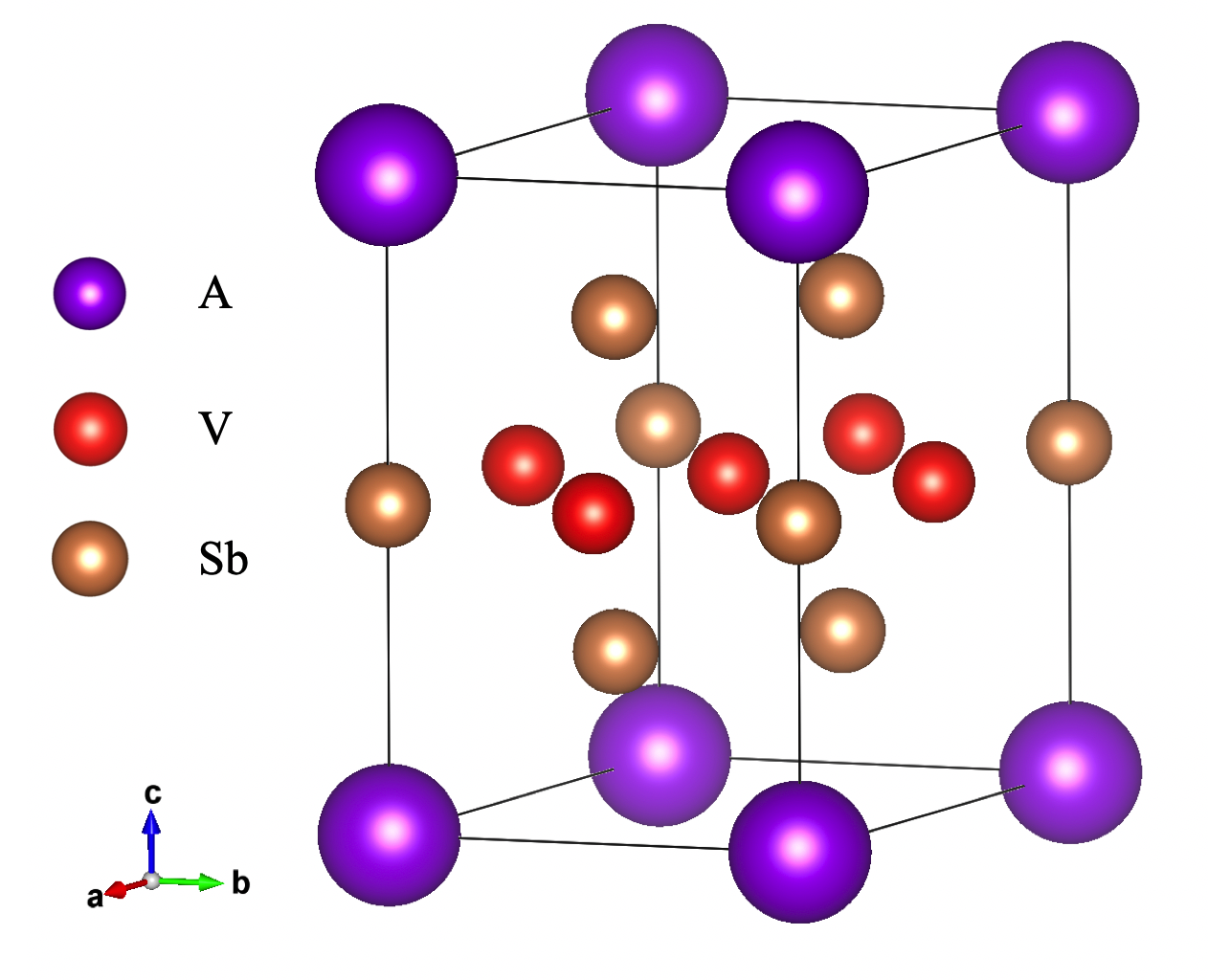}
			\label{fig.kagome lattice 1}
		}
		\subfigure[]
		{
			\includegraphics[width=1.3in]{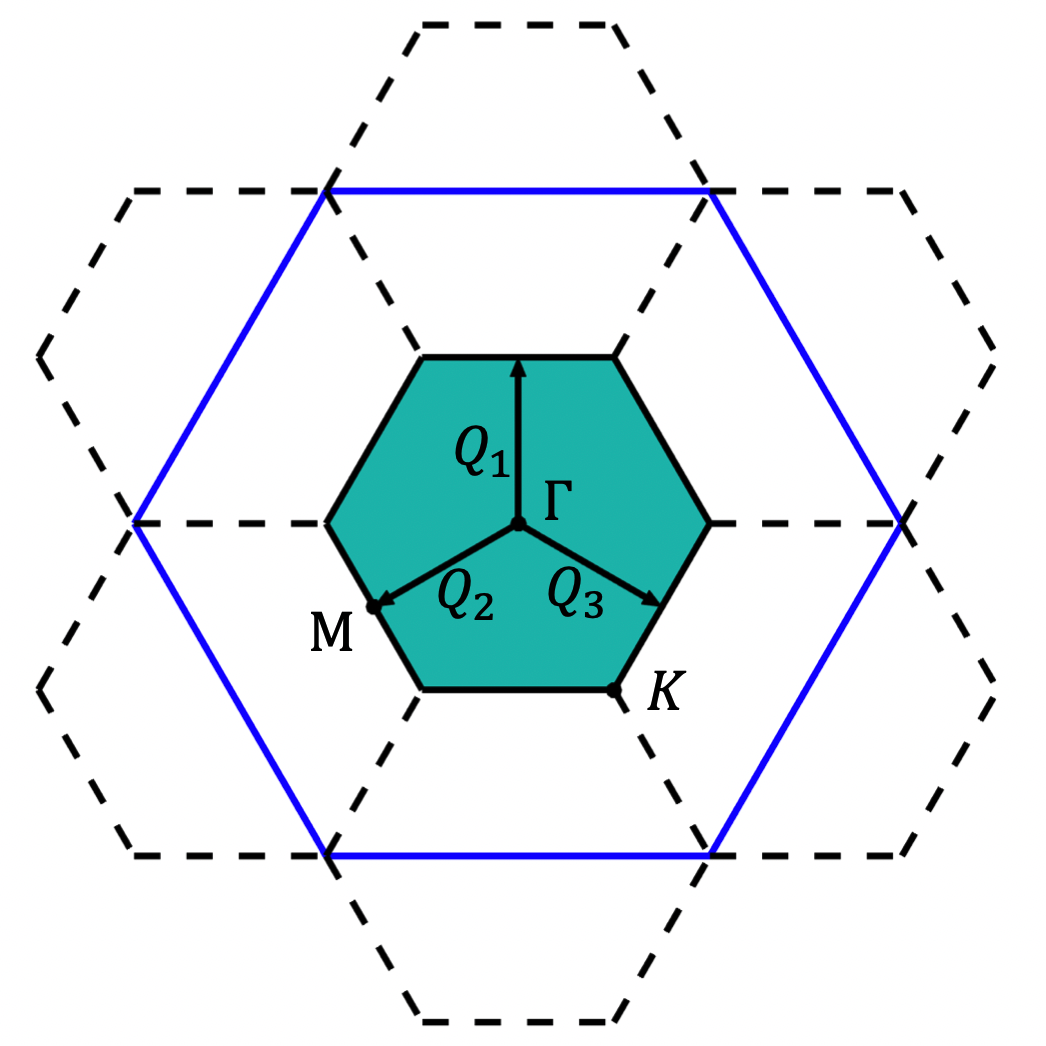}
			\label{fig.BZ}
		}
		\subfigure[]
		{
			\includegraphics[width=1.47in]{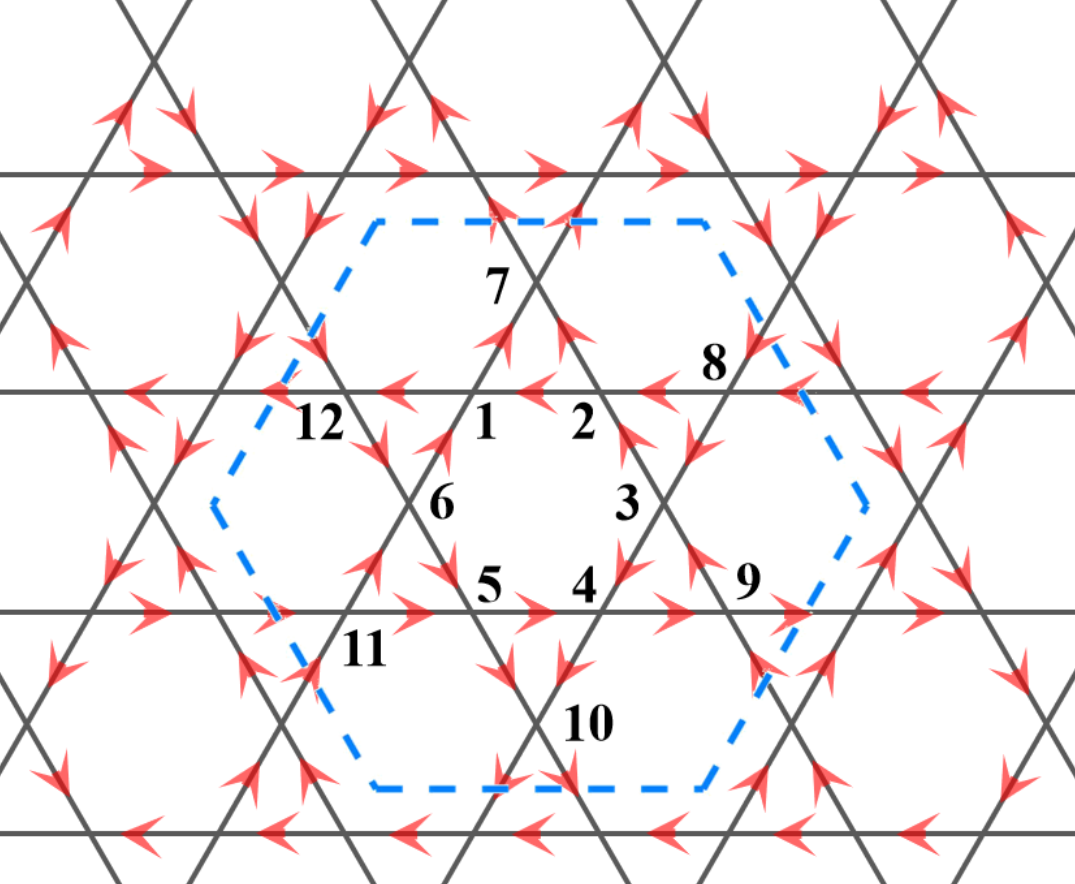}
			\label{fig.lattice_structure}
		}
		\subfigure[]
		{
			\includegraphics[width=1.53in]{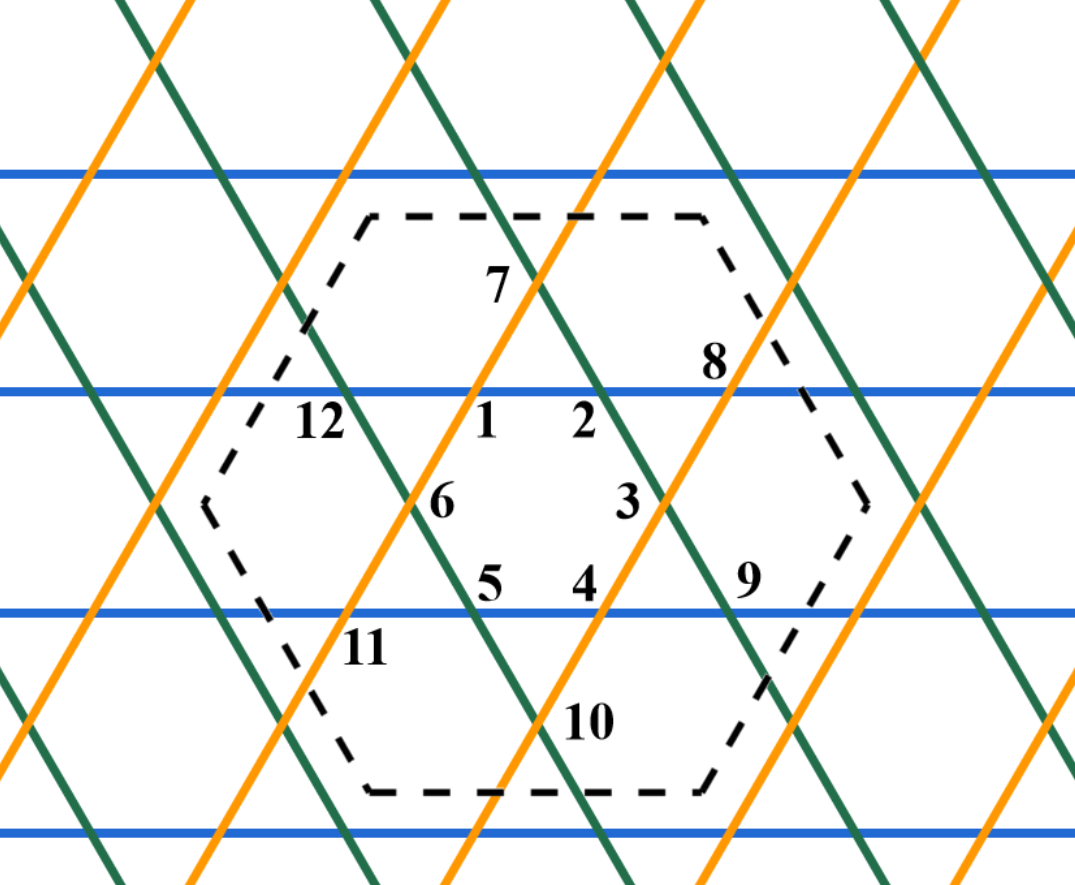}
			\label{fig.d-wave_structure}
		}
		\caption{(a) The lattice structure of $A \mathrm{V_3Sb_5} $. (b) The Brillouin zone of the kagome lattice with $ 2\times 2$ modulation which shrinks by 1/4. $Q$s are related to van Hove singularities, which can be written as $Q_1 = (0,\pi/\sqrt{3})$, $Q_2=(-\pi/2,-\pi/2\sqrt{3})$, $Q_3=(\pi/2,-\pi/2\sqrt{3})$. (c) Schematic of chiral flux phase model \cite{Feng2021Chiral}. (d)  Schematic of the $ d $+i$ d $-wave pairing. Different colors represent different phases. Blue lines represent $ \phi=0 $, yellow lines represent $ \phi=2\pi/3 $ and green lines represent $ \phi=4\pi/3 $. If there is a $ d $-i$ d $ pairing, the order of the definition reverses.}
		\label{fig.model}
	\end{figure}
	
	 {
		$ A\mathrm{V_3Sb_5} $ exhibits a CDW that can be classified into two types: in-plane CDW, and out-of-plane CDW that also known as $ c $-axis modulation. The $ 2\times 2 $ modulation of the CDW has been verified by several experiments \cite{Jiang2021Unconventional, Zhao2021Cascade, Liang2021Three, Chen2021Roton, Li2022Rotation, Uykur2021Low, Wang2021Electronic, mielke2021timereversal, Wang2021Charge}. Muon spin spectroscopy has detected a magnetic response of the chiral charge order, indicating time-reversal symmetry breaking (TRSB) \cite{Jiang2021Unconventional}. It has been confirmed by other experiments \cite{Li2022Rotation, Shumiya2021Intrinsic, Wang2021Electronic, mielke2021timereversal, yu2021evidence}. Several theories have attempted to explain the origin of TRSB, with the chiral flux phase (CFP) being the most successful in carrying nontrivial topology and naturally explaining TRSB \cite{Feng2021Chiral}. Since we are considering the quasi-two-dimensional properties of the kagome lattice, we take the $ 2\times 2 $ modulation into account only. 
	}

     {
   Before we begin the analysis of the model, we have to emphasize that the different origins of the CDW might lead to different physics and, consequently, different models. There are some experiments that point out that CDW might be related to the displacement of atoms \cite{ptok2021dynamical}. Many studies have observed the CDW transition in the $\mathrm{AV_3Sb_5 (A=K, Rb, Cs)}$ system \cite{ortiz2020cs, Ortiz2021Superconductivity, Yin2021Superconductivity}, with the indicator being a change in specific heat. This characteristic highlights the significant contribution of the electronic structural phase transition. What’s more, our paper primarily explores the relationship between the system’s topological properties and transport properties, the breaking of time-reversal symmetry, as one of the most critical concepts in topological physics, should be considered. Therefore we choose the chiral flux phase model to describe the CDW in the system.}
 
	The $2\times2$ modulation charge order results in an enlarged unit cell that is four times larger than the previous one, while the Brillouin zone shrinks by 1/4 [as shown in Fig.\ref{fig.BZ}]. Several theories tried to explain the properties of the $2\times2$ CDW, including those presented in Refs. \cite{Denner2021Analysis, Feng2021Chiral, Miao2021Geometry, Feng2021Low}. Among them, the CFP model (shown in Fig.\ref{fig.lattice_structure}) is the most convinc- ing model for capturing the chiral CDW characteristic, which has been confirmed by muon spin spectroscopy measurements \cite{Jiang2021Unconventional}. The CFP Hamiltonian can be expressed in real space as \cite{Feng2021Chiral}
	
	\begin{equation}\label{key}
		\hat{H}_{CDW} = -i \xi \sum_{\mathbf{R}} \Delta_{CFP}(\mathbf{R})\cdot O(\mathbf{R})+h.c.\quad ,
	\end{equation}  
	 {where $\Delta_{CFP}(\mathbf{R})_i=\left(\cos\left(\mathbf{Q}_a\cdot \mathbf{R}\right),\ \cos\left(\mathbf{Q}_b\cdot \mathbf{R}\right),\ \cos\left(\mathbf{Q}_c\cdot \mathbf{R}\right)\right)$ and $ O(\mathbf{R})=\left(c_A^{\dagger}c_B,\ c_B^\dagger c_C,\ c_C^\dagger c_A\right)$} 
	are three-dimensional vectors. The wave vectors $\mathbf{Q}_i\ (i=a,b,c)$ are related to van Hove singularities at the three equivalent $ M $ points on the boundary of the Brillouin zone, as shown in Fig. \ref{fig.BZ}.

	For the $2\times2$ CDW modulation kagome lattice, there would be about 120 free parameters in the SOC term if no ap- proximations were made. Even after considering time-reversal symmetry and point group symmetry, there might still be five free parameters left, which deviates from our original goal of constructing an effective model. To simplify the model, we adopt the Rashba model \cite{PhysRevB.64.121202} and rewrite it into a lattice model with sixfold symmetry, given by
	\begin{equation}\label{formula.soc}
		\hat{H}_{SOC}(\mathbf{r})=\lambda\sum_{<i,j>}\sum_{<\alpha,\alpha'>}c_{i\alpha}^\dagger  \left(R_{\pi/2}\mathbf{e}_{i\alpha,j\alpha'}\cdot \boldsymbol{\sigma}\right) c_{j\alpha'}\quad ,
	\end{equation}
	where $ c_{i\alpha}^\dagger = \left(c_{i\alpha,\uparrow}^\dagger, c_{i\alpha,\downarrow}^\dagger\right) $, $ \lambda $ represents the Rashba spin-orbit coupling strength, and $ \mathbf{e}_{i\alpha,j\alpha'}=\mathbf{e}_{i\alpha}-\mathbf{e}_{j\alpha'} $ is the unit vector from site $ i\alpha $ to site $ j\alpha' $, which is a constant value when the unit cell indexes $ i,j $ are specially chosen. 
	 {$ R_{\pi/2} $ is the 3-D in-plane rotation matrix with rotation angle $ \pi/2 $.} We consider the nearest-neighbor tight-binding model with a periodic bound- ary condition, thus it is not necessary to involve the unit cell index. The vector $ \mathbf{e}_{i\alpha,j\alpha'} \equiv \mathbf{e}_{\alpha,\alpha'}=\mathbf{e}_{\alpha}-\mathbf{e}_{\alpha'} $ is only dependent on the sublattice indexes. Hence, we can explicitly see that the Fourier transformation form of the SOC Hamiltonian in $ \mathbf{k} $ space is independent of the real space index (see Appendix \ref{App.A}). 
	
	In order to simplify analysis and numerical calculations, a Fourier transform is often employed to convert the real-space Hamiltonian into momentum space, using the basis $ c_{\mathbf{k}}^\dagger = \left(c_{\mathbf{k}1,\uparrow}^\dagger, c_{\mathbf{k}2,\uparrow}^\dagger \cdots, c_{\mathbf{k}12,\uparrow}^\dagger, c_{\mathbf{k}1,\downarrow}^\dagger, c_{\mathbf{k}2,\downarrow}^\dagger,\cdots , c_{\mathbf{k}12,\downarrow}^\dagger \right) $. In this space, the Hamiltonian can be expressed as $ \hat{H}_{0}=c_{\mathbf{k}}^\dagger \mathcal{H}_{0} c_{\mathbf{k}}$, where
	\begin{equation}\label{key}
		\mathcal{H}_{0}=
		\begin{pmatrix}
			\mathcal{H}_{TB}+\mathcal{H}_{CDW}& \mathcal{H}_{SOC}^{\uparrow\downarrow}\\
			\mathcal{H}_{SOC}^{\downarrow\uparrow}& \mathcal{H}_{TB}+\mathcal{H}_{CDW}
		\end{pmatrix}\quad .
	\end{equation}
	It should be noted that $ \mathcal{H}_{TB}+\mathcal{H}_{CDW} $ is identical for both spin-up and spin-down, as magnetism is not considered in this model. $ \mathcal{H}_{SOC}^{\uparrow\downarrow} $ represents the SOC Hamiltonian with the basis $ c_{\mathbf{k}\alpha \uparrow}^\dagger c_{\mathbf{k}\alpha'\downarrow} $, and the naming convention for $ \mathcal{H}_{SOC}^{\downarrow\uparrow} $ follows the same rule.
	
	It is straightforward to verify that our model exhibits a six-fold rotation symmetry, denoted by $ C_6 $. Specifically, the sixfold rotation symmetry of $ \mathcal{H}_{CDW} $ has been demonstrated in a previous study \cite{Feng2021Chiral}. To establish the existence of the sixfold rotation symmetry in $ \mathcal{H}_{TB} $ and $ \mathcal{H}_{SOC} $, we need to demonstrate that their forms preserve the symmetry. The sixfold rotation symmetry of $ \mathcal{H}_{TB} $ can be revealed by its form as shown in Eq.(\ref{formula.tb}). Notably, when we rotate the real space, it is equivalent to exchange the numbering rules while maintaining Eq. (\ref{formula.tb}). Furthermore, the sixfold symmetry of $ \mathcal{H}_{SOC} $ is guaranteed by the fact that $ \boldsymbol{\sigma} $ can be treated as a series of constant matrices under space rotation transformations. Therefore the rotations are equivalent to exchange the numbering rules again in ac- cordance with Eq. (\ref{formula.soc}).
	
	After discussing the geometric and electric properties, we will now delve into the model of superconductivity. While the pairing symmetry of $ A\mathrm{V_3Sb_5} $ ($ A $=K, Rb, Cs) has not yet been confirmed \cite{Xu2021Multiband, Liang2021Three, Chen2021Roton, Liang2021Three, Wang2021Electronic, zhao2021nodal}, we can construct some possible SC pairing symmetries to gain insight into the transition properties of the SC states. Additionally, we would like to highlight an observable value that can potentially distinguish between different pairing symmetries in experiments. Note that, although there are some experiments interpreting their findings as indicative of $ p $-wave occurrence \cite{Liang2021Three,Wang2021Electronic}, a greater number of studies point towards a spin-singlet pairing. There- fore our paper highlights the three most plausible spin-singlet superconducting scenarios and distinguishes among them us- ing the thermal Hall effect.
	
	The most probable SC pairing symmetry is $ s $-wave pairing, also known as conventional SC. The corresponding Hamiltonian can be expressed as
	\begin{equation}\label{key}
		\begin{aligned}
			\hat H_{s-wave}=&\frac{\Delta}{2}\sum_{\mathbf{k}, \alpha}c_{\mathbf{k}\alpha,\uparrow}^\dagger c_{-\mathbf{k}\alpha,\downarrow}^\dagger + h.c.\\
			-&\frac{\Delta}{2}\sum_{\mathbf{k}, \alpha}c_{\mathbf{k}\alpha,\downarrow}^\dagger c_{-\mathbf{k}\alpha,\uparrow}^\dagger + h.c.\quad ,
		\end{aligned}
	\end{equation}
	where $ \Delta $ represents the SC gap function, which is a constant for $ s $-wave pairing. The negative sign in the second term arises from the anti-commutation relation of the Fermion creation and annihilation operators $ \left\{c_{\mathbf{k},\alpha}^\dagger,c_{\mathbf{k}',\alpha'}^\dagger\right\} = \left\{c_{\mathbf{k},\alpha},c_{\mathbf{k}',\alpha'}\right\} = 0$ (see Appendix \ref{App.2}).
	
	Another possible spin-singlet pairing is $ d $-wave pairing, with angular momentum $ l = 2 $ and even spatial wave function. Based on the irreducible representations of the finite subgroups of SO(3), there are two possible $ d $-wave pairings, namely $d_{x^2-y^2}$-wave and $d_{xy}$-wave. When the momentum $\left(k_x,k_y\right)$ rotates by $\pi/2$ to become $\left(-k_y,k_x\right)$, the gap function $\Delta_{d-wave}$ changes sign to become $-\Delta_{d-wave}$, and when $\left(k_x,k_y\right)\rightarrow\left(-k_x,-k_y\right)$, $\Delta_{d-wave}\rightarrow \Delta_{d-wave}$. Thus, as $\mathbf{k}$ rotates once in $\mathbf{k}$-space, $\Delta_{d-wave}$ undergoes two periods.

	On a kagome lattice, it is more convenient to consider a complex $ d $-wave pairing or $ d  $+i$ d $-wave pairing. To construct a SC gap function using the tight-binding model, a $ d $-wave pairing SC is transformed from real space to $ \mathbf{k} $ space by assuming an extra phase when pairing in different directions \cite{PhysRevB.77.235420}. As a result, the real space $ d  $+i$ d  $-wave SC can be determined and written in a Fourier transformation form as follows.
	
	\begin{equation}\label{key}
		\begin{aligned}
			\hat H_{d+id-wave}=&\frac{\Delta}{2}\sum_{\mathbf{k}, \alpha}e^{i2\theta_{i\alpha,j\alpha}}e^{i\mathbf{k}\cdot\mathbf{e}_{i\alpha,j\alpha}}c_{\mathbf{k}\alpha,\uparrow}^\dagger c_{-\mathbf{k}\alpha,\downarrow}^\dagger + h.c.\\
			-&\frac{\Delta}{2}\sum_{\mathbf{k}, \alpha}e^{i2\theta_{i\alpha,j\alpha}}e^{i\mathbf{k}\cdot\mathbf{e}_{i\alpha,j\alpha}}c_{\mathbf{k}\alpha,\downarrow}^\dagger c_{-\mathbf{k}\alpha,\uparrow}^\dagger + h.c.\quad ,
		\end{aligned}
	\end{equation}
	where $ \theta_{i\alpha,j\alpha} $ is the angle between $ \mathbf{e}_{i\alpha} $ and $ \mathbf{e}_{j\alpha} $, and it depends on the sublattice index $ \alpha $. This is independent of the unit cell indexes $ i,j $ due to the periodic boundary condition. Note that the gap function of $ d $+i$ d $-wave pairing is given by $ {\Delta}(\mathbf{k}) = \Delta e^{i2\theta_{i\alpha,j\alpha}}e^{i\mathbf{k}\cdot\mathbf{e}_{i\alpha,j\alpha}} $, which is an even function of $ \mathbf{k} $. It can be confirmed that under the transformation of $ \mathbf{k} \rightarrow -\mathbf{k}$, $\Delta(\mathbf{k})$ remains unchanged. The sign-change between the first and second terms occurs for the same reason as the s-wave pairing. The phase of the SC gap function for the $ d  $+i$ d $-wave pairing can be represented by Fig. \ref{fig.d-wave_structure}, where the blue lines represent $ \phi=2\theta=0 $, yellow lines represent $ \phi=2\pi/3 $, and green lines represent $ \phi=4\pi/3 $. It is important to note that the $ d  $+i$ d  $-wave pairing here refers to a complex $ d  $-wave SC state in the sense of real space and atomic level, as opposed to a simple $ d $-wave pairing.
	
	The $ d  $-i$ d  $-wave pairing state can be seen as the opposite SC pairing state of the $ d  $+i$ d  $-wave  pairing state when the normal state Hamiltonian $\hat{H}_0$ is topologically trivial. However, when $\hat{H}_0$ is topologically non-trivial, there is a significant difference between the d+id-wave and $ d  $-i$ d  $-wave  pairing states. Therefore, it is necessary to consider the $ d  $-i$ d  $-wave  SC pairing state, which can be obtained by transforming $\theta_{i\alpha,j\alpha}\rightarrow-\theta_{i\alpha,j\alpha}$, resulting in the Hamiltonian expression
	\begin{equation}\label{key}
		\begin{aligned}
			\hat H_{d-id-wave}=&\frac{\Delta}{2}\sum_{\mathbf{k}, \alpha}e^{-i2\theta_{i\alpha,j\alpha}}e^{i\mathbf{k}\cdot\mathbf{e}_{i\alpha,j\alpha}}c_{\mathbf{k}\alpha,\uparrow}^\dagger c_{-\mathbf{k}\alpha,\downarrow}^\dagger + h.c.\\
			-&\frac{\Delta}{2}\sum_{\mathbf{k}, \alpha}e^{-i2\theta_{i\alpha,j\alpha}}e^{i\mathbf{k}\cdot\mathbf{e}_{i\alpha,j\alpha}}c_{\mathbf{k}\alpha,\downarrow}^\dagger c_{-\mathbf{k}\alpha,\uparrow}^\dagger + h.c.\quad ,
		\end{aligned}
	\end{equation}
	where the parameters in the $ d $-i$ d $-wave pairing state Hamiltonian are defined in exactly the same way as in the $ d $+i$ d $-wave pairing state.
	
	In the customary approach, the Hamiltonian is expressed in Bogoliubov-de Gennes (BdG) form in the Nambu representation, given by
	\begin{equation}\label{key}
		\begin{aligned}
			\left(c_{\mathbf{k}}^\dagger, c_{-\mathbf{k}}\right)_r=
			\left(\right.&\left.c_{\mathbf{k}1,\uparrow}^\dagger,\cdots,c_{\mathbf{k}12,\uparrow}^\dagger,c_{\mathbf{k}1,\downarrow}^\dagger,\cdots,c_{\mathbf{k}12,\downarrow}^\dagger,\right.\\
			&\left.c_{-\mathbf{k}1,\uparrow},\cdots,c_{-\mathbf{k}12,\uparrow},c_{-\mathbf{k}1,\downarrow},\cdots,c_{-\mathbf{k}12,\downarrow}\right)\quad ,
		\end{aligned}
	\end{equation}
	where the subscript $r$ denotes the rearrangement of the basis. The SC Hamiltonian in the Nambu representation can be written as
	\begin{equation}\label{key}
		\hat{H}_{SC} = 	\left(c_{\mathbf{k}}^\dagger, c_{-\mathbf{k}}\right)_r 
		\begin{pmatrix}
			0&\mathcal{H}_{SC}\\
			\mathcal{H}_{SC}^\dagger&0
		\end{pmatrix}
		\begin{pmatrix}
			c_{\mathbf{k}}\\
			c_{-\mathbf{k}}^\dagger
		\end{pmatrix}_r\quad .
	\end{equation}
	Thus, the entire Hamiltonian can be expressed as
	\begin{equation}\label{key}
		\mathcal{H} = 
		\begin{pmatrix}
			\mathcal{H}_{0}(\mathbf{k})&\mathcal{H}_{SC}\\
			\mathcal{H}_{SC}^\dagger& -	\mathcal{H}_{0}^*(-\mathbf{k})
		\end{pmatrix}\quad .
	\end{equation}
	 {where we wrote the Hamiltonian into BdG form. Note that $ \mathcal{H}_{SC} $ is the superconducting term which might be s-wave, d+id-wave or d-id-wave pairing superconducting term respectively.}
	
	\section{Method}\label{Sec.Method}

	In this paper, we investigate the topological properties and quasiparticle transport of superconducting pairing states with chiral CDW and SOC on a kagome lattice. We aim to distinguish different superconducting pairing symmetries by comparing the thermal Hall conductivity curves of different SC pairings, and we argue that the differences can be at- tributed to topology.
	
	The $ Z $ invariant, or the Chern number, serves as a good topological number for systems with particle-hole symme- try in the absence of time-reversal symmetry. The presence of the CFP term breaks time-reversal symmetry, and as the Hamiltonian is in a BdG form, the system naturally possesses a particle-hole symmetry. We explore three different types of superconducting pairing symmetries: $ s $-wave, $ d  $+i$ d  $-wave, and $ d  $-i$ d  $-wave pairings. When we apply the transformation $ \mathbf{k} \rightarrow -\mathbf{k}^* $, the gap functions of spin-singlet pairings switch to their negative counterparts. Therefore, the spin-singlet pairings belong to D class, which can be characterized by a $ Z $ invariant in a two-dimensional system \cite{PhysRevB.55.1142}. In our model, the normal state Hamiltonian is topologically nontrivial, which makes the situation particularly intriguing.
	
	In addition to the topological analysis, we also investigate quasiparticle transport, specifically the thermal Hall effect. In the semiclassical theory, the low-temperature Hall conductivity is mainly influenced by the Berry curvature near the Fermi surface \cite{Wang2021Berry}. Therefore we can utilize the calculation of the Berry curvature to understand the intrinsic thermal Hall conductivity and attempt to connect it with the topological number of the system, the Chern number, through the Berry curvature.
	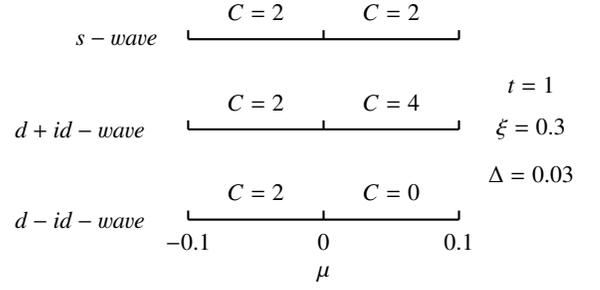
\begin{figure}[t]
		\centering
		\begin{tikzpicture}[scale = 1.2]
			\draw[-,thick](0,3)--(3,3) ;
			\draw[-,thick](0,2)--(3,2) ;
			\draw[-,thick](0,1)--(3,1) ;
			\draw[-,thick](0,3)--(0,3.1);
			\draw[-,thick](3,3)--(3,3.1);
			\draw[-,thick](0,2)--(0,2.1);
			\draw[-,thick](3,2)--(3,2.1);
			\draw[-,thick](0,1)--(0,1.1);
			\draw[-,thick](3,1)--(3,1.1);
			\draw[-,thick](1.5,1)--(1.5,1.1);
			\draw[-,thick](1.5,2)--(1.5,2.1);
			\draw[-,thick](1.5,3)--(1.5,3.1);
			\node at (-0.8,3){$ s-wave $};
			\node at (-1.2,2){$ d+id-wave $};
			\node at (-1.2,1){$ d-id-wave $};
			\node at (0.75,3.3){$ C=2 $};
			\node at (2.25,3.3){$ C=2 $};
			\node at (0.75,2.3){$ C=2 $};
			\node at (2.25,2.3){$ C=4 $};
			\node at (0.75,1.3){$ C=2 $};
			\node at (2.25,1.3){$ C=0 $};
			\node at (1.5,0.4){$ \mu $};
			\node at (0,0.75){$ -0.1 $};
			\node at (1.5,0.75){$ 0 $};
			\node at (3,0.75){$ 0.1 $};
			\node at (3.8,2.5){$ t = 1 $};
			\node at (3.8,2.0){$ \xi = 0.3 $};
			\node at (3.8,1.5){$ \Delta = 0.03 $};
		\end{tikzpicture}
		\caption{The topological phase diagrams of the superconducting states without SOC are presented. It is worth noting that although the phase diagram is obtained for $ \Delta=0.03 $, it has been verified that the topological phase diagram remains unchanged even when $ \Delta $ ranges from 0.01 to 0.03. Additionally, it should be mentioned that the case $ \mu=0 $ corresponds to the left portion of the phase diagrams. Please note that the region where $ \mu<0 $ is not a superconducting state. We only marked it in this way to indicate the appearance of topological superconductivity in certain parameters.}
		\label{fig.non_soc_phase_map}
	\end{figure}
	Berry curvature, which is one of the most important topological representations, is derived from the Berry phase $ \gamma_n = (1/2)\int_{\mathcal{S}} \mathrm{d}R^{\mu}\land \mathrm{d}R^{\nu} \Omega_{\mu\nu}^n(\mathbf{R}) $. The Berry phase is an observable quantity that is also known as a geometric phase, so it must be gauge-invariant module $ 2\pi $. Therefore, the Berry curvature must be able to be written in a gauge-invariant form \cite{RevModPhys.82.1959}
	
	\begin{equation}\label{key}
		\Omega_{\mu\nu}^n = i \sum_{n'\neq n}\frac{\bra{n}\partial H/\partial R^\mu \ket{n'}\bra{n'}\partial H/\partial R^\nu\ket{n}-(\mu\leftrightarrow \nu)}{\left(\epsilon_n-\epsilon_{n'}\right)^2}\quad ,
	\end{equation}
	where $ \ket{n},\ket{n'} $ are both the eigenstates of the Hamiltonian, and $ \epsilon_n,\epsilon_{n'} $ are the eigenvalues of the Hamiltonain. The Chern number is calculated by integrating the Berry curvature divided by $ 2\pi $, 
	\begin{equation}\label{key}
		C^n=\frac{1}{2\pi}\int_{BZ}\Omega_{k_x,k_y}^n\left(\mathbf{k}\right)\mathrm{d}^2\mathbf{k}\quad ,
	\end{equation}
	where $ C^n $ represents the Chern number for the $ n^{th} $ band. $ BZ $ represents the integral are done inside the Brillouin region, and $ \Omega_{k_x,k_y}^n\left(\mathbf{k}\right) $ represents the Berry curvature respect to the 2-dimensional momentum space. 
	
	When energy degeneracy occurs, the gauge-invariant form of Berry curvature is no longer well-defined, rendering $C^n$ ill-defined as well. However, we can still describe the topological properties of superconducting states through the Chern number $C=\sum_{n\in occ}C^n$. To achieve this, we introduce the pseudo Berry curvature as follows
	\begin{equation}\label{key}
		\Omega_{\mu\nu}^{*n} = i \sum_{n'\notin occ}\frac{\bra{n}\partial H/\partial R^\mu \ket{n'}\bra{n'}\partial H/\partial R^\nu\ket{n}-(\mu\leftrightarrow \nu)}{\left(\epsilon_n-\epsilon_{n'}\right)^2}\quad ,
	\end{equation}
	where $ n\in occ $ and $ n'\notin occ $. Note that $\Omega_{\mu\nu}^{*n}$ is not the Berry curvature, and its integral divided by $2\pi$ is not the Chern number for the $n^{th}$ band. Nevertheless, it can be proven that the Chern number is 
	\begin{equation}\label{formula.chern_number}
		C = \frac{1}{2\pi}\sum_{n\in occ}\int_{BZ}\Omega_{k_x,k_y}^{*n}\left(\mathbf{k}\right)\mathrm{d}^2\mathbf{k}\quad .
	\end{equation}
	where $BZ$ represents the integration over the Brillouin zone, and $\Omega_{k_x,k_y}^{*n}(\mathbf{k})$ represents the pseudo Berry curvature with respect to the 2-dimensional momentum space. (See Appendix \ref{App.C} for a detailed proof.)

	After calculating the Berry curvature, we can determine the quasiparticle transport, particularly the thermal Hall effect which we investigate in this article.

	\begin{figure*}[t]
		\centering
		\subfigure[]
		{
			\includegraphics[width=1.64in]{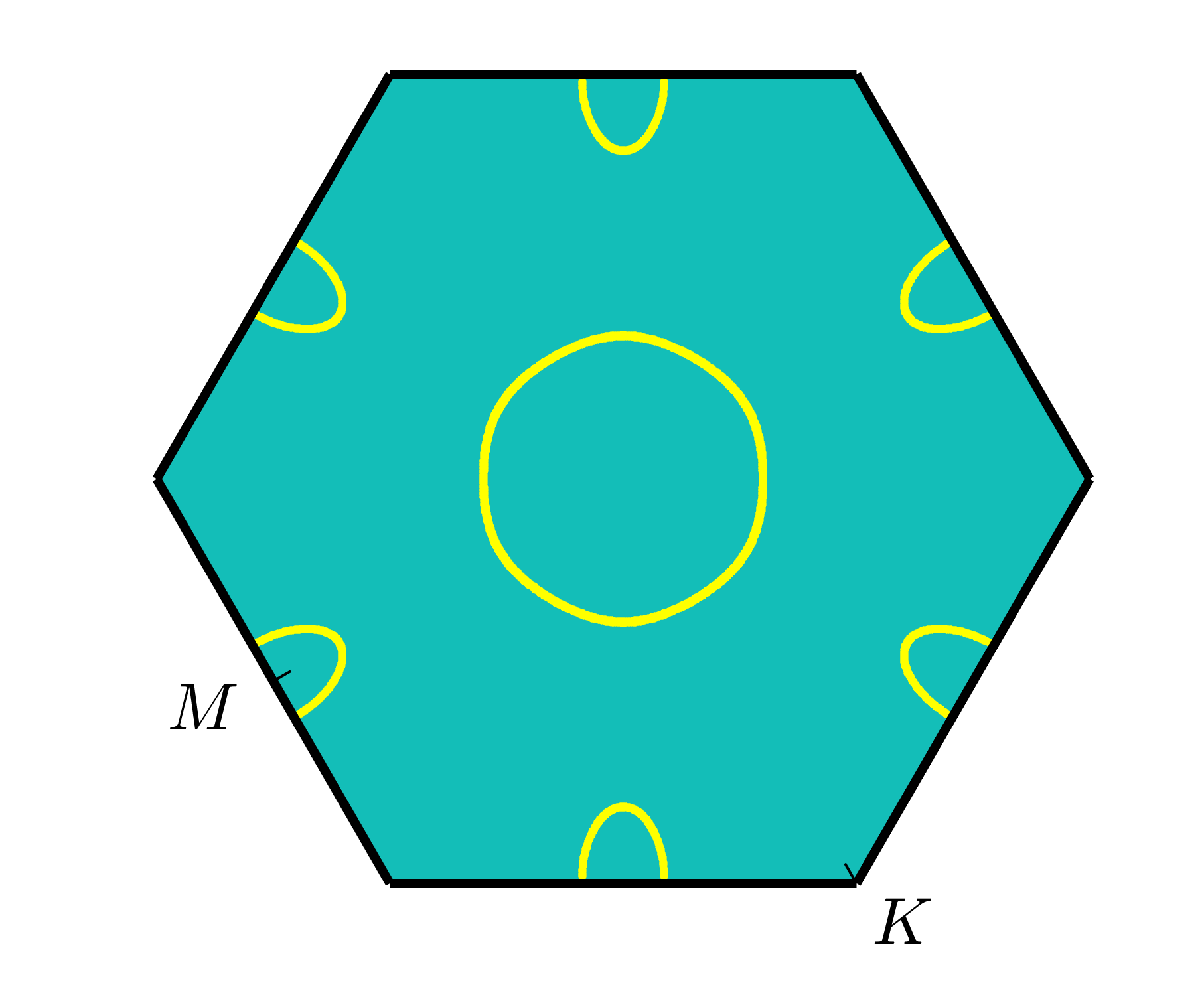}
			\label{fig.Ferimi_0_0.3_0.1}
		}
		\subfigure[]
		{
			\includegraphics[width=1.64in]{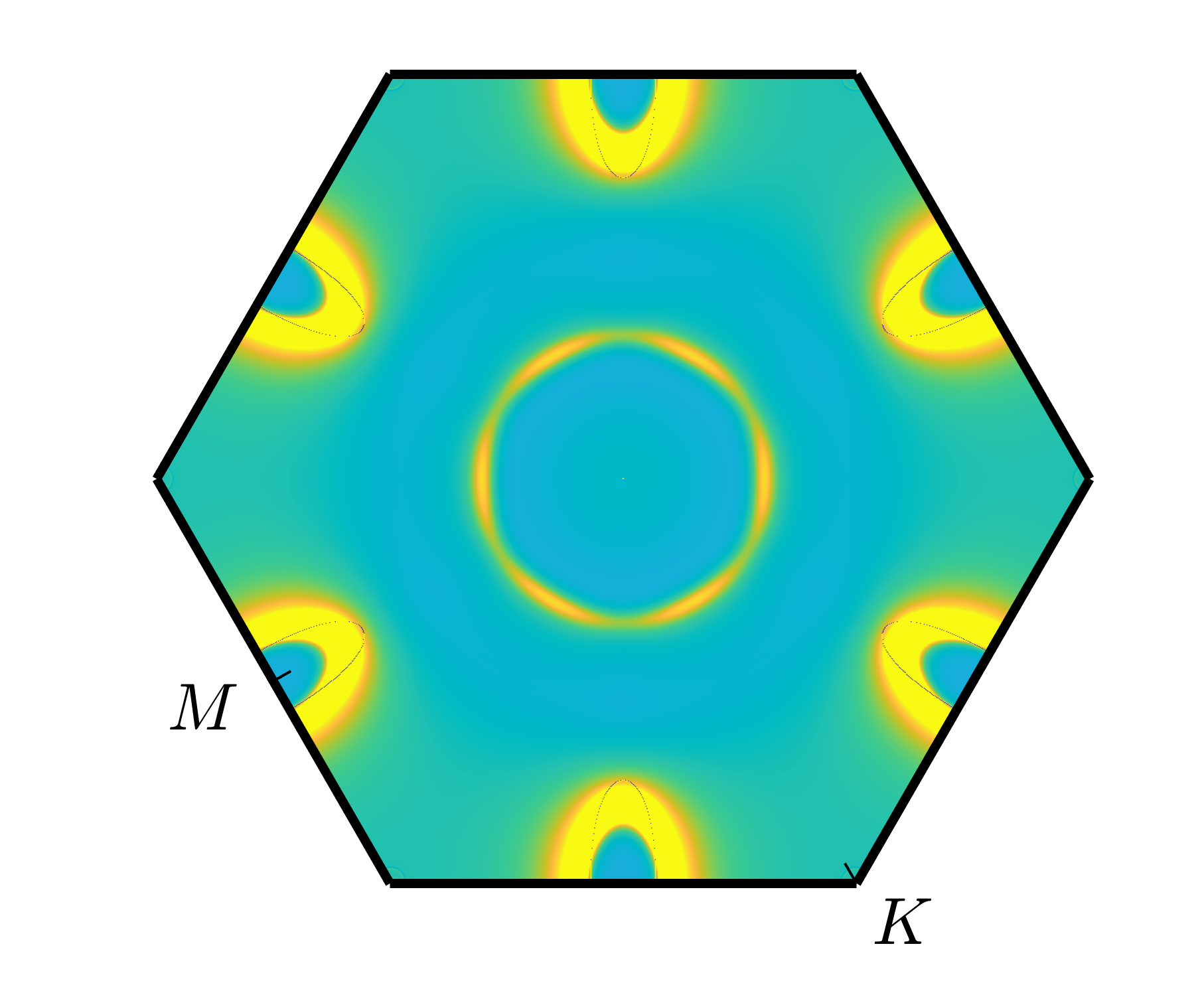}
			\label{fig.Berry_S_30_3_10}
		}
		\subfigure[]
		{
			\includegraphics[width=1.64in]{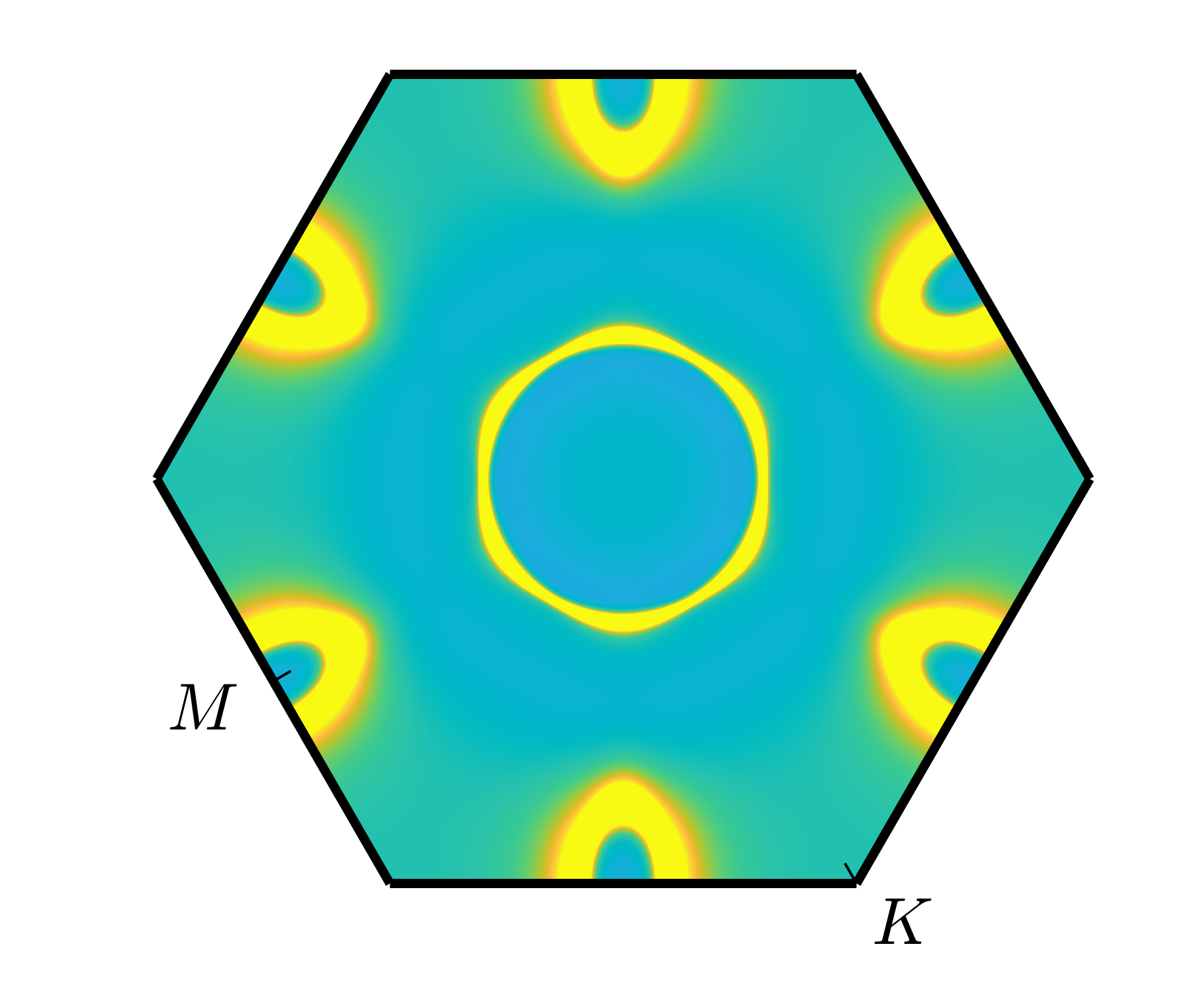}
			\label{fig.Berry_Dp_30_3_10}
		}
		\subfigure[]
		{
			\includegraphics[width=1.64in]{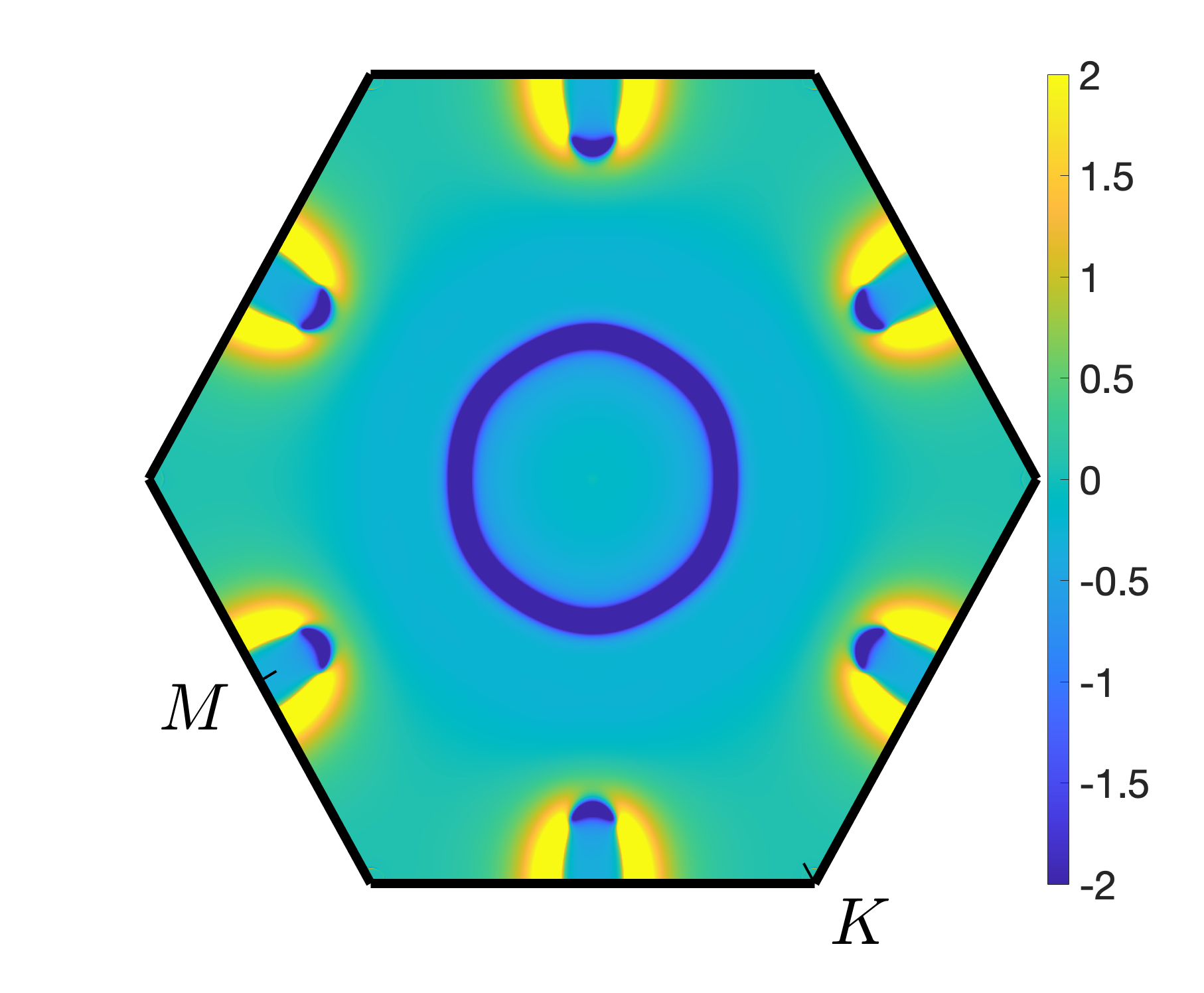}
			\label{fig.Berry_Dm_30_3_10}
		}
		\subfigure[]
		{
			\includegraphics[width=1.64in]{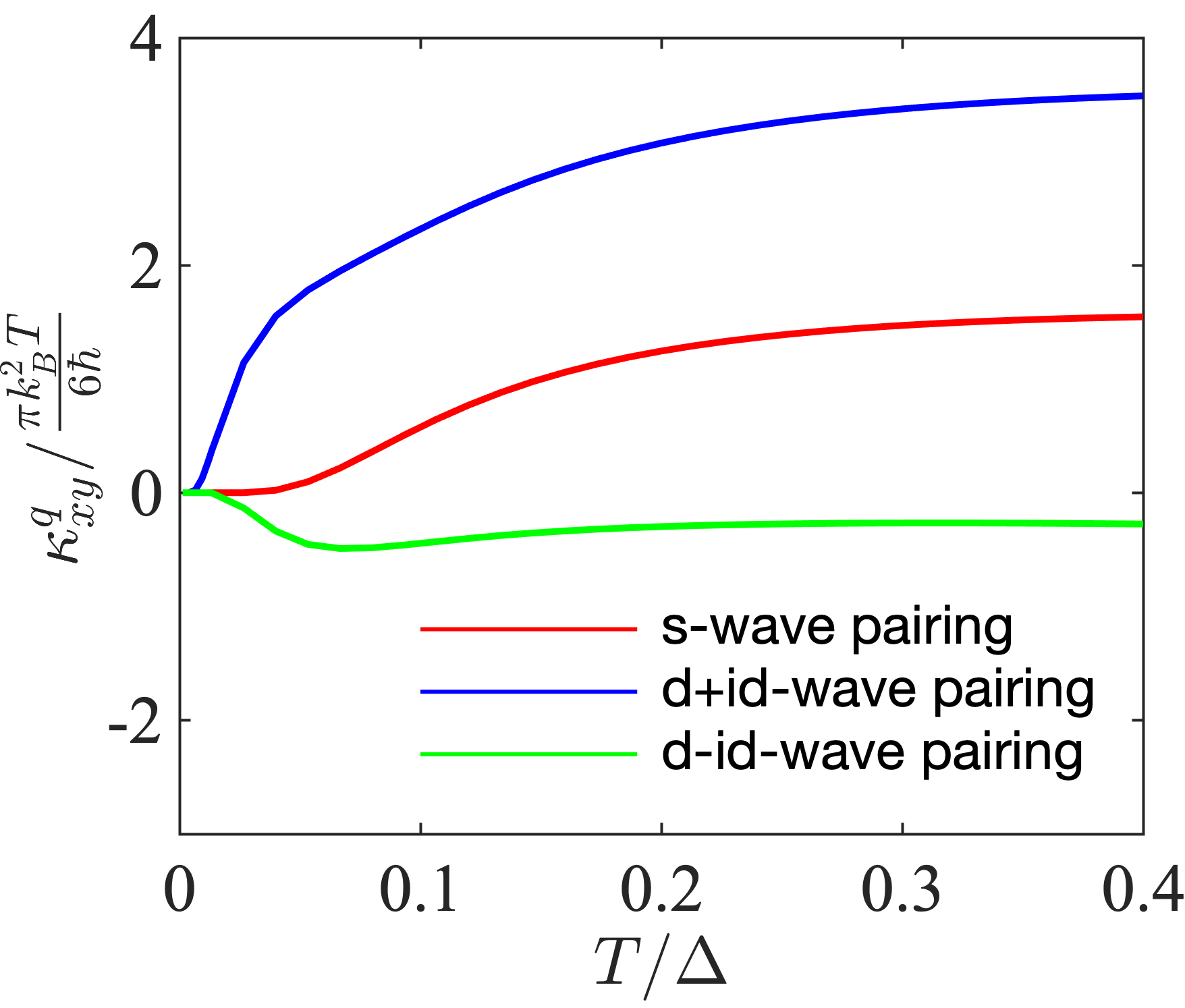}
			\label{fig.SC_30_3_10}
		}
		\subfigure[]
		{
			\includegraphics[width=1.64in]{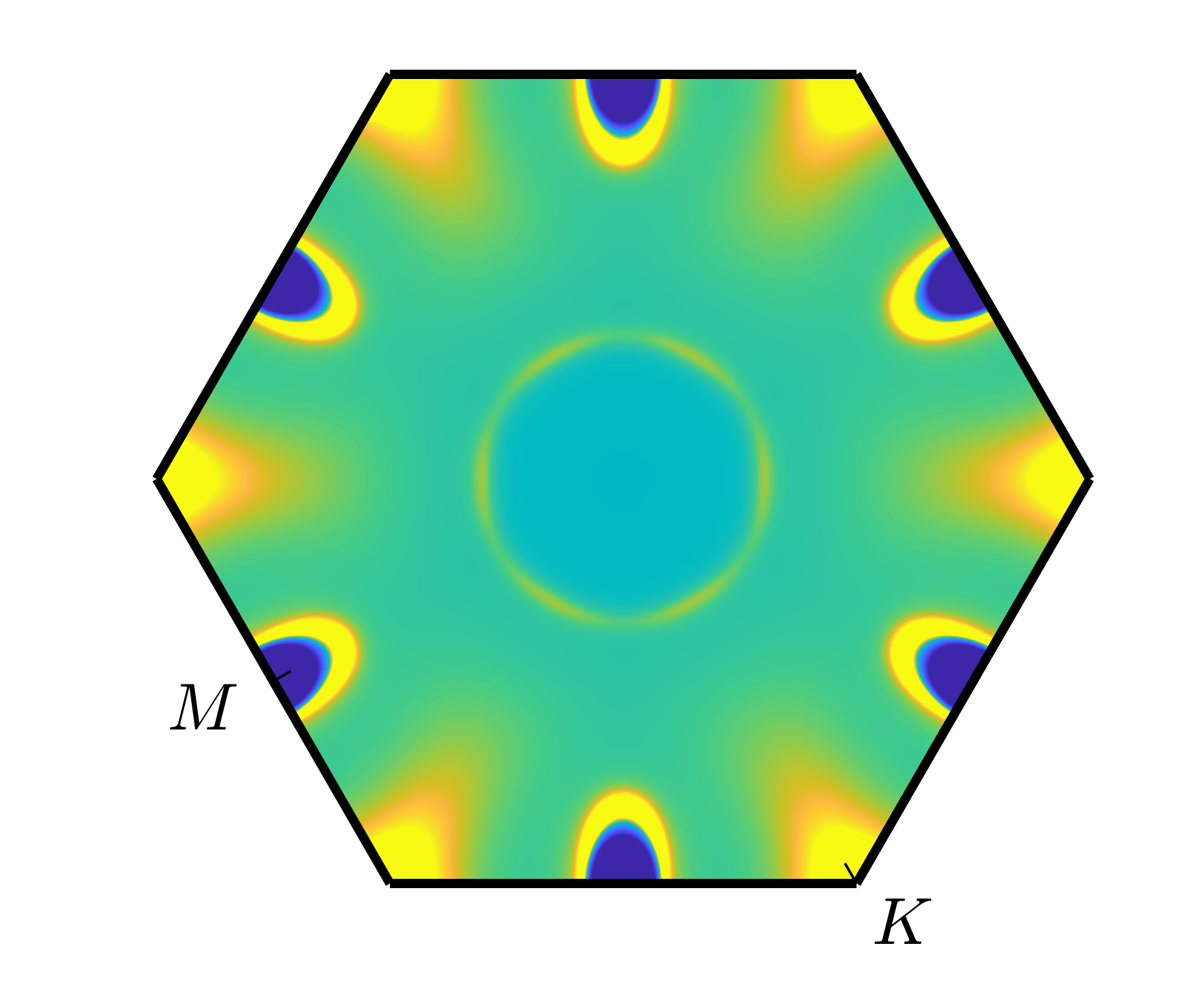}
			\label{fig.Berry13_S_30_3_10}
		}
		\subfigure[]
		{
			\includegraphics[width=1.64in]{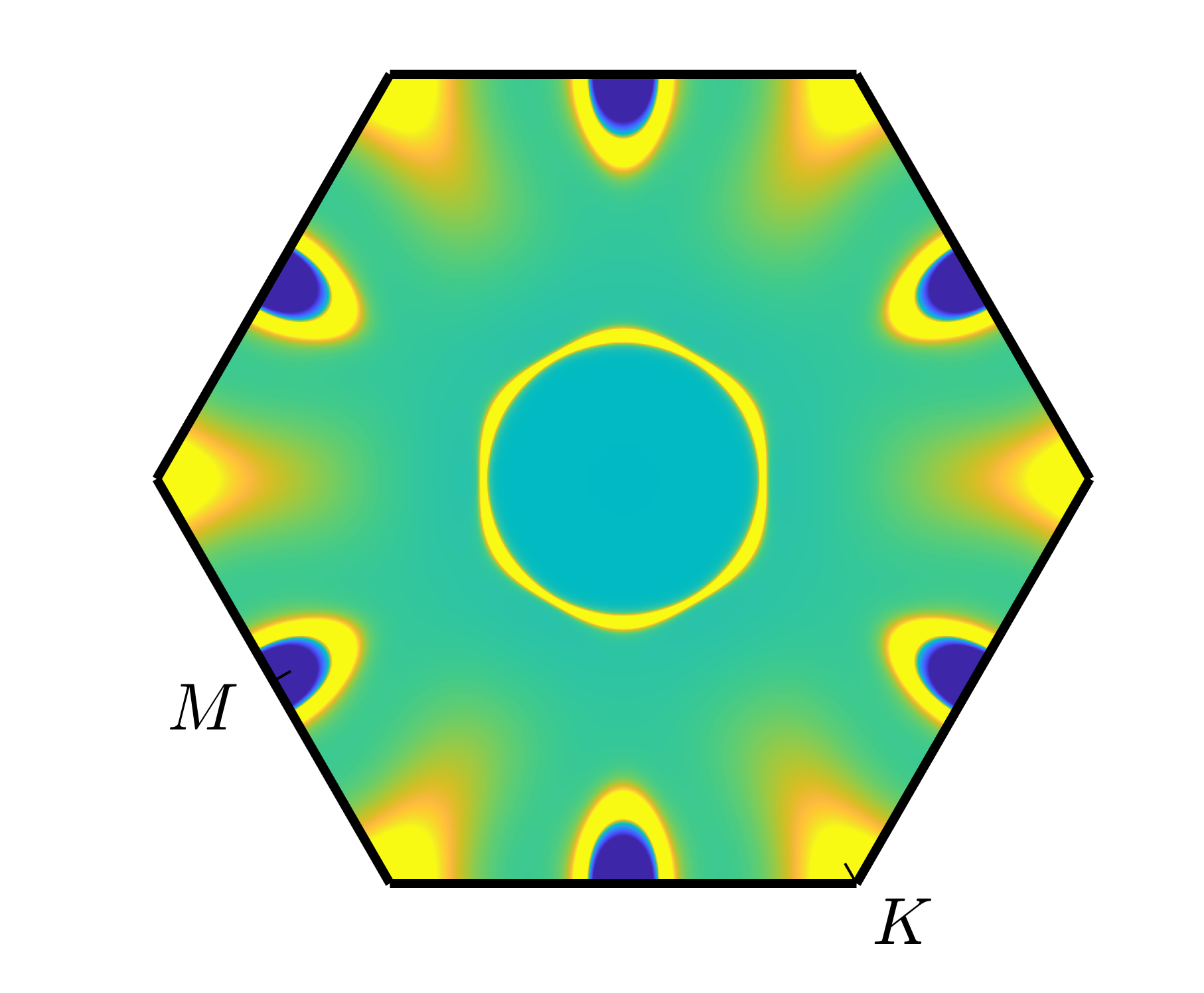}
			\label{fig.Berry13_Dp_30_3_10}
		}
		\subfigure[]
		{
			\includegraphics[width=1.64in]{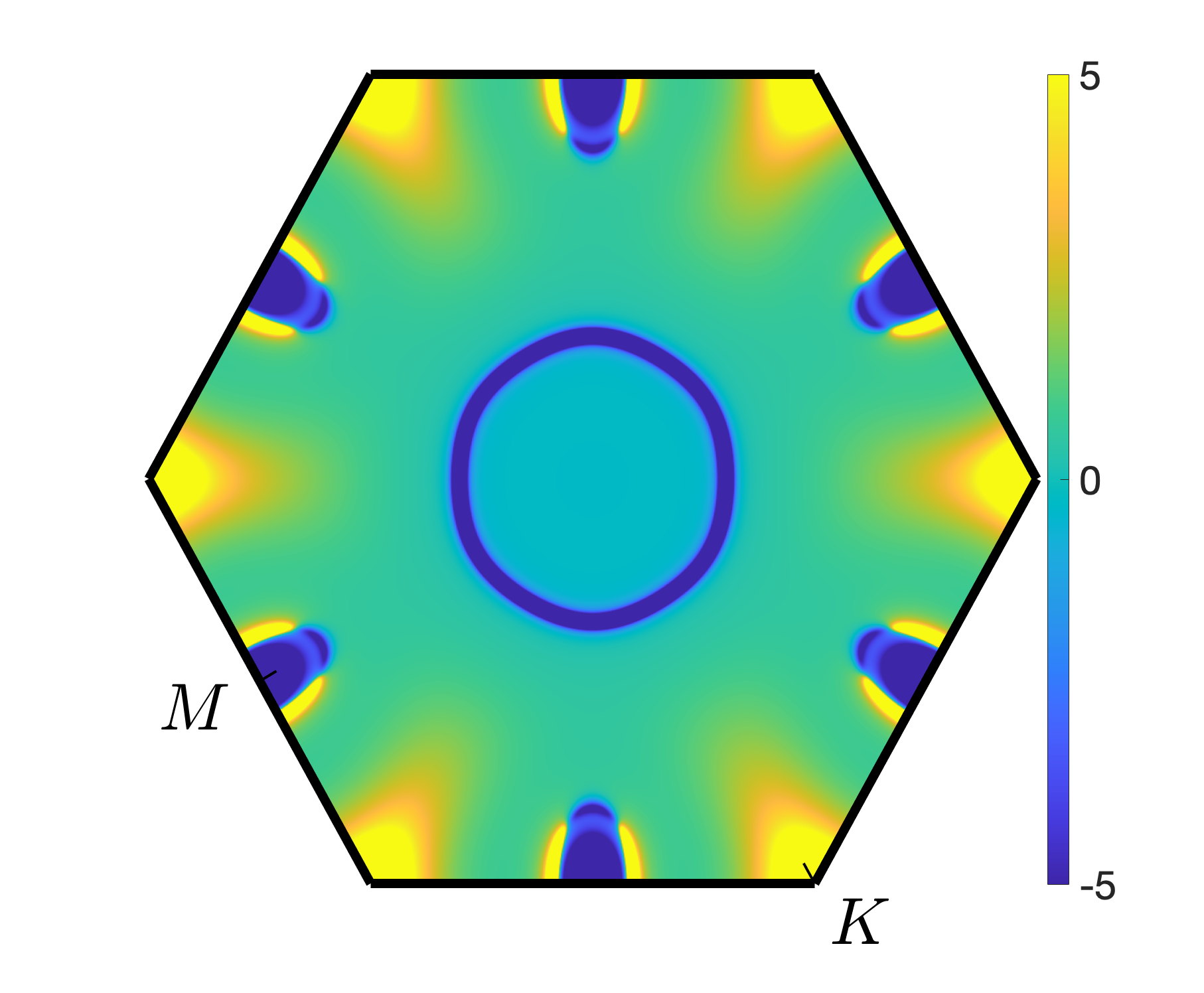}
			\label{fig.Berry13_Dm_30_3_10}
		}
		\caption{(a) Fermi surface of the model, which mainly distributes around $ \Gamma $ point and beside $ M $ points. [(b-d)] The summation of Berry curvature of the all occupied bands for $ s $-wave, $ d $+i$ d $-wave, $ d $-i$ d $-wave pairings, which represent the topological properties and relate to Chern number. (e) Thermal Hall conductivity curves for $ s $-wave, $ d $+i$ d $-wave and $ d $-i$ d $-wave pairing. We can tell the differences of curves of the three pairing symmetry without considering the quantitative features. [(f-h)] Berry curvature of the 13th band for $ s $-wave, $ d $+i$ d $-wave and $ d $-i$ d $-wave pairing, which connect tightly to the properties of thermal Hall conductivity curves. All of the results are calculated at the parameters $ (\xi,\Delta,\mu)=(0.3,0.03,0.1) $.}
        \label{fig.3}
	\end{figure*}

	The thermal Hall effect is a crucial observable effect that is possible to distinguish different superconducting pairings. The intrinsic anomalous Hall effect (AHE) is governed by the Berry curvature \cite{RevModPhys.82.1539}, which can also be obtained through semiclassical theory that considers wave-packet dynamics \cite{PhysRevB.53.7010, PhysRevB.59.14915}. However, neither quantum nor semiclassical theory involves the superconductivity that relies on the effective attraction between electrons. Therefore, we employ the semiclassical theory for superconductors \cite{Wang2021Berry} to derive the thermal Hall conductivity given by
	\begin{equation}\label{formula.thermal hall}
		\kappa_{xy}^{q} = \frac{2}{T} \int \frac{\mathrm{d}^2 k}{(2\pi)^2} \left(\boldsymbol\Omega_{\mathbf{k}}\right)_{xy} \int^{\infty}_{E_{\mathbf{k}}}f'(\eta , T) \eta^2\mathrm{d}\eta \quad ,
	\end{equation}
	where we set $ k_B=1 $ for convenience. The factor 2 comes from the spin contribution. $ \boldsymbol{\Omega} _{\mathbf{k}} $ represents the Berry curvature with the subscript $ xy $ indicating the flat where the Hall conductivity locates. $ f(E,T) $ is the Fermi-Dirac distribution and $ f' $ is its derivative with respect to $E$. The zero-temperature thermal Hall conductivity can be written as $ \kappa_0 = \pi C_1k_B^2T/6\hbar $, where $ C_1 $ is the first Chern number of the system. However, the low-temperature Hall conductivity depends on both the Berry curvature and energy band structures. Thus, to understand the low-temperature results, we must combine these two aspects.

	\section{ANALYSIS AND RESULTS WITHOUT SOC}\label{Sec.noSOC}
	
	We present a model that describes superconducting states on a kagome lattice with chiral charge density wave, and we analyze the topological phase diagrams and quasiparticle transport in the absence of SOC. In the absence of SOC, the system retains spin symmetry. The Hamiltonian in the case without SOC can be written as shown below
	\begin{widetext} 
		\begin{eqnarray}  \label{formula.nonsoc}
			\mathcal{H}=
			\begin{pmatrix}
				\mathcal{H}_{TB}\left(\mathbf{k}\right)+\mathcal{H}_{CDW}\left(\mathbf{k}\right)&0&0&\Delta\\
				0&\mathcal{H}_{TB}\left(\mathbf{k}\right)+\mathcal{H}_{CDW}\left(\mathbf{k}\right)&-\Delta&0\\
				0&-\Delta^\dagger&-\left[\mathcal{H}_{TB}\left(-\mathbf{k}\right)+\mathcal{H}_{CDW}\left(-\mathbf{k}\right)\right]^*&0\\
				\Delta^\dagger&0&0&-\left[\mathcal{H}_{TB}\left(-\mathbf{k}\right)+\mathcal{H}_{CDW}\left(-\mathbf{k}\right)\right]^*
			\end{pmatrix},
		\end{eqnarray}  
	\end{widetext}
	where $\Delta$ represents a part of the Hamiltonian for the superconducting term. Without SOC, the Hamiltonian can eliminate the influence of spins and be reduced to half of its original dimensions. One of the reduced matrices comes from the first and fourth row, and the other comes from the second and third row. It can be easily proved that the two matrices are the same by performing
	a unitary transformation with $\sigma_z$. The reduced Hamiltonian can be written as shown below
	\begin{equation}\label{formula.reduced_hamiltonian}
		\mathcal{H}=
		\begin{pmatrix}
			\mathcal{H}_{TB}\left(\mathbf{k}\right)+\mathcal{H}_{CDW}\left(\mathbf{k}\right)&\Delta\\
			\Delta^\dagger&-\left[\mathcal{H}_{TB}\left(-\mathbf{k}\right)+\mathcal{H}_{CDW}\left(-\mathbf{k}\right)\right]^*
		\end{pmatrix},
	\end{equation}
	which is expressed using the basis of $\left(C_{\mathbf{k}1}^\dagger,\cdots,C_{\mathbf{k}12}^\dagger,C_{-\mathbf{k}1},\cdots,C_{-\mathbf{k}12}\right) $.
	
	The topological phase diagram was calculated using Eq. (\ref{formula.chern_number}), as shown in Fig. \ref{fig.non_soc_phase_map}. Although the phase diagram was calculated for the parameters $(\xi,\Delta)=(1,0.3,0.03)$, we verified that the phase diagram is valid for $\Delta\in[0.01,0.03]$, indicating that the superconducting gap does not affect the system's topological properties.

	For $\mu\in [-0.1,0]$, the Chern number for all pairing symmetries is 2. This is because the system becomes an insulator for $\mu\in [-0.1,0]$, and the sum of the Chern numbers of all occupied bands equals 2. Since the Fermi surface is already gapped, the superconducting pairing symmetry does not contribute to the system’s topology, and hence all supercon- ducting pairing symmetries are topologically trivial. However, for $\mu\in \left(\left.0,0.1\right]\right.$, the system becomes a metal, and the Chern number for s-wave pairing remains 2, while the Chern num ber for $ d  $+i$ d  $-wave pairing increases by 2 to become 4, and the Chern number for $ d $ -i$ d  $-wave pairing decreases by 2 to become 0. $ s $-wave superconducting pairing is topologically trivial and does not affect the system’s Chern number. On the other hand, complex $ d $-wave contributes to a Chern number of 2, where $ d $+i$ d $-wave and $ d $-i$ d $-wave possess opposite angular momenta, with the former SC pairing contributing +2 and the latter contributing -2.
	
	Strictly speaking, there are no topological superconducting terms when $\mu\in[-0.1,0]$. That is, when the normal states of the system are insulating, there are no differences in topology among $ s $-wave, $ d $+i$ d $-wave, and $ d $-i$ d $-wave. Our objective is to calculate the thermal Hall conductivity curves, which depend on the system’s topological properties, to distinguish different SC pairings. If there are no differences among the three pairing symmetries, we cannot differentiate them. Hence, we study the parameter regions where $\mu\in \left(\left.0,0.1\right]\right.$, and we take $\mu=0.1$ as an example.

	The Fermi surface of the model at parameters $ (\xi,\Delta,\mu)=(0.3,0.03,0.1) $ is shown in Fig.\ref{fig.Ferimi_0_0.3_0.1}, mainly distributed around the $ \Gamma $ and $ M $ points. The Berry curvatures for the three different pairing symmetry superconducting states are shown in Fig.\ref{fig.Berry_S_30_3_10}-\ref{fig.Berry_Dm_30_3_10}, the systems representing the topology of the whole system. The distribution of Berry curvatures of the three different pairing symmetry states matches that of the Fermi surface because the superconducting terms gap the Fermi surface, making the system fully gapped and topologically nontrivial, while having topological superconducting term. Focusing on the differences among the three pairing states, we can see that the main difference is the Berry curvature on the ringlike regions around the $ \Gamma $ point. Berry curvature on the ringlike region of $ s $-wave and $ d  $+i$ d  $-wave pairings are both positive, but that of $ s $-wave is significantly smaller than that of $ d  $+i$ d  $-wave, while Berry curvature on the ringlike region of $ d  $-i$ d  $-wave pairing is negative. The results make sense because complex $ d $-wave superconducting states contribute a ringlike region of Berry curvature around the $\Gamma$ point in the hexagonal lattice \cite{Wang2021Berry}.
	
	\begin{figure}[t]
		\centering
		\subfigure[]
		{
			\includegraphics[width=1.6in]{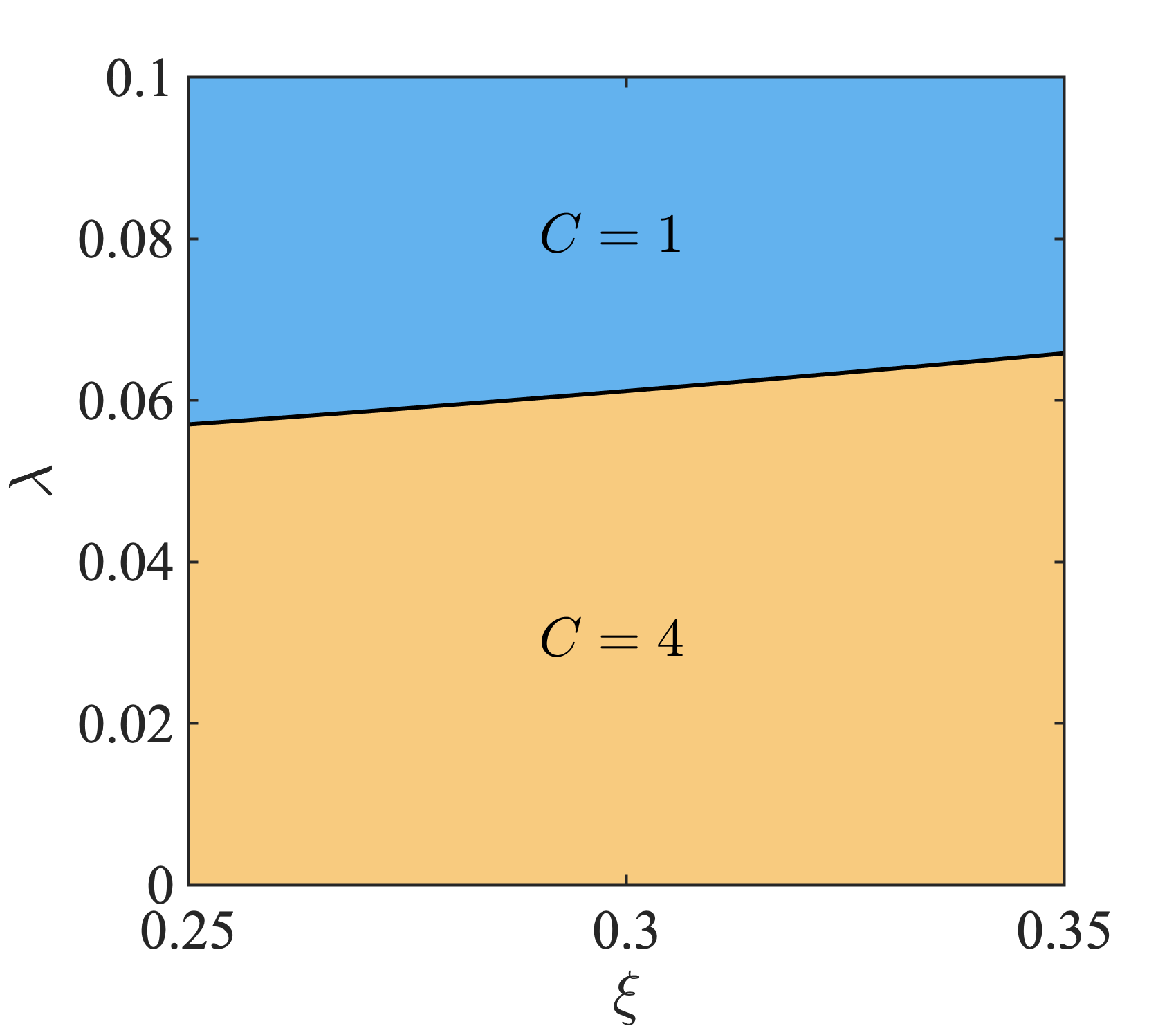}
			\label{fig.PhaseDiagramS}
		}\\
		\subfigure[]
		{
			\includegraphics[width=1.6in]{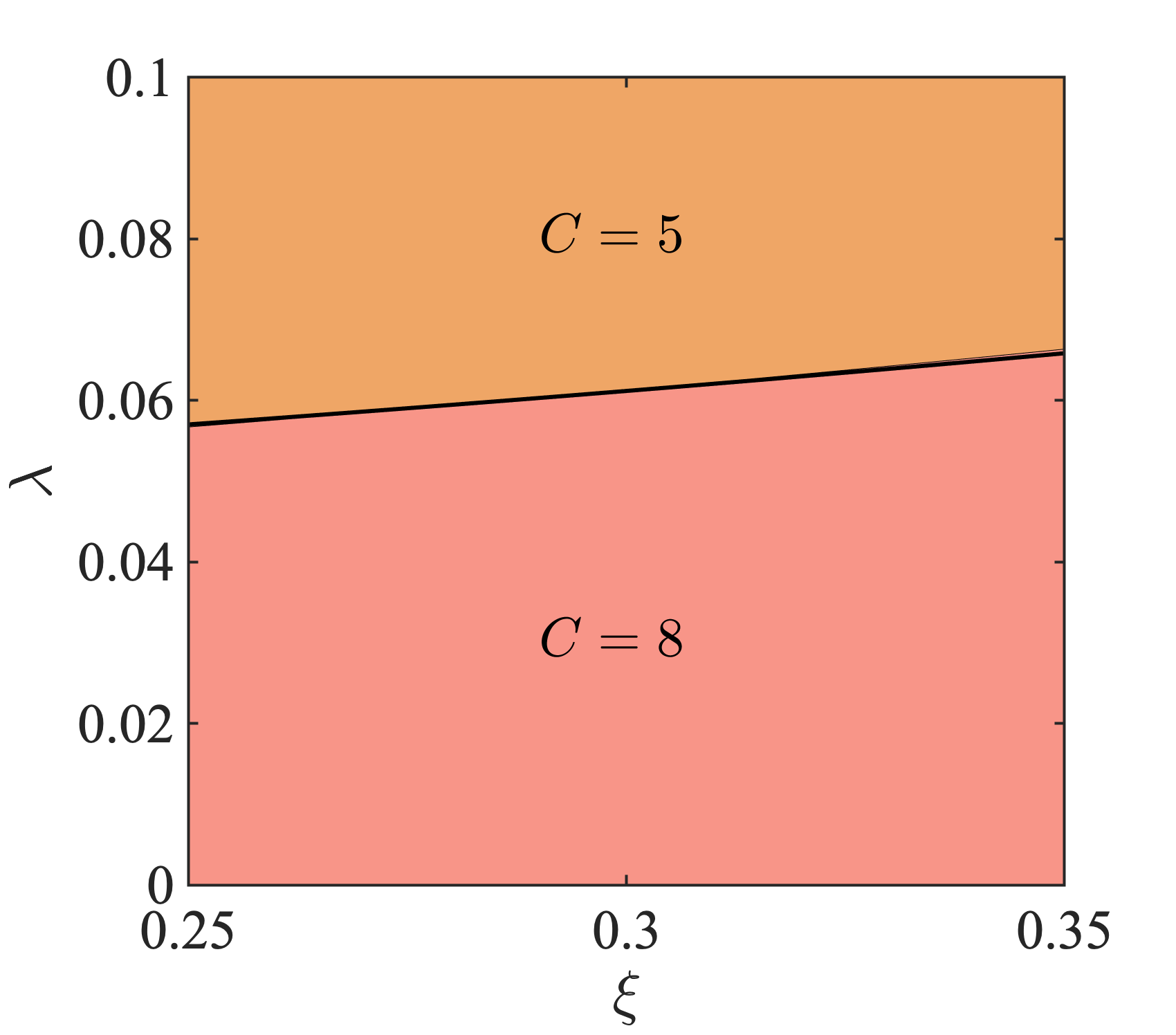}
			\label{fig.PhaseDiagramDp}
		}
		\subfigure[]
		{
			\includegraphics[width=1.6in]{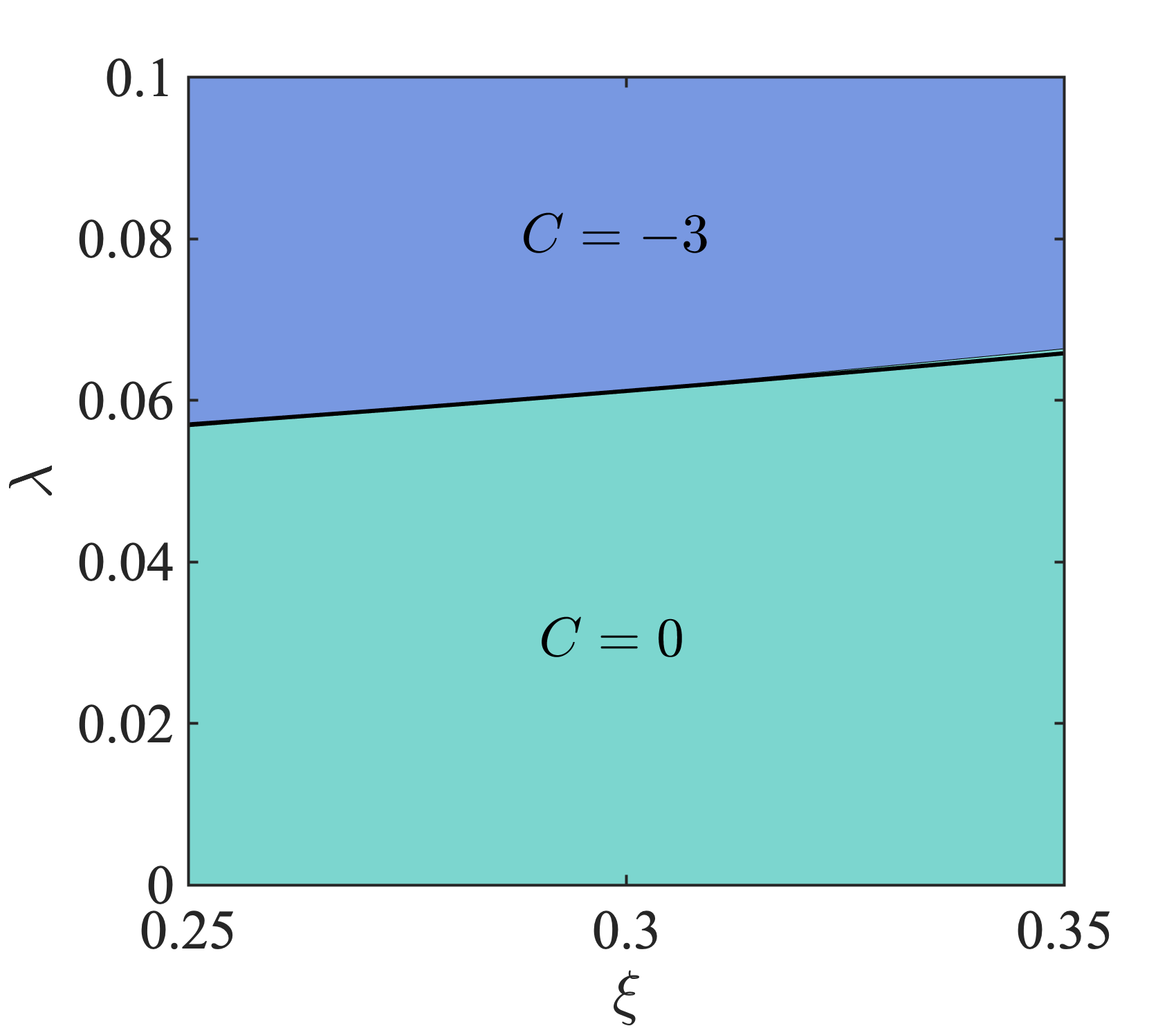}
			\label{fig.PhaseDiagramDm}
		}
		\caption{(a) Topological phase diagram for $ s $-wave pairing SC states. (b) Topological phase diagram for $ d $+i$ d $-wave pairing SC states. (c) Topological phase diagram for $ d $-i$ d $-wave pairing SC states. All of them are calculated at the parameter $ (\Delta,\mu)=(0.03,0.1) $. }
	\end{figure}
	
	\begin{figure*}[t]
		\centering
		\subfigure[]
		{
			\includegraphics[width=1.64in]{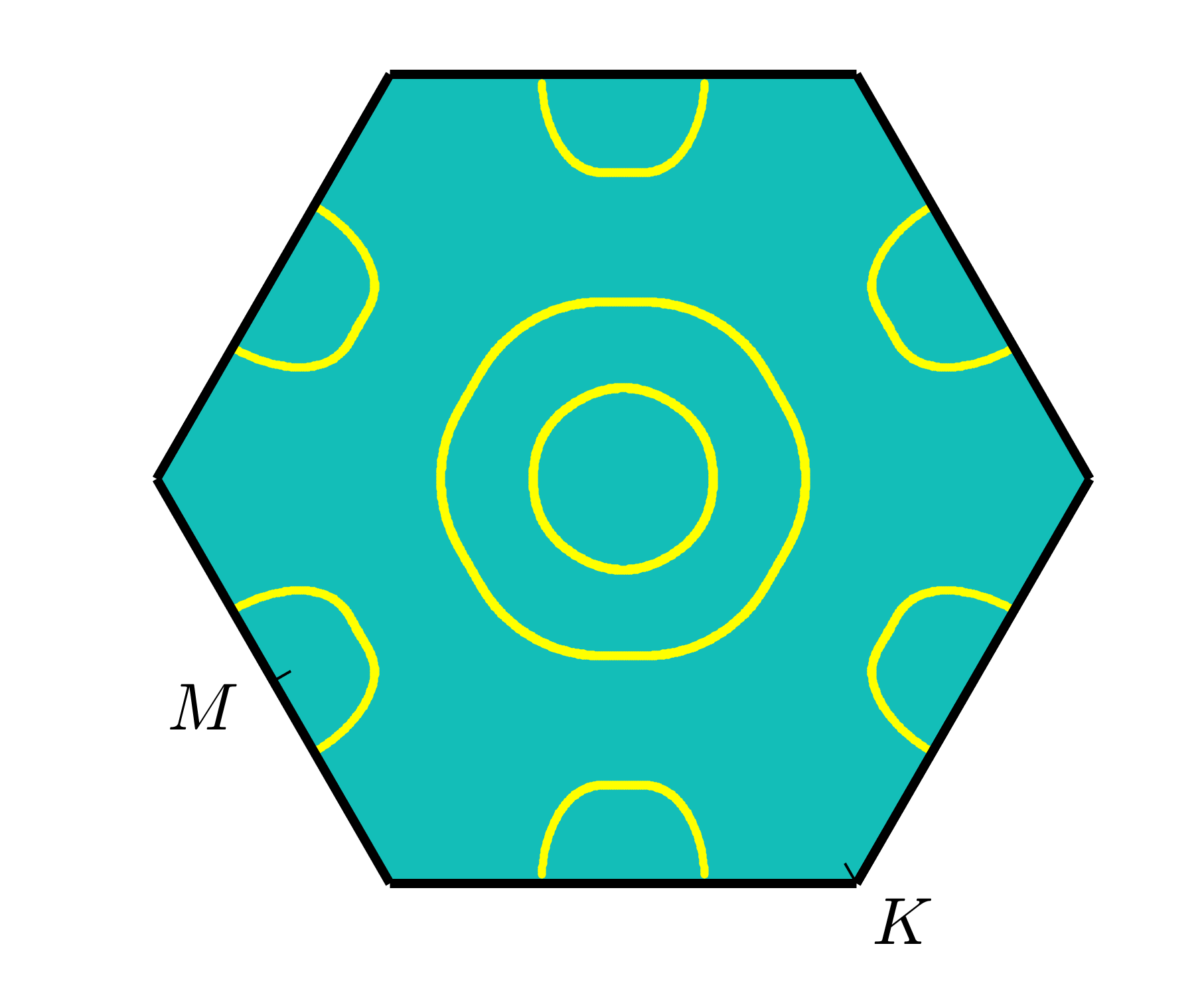}
			\label{fig.Ferimi_0.1_0.3_0.1}
		}
		\subfigure[]
		{
			\includegraphics[width=1.64in]{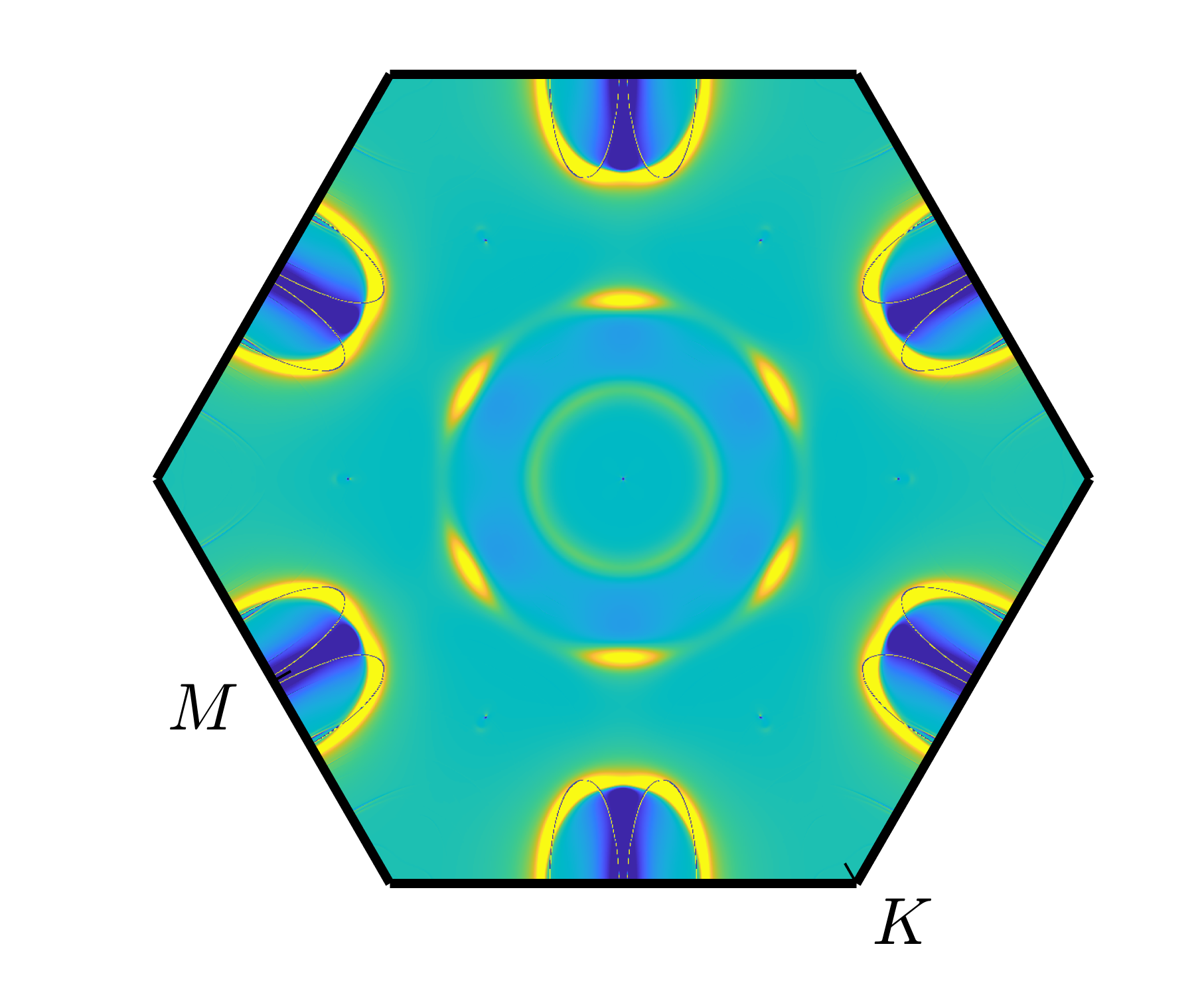}
			\label{fig.Berry_S_10_30_3_10}
		}
		\subfigure[]
		{
			\includegraphics[width=1.64in]{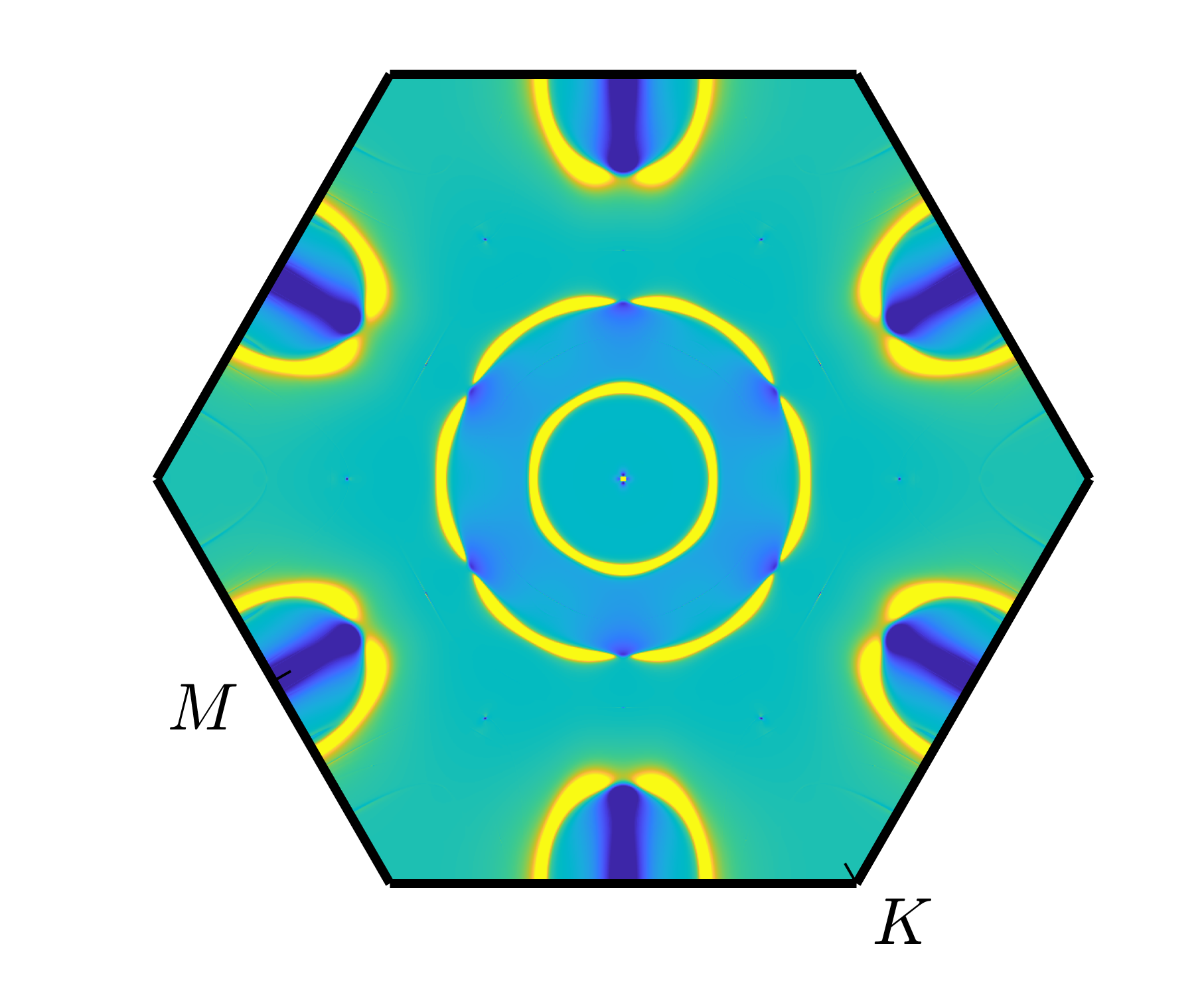}
			\label{fig.Berry_Dp_10_30_3_10}
		}
		\subfigure[]
		{
			\includegraphics[width=1.64in]{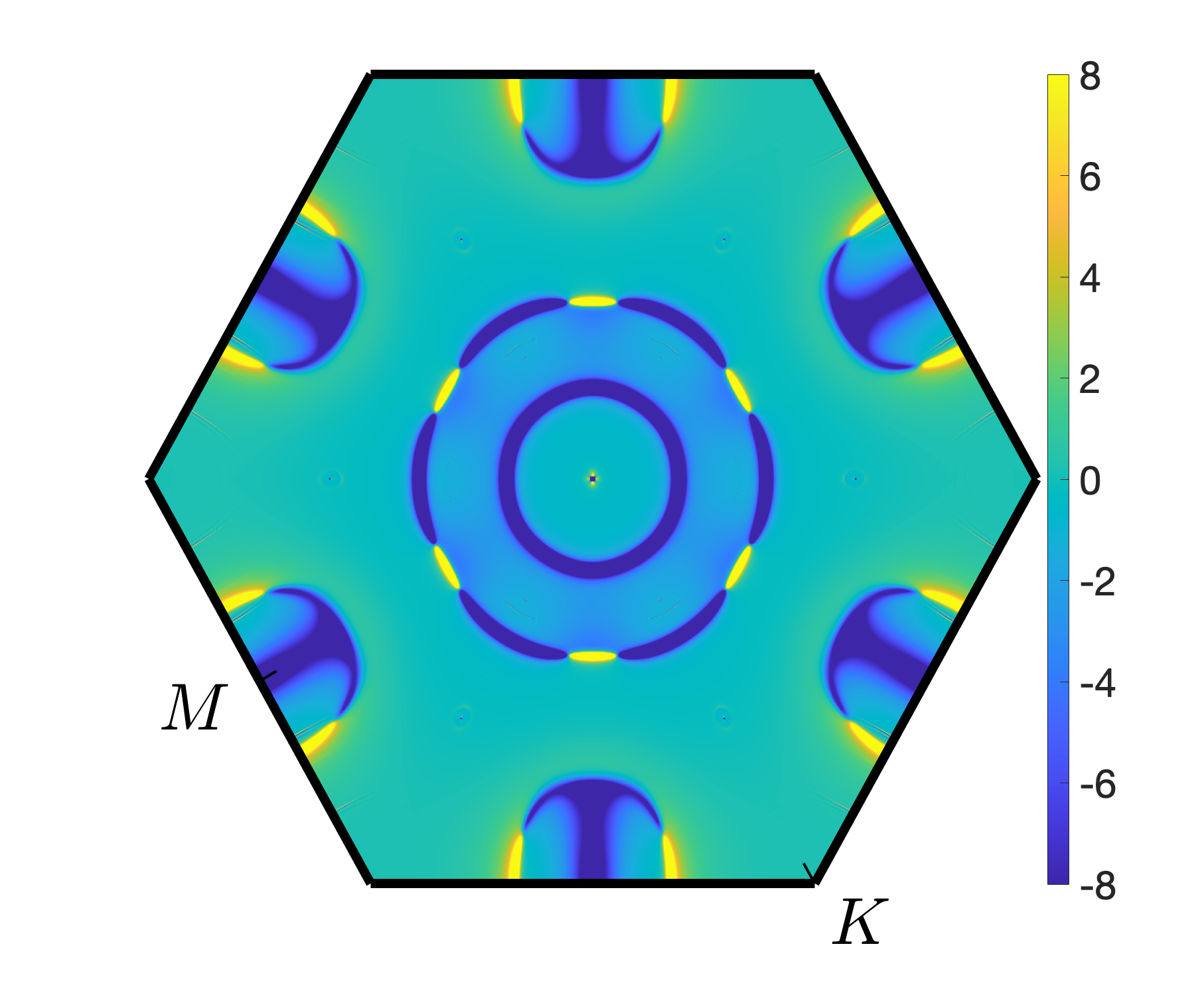}
			\label{fig.Berry_Dm_10_30_3_10}
		}
		\subfigure[]
		{
			\includegraphics[width=1.64in]{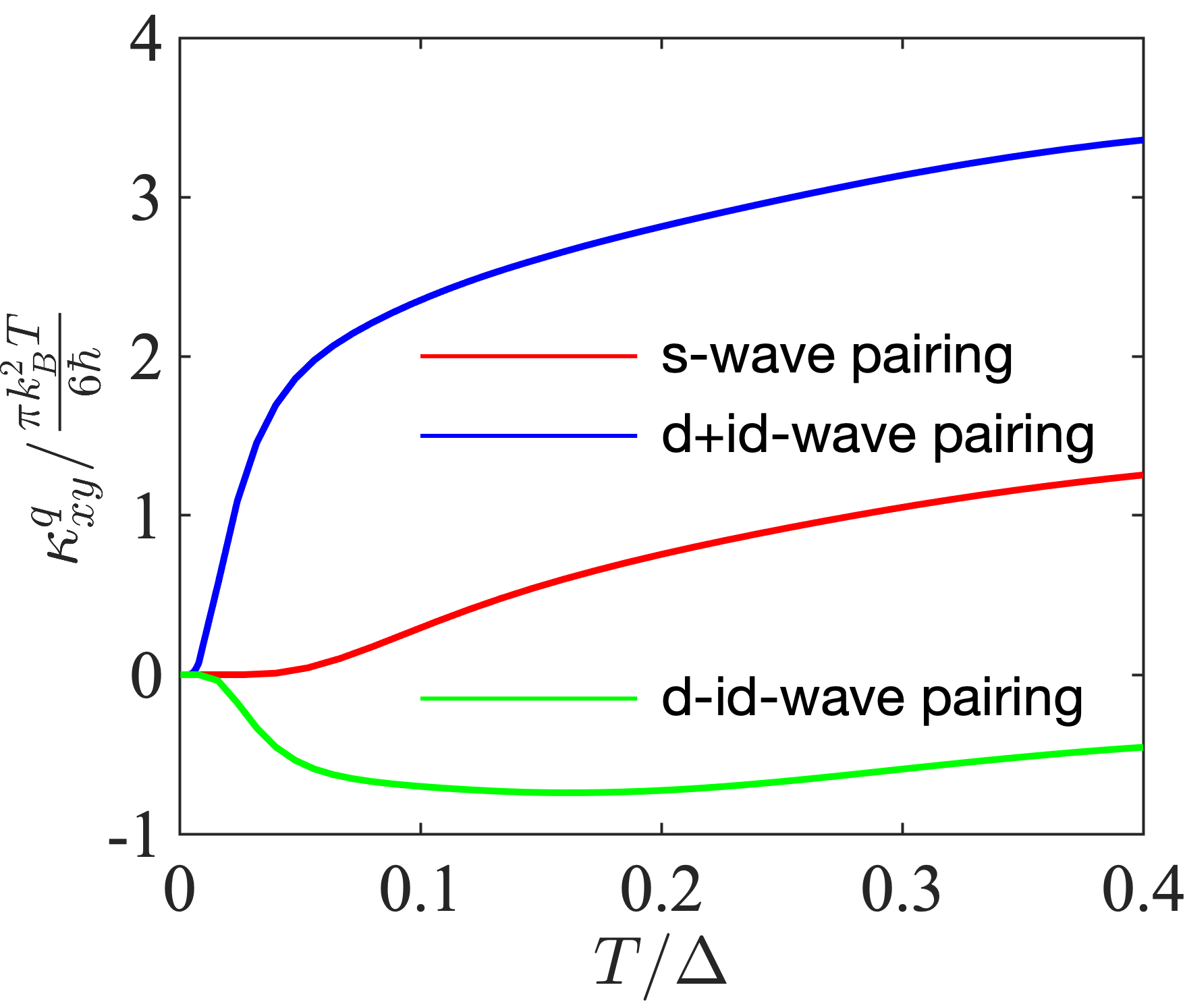}
			\label{fig.SC_10_30_3_10}
		}
		\subfigure[]
		{
			\includegraphics[width=1.64in]{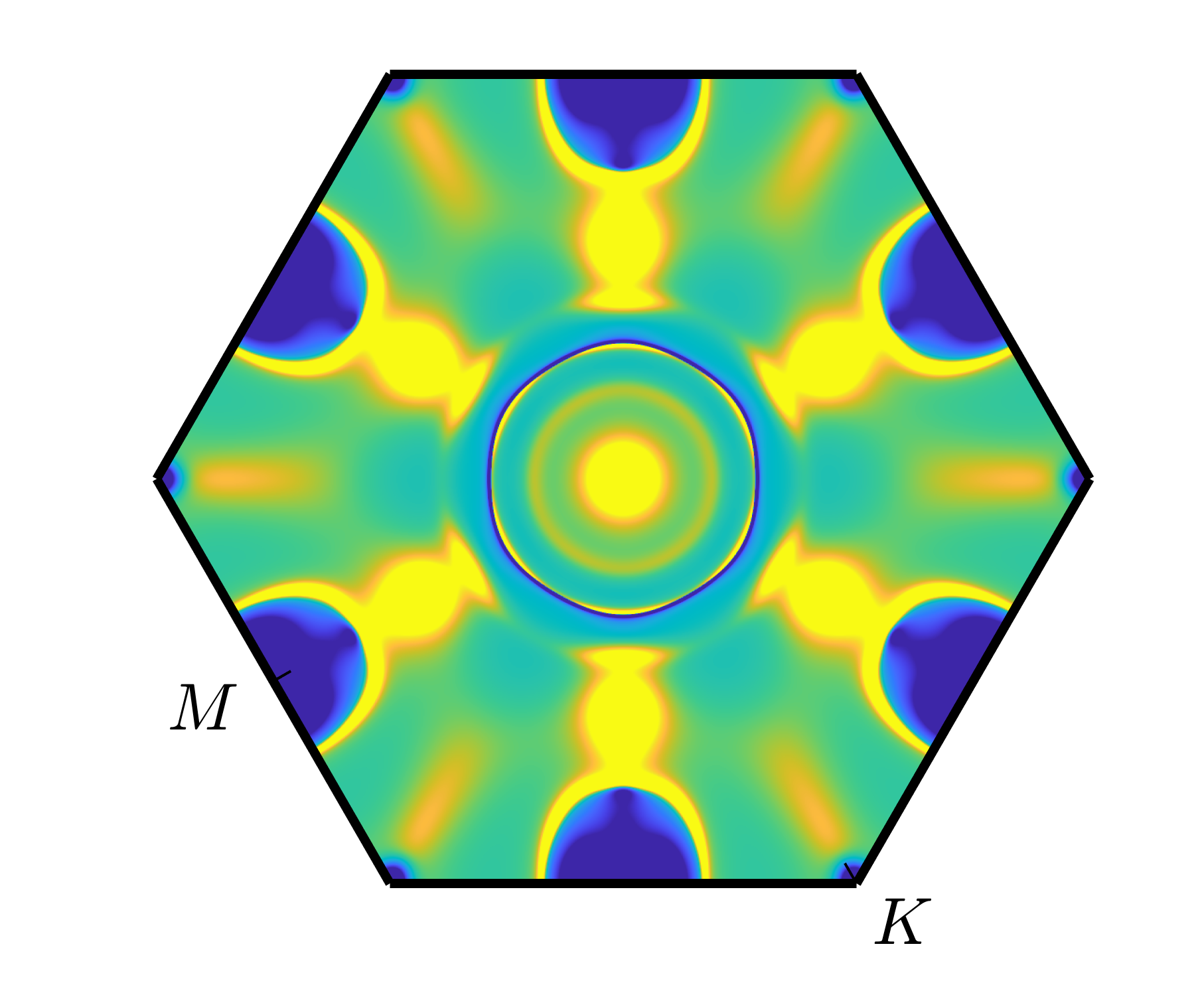}
			\label{fig.Berry25_S_10_30_3_10}
		}
		\subfigure[]
		{
			\includegraphics[width=1.5in]{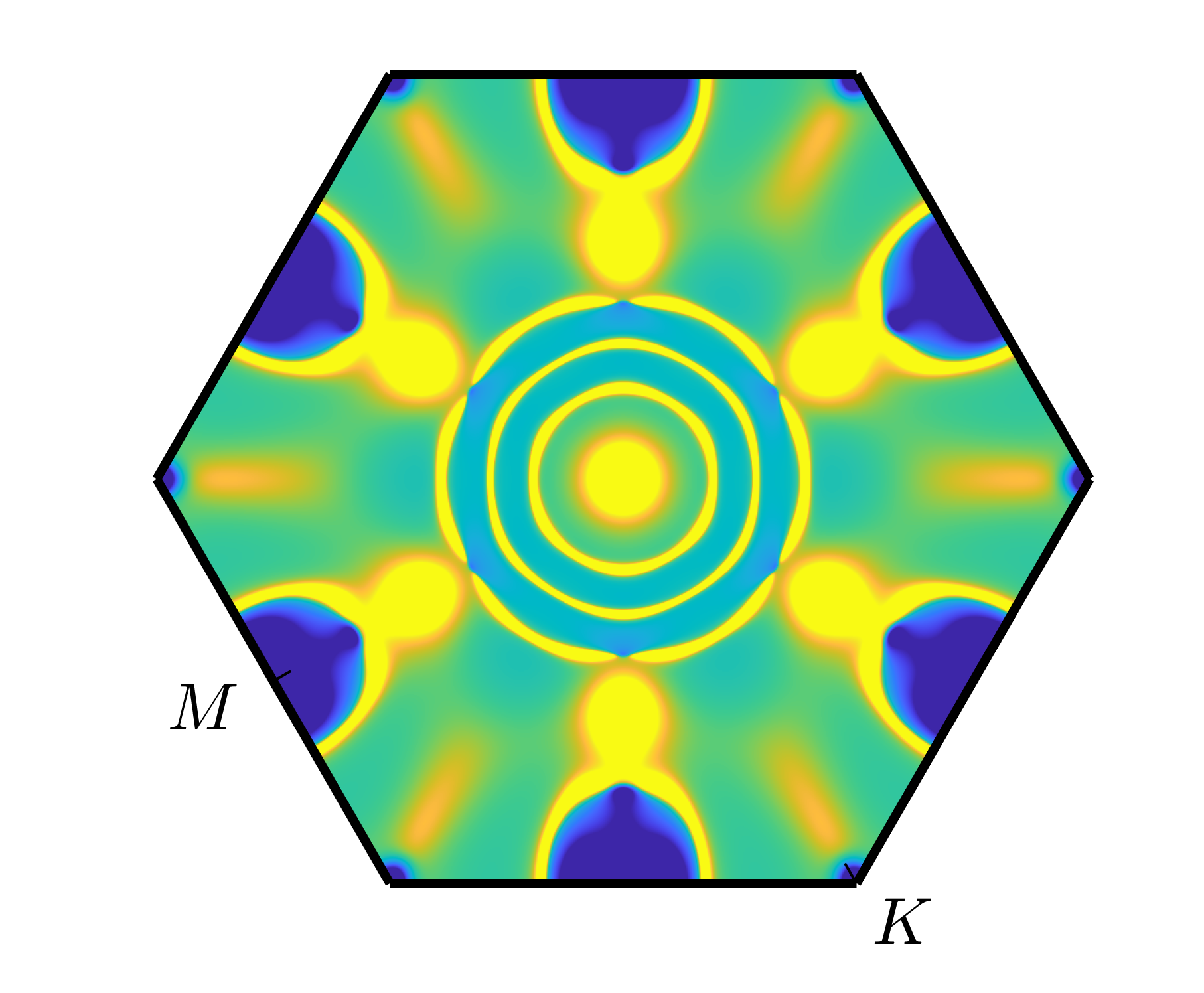}
			\label{fig.Berry25_Dp_10_30_3_10}
		}
		\subfigure[]
		{
			\includegraphics[width=1.5in]{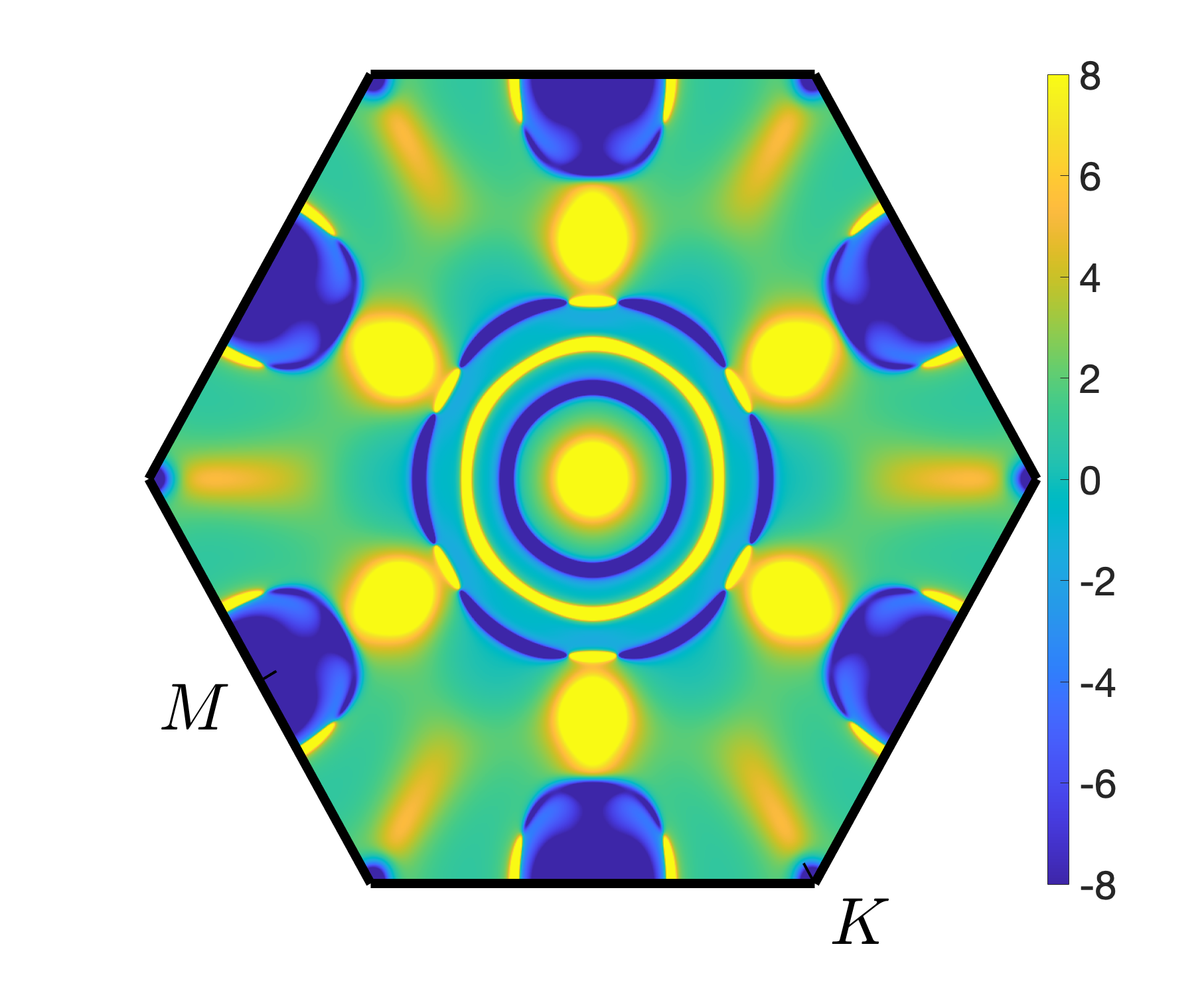}
			\label{fig.Berry25_Dm_10_30_3_10}
		}
		\caption{(a) Fermi surface at the parameter $ (\xi,\lambda,\Delta,\mu)=(0.3,0.1,0.03,0.1) $, where the ringlike part around $ \Gamma $ point is broken into two pieces by SOC term. [(b-d)] Berry curvature related to Chern number for $ s $-wave, $ d $+i$ d $-wave and $ d $-i$ d $-wave pairing symmetry superconducting states, which is the summation of the occupied bands. (e) Thermal Hall conductivity curves at the parameter $ (\xi,\lambda,\Delta,\mu)=(0.3,0.1,0.03,0.1) $. [(f-h)] Berry curvature connectied to thermal Hall conductivity for $ s $-wave, $ d $+i$ d $-wave and $ d $-i$ d $-wave pairing symmetry superconducting states, which belongs to the 25th band calculating from the highest energy band (counting from the highest energy band).}
        \label{fig.5}
	\end{figure*} 
	
	As Eq. (\ref{formula.thermal hall}) states, thermal Hall conductivity depends on Berry curvature. Due to the Fermi-Dirac distribution, the low temperature portion of the thermal Hall conductivity curve is primarily determined by the highest occupied band, which is the band closest to zero energy. Therefore we calculated the Berry curvatures of the highest occupied band [shown in Fig.\ref{fig.Berry13_S_30_3_10}-\ref{fig.Berry13_Dm_30_3_10}] and the thermal Hall conductivity curves [shown in Fig. \ref{fig.SC_30_3_10}]. The qualitative differences in the curves make it easy to distinguish among different pairing symmetry states, which can be inferred from the Berry curvatures of the highest band. The thermal Hall conductivity of the $ s $-wave pairing state, represented by the red curve, is nearly zero near $ 0K $ and gradually increases with temperature, while that of the $ d $+i$ d $-wave pairing state, represented by the blue curve, increases rapidly near $ 0K $. On the other hand, the thermal Hall conductivity of the $ d $-i$ d $-wave pairing state, represented by the green curve, decreases to a negative value as the temperature increases. Although all three types of SC pairing symmetry states have nonzero Berry curvature at the K and M points, they are almost identical at these points. The primary difference among these three states is the ringlike region around the $\Gamma$ point, which we have explained arises from different topological superconducting states. 
	
	Therefore different superconducting states can be qualitatively distinguished by examining the thermal Hall conductivity curves since different pairing symmetry SC states contribute different topological properties, or more precisely, different Berry curvature in the ringlike region around the $ \Gamma $ point.

	\section{ANALYSIS AND RESULTS WITH-SOC}\label{Sec.SOC}
	
	In this section, we investigate the impact of SOC on a model of superconducting states on a kagome lattice with chiral CDW, where the spin symmetry is absent. We analyze two scenarios: (i) the normal state is a metal, such that a small SOC can split the Fermi surface, resulting in a nonzero Chern number and providing different topological superconducting states; (ii) the normal state is an insulator, requiring a large SOC to break the spin symmetry violently. In this case, a Fermi surface will emerge, and the superconducting terms can gap the new Fermi surface and produce topological supercon- ducting states.
	
	When $ \lambda\neq 0 $, the Hamiltonian can no longer be reduced to a block diagonal form, so we must consider the complete Hamiltonian. It is important to note that Fig. \ref{fig.non_soc_phase_map} depicts the phase diagram for the reduced Hamiltonian in Eq. (\ref{formula.reduced_hamiltonian}). However, we must consider Eq. (\ref{formula.reduced_hamiltonian}) as a part of Eq.(\ref{formula.nonsoc}), so the complete Hamiltonian's topological phase diagrams only need to double the Chen number in Fig. \ref{fig.non_soc_phase_map}.
	
	In the first scenario with SOC, we gradually increase the strength coefficient $ \lambda\in [0,0.1] $ in Eq. (\ref{formula.nonsoc}), the Hamiltonian is given by
	\begin{widetext} 
		\begin{eqnarray}
			\mathcal{H}=
			\begin{pmatrix}
				\mathcal{H}_{TB}\left(\mathbf{k}\right)+\mathcal{H}_{CDW}\left(\mathbf{k}\right)&\mathcal{H}_{SOC}^{\uparrow\downarrow}\left(\mathbf{k}\right)&0&\Delta\\
				\mathcal{H}_{SOC}^{\downarrow\uparrow}\left(\mathbf{k}\right)&\mathcal{H}_{TB}\left(\mathbf{k}\right)+\mathcal{H}_{CDW}\left(\mathbf{k}\right)&-\Delta&0\\
				0&-\Delta^\dagger&-\left[\mathcal{H}_{TB}\left(-\mathbf{k}\right)+\mathcal{H}_{CDW}\left(-\mathbf{k}\right)\right]^*&-\mathcal{H}_{SOC}^{\uparrow\downarrow}\left(-\mathbf{k}\right)^*\\
				\Delta^\dagger&0&-\mathcal{H}_{SOC}^{\downarrow\uparrow}\left(-\mathbf{k}\right)^*&-\left[\mathcal{H}_{TB}\left(-\mathbf{k}\right)+\mathcal{H}_{CDW}\left(-\mathbf{k}\right)\right]^*
			\end{pmatrix},
		\end{eqnarray}  
	\end{widetext}
	where $ \mathcal{H}_{SOC}\left(\mathbf{k}\right) $ depends on the strength coefficient $ \lambda $. We choose a small $ \lambda $ to observe how SOC breaks the spin symmetry and splits the Fermi surface shown in Fig. \ref{fig.Ferimi_0_0.3_0.1} more clearly.
	 
	We have computed phase diagrams at the fixed parameter values $ (\Delta,\mu)=(0.03,0.1) $, which are presented in Fig. \ref{fig.PhaseDiagramS}-\ref{fig.PhaseDiagramDm}. To identify the topological phase transition boundaries, we observed that the gap closing alwaysTo identify the topological phase transition boundaries, we observed that the gap closing always accompanied the topological phase transition. The phase transition boundary is discerned as a diagonal slash from the bottom left to the top right, indicating that it is more challenging to gap the Fermi surface for larger values of $ \xi $. Furthermore, we note that the phase transition is primarily driven by the SOC term. By increasing the value of $ \lambda $ from $ 0 $ to $ 0.1 $, the SOC term re-gaps the Fermi surface and causes a reduction of Chern number by 3. For our analysis of the impact of the SOC term on the topological properties and thermal Hall conductivity of the system, we chose the parameter values $(\xi,\lambda,\Delta,\mu)=(0.3,0.1,0.03,0.1)$.
	
	In Fig. \ref{fig.Ferimi_0.1_0.3_0.1} and Fig. \ref{fig.Berry_S_10_30_3_10}-\ref{fig.Berry_Dm_10_30_3_10}, we present the Fermi surface and the Berry curvatures of the summation of the occupied bands related to the Chern numbers for the three pairing symmetry SC states. In the ringlike region around the $ \Gamma $ point, which is contributed by the SOC term, both the Fermi surface and the Berry curvature are split into two pieces. Furthermore, we observe changes in the shapes of the Berry curvature around the $ M $ points.

	The thermal Hall conductivity curves for three different SC states with distinct pairing symmetries are displayed in Fig. \ref{fig.SC_10_30_3_10}. It is observed that, despite the variation of $ \lambda $ from $ 0 $ to $ 0.1 $, the curves exhibit little change. This is due to two factors that are consistent with the scenario without SOC. First, the changes in the Chern numbers of all three SC states with distinct pairing symmetries are identical, which implies that the SOC term’s impact on the topological properties is uniform. Second, the Berry curvature, which is crucial for determining the thermal Hall conductivity, is illustrated in Fig. \ref{fig.Berry25_S_10_30_3_10}-\ref{fig.Berry25_Dm_10_30_3_10}. Although the pattern of Berry curvature in the presence of SOC is more intricate than that without it, the principal differences among the three SC states are found in the ringlike region around the $ \Gamma $ point. To be specific, there are only two rings that differ in the ringlike region of the Berry curvature shown in Figs. \ref{fig.Berry25_S_10_30_3_10}-\ref{fig.Berry25_Dm_10_30_3_10} because the Fermi surface [Fig.\ref{fig.Ferimi_0.1_0.3_0.1}] is only divided into two pieces in this region. The rings of $ s $-wave pairing are fainter, which means that the Berry curvature here is small, whereas the rings of $ d $+i$ d $-wave pairing are significantly thicker. In contrast, the Berry curvature in the rings of $ d $-i$ d $-wave pairing is negative.  {The explanation of the similiarity between Fig. \ref{fig.SC_30_3_10} and Fig. \ref{fig.SC_10_30_3_10} is given in Appendix \ref{App.E}.}
	
	Consequently, the differences in the thermal Hall conduc- tivity curves arise from the distinct pairing symmetry SC states rather than from the SOC term, which is the same as the situation in the absence of SOC. It is noteworthy that the strength of the SOC term is relatively small compared to the strength of the CDW and SC terms and can therefore be treated as a perturbation. Thus, in this case, the differences in the thermal Hall conductivity curves are protected by topology provided by the topological SC state.

	Up to this point, we have focused on the scenario where $ \mu=0.1 $, in which case the system is a superconductor even if $ \lambda=0 $, due to the fact that the normal state without SOC term $ \hat{H}_{TB}+\hat{H}_{CDW} $ is metallic. Therefore the breaking of spin symmetry in a superconducting state by the SOC term can be considered as a perturbation.
	
	In contrast, when $ \mu\le0 $ in the absence of SOC, the normal state is an insulator, and thus it cannot be considered a super- conducting state. However, if we examine the situation where $ \lambda $ is large enough to shift at least one band across the Fermi energy level, the system can undergo a phase transition and become a metallic state. In such a scenario, topological super- conducting states may exist as well. We wish to investigate whether it is possible to differentiate among various pairing symmetry SC states based on their thermal Hall conductivity curves in this context. For the sake of convenience, we will assume $ \mu=0 $ as an illustrative example.

\begin{figure}[t]
		\centering
		\subfigure[]
		{
			\includegraphics[width=1.6in]{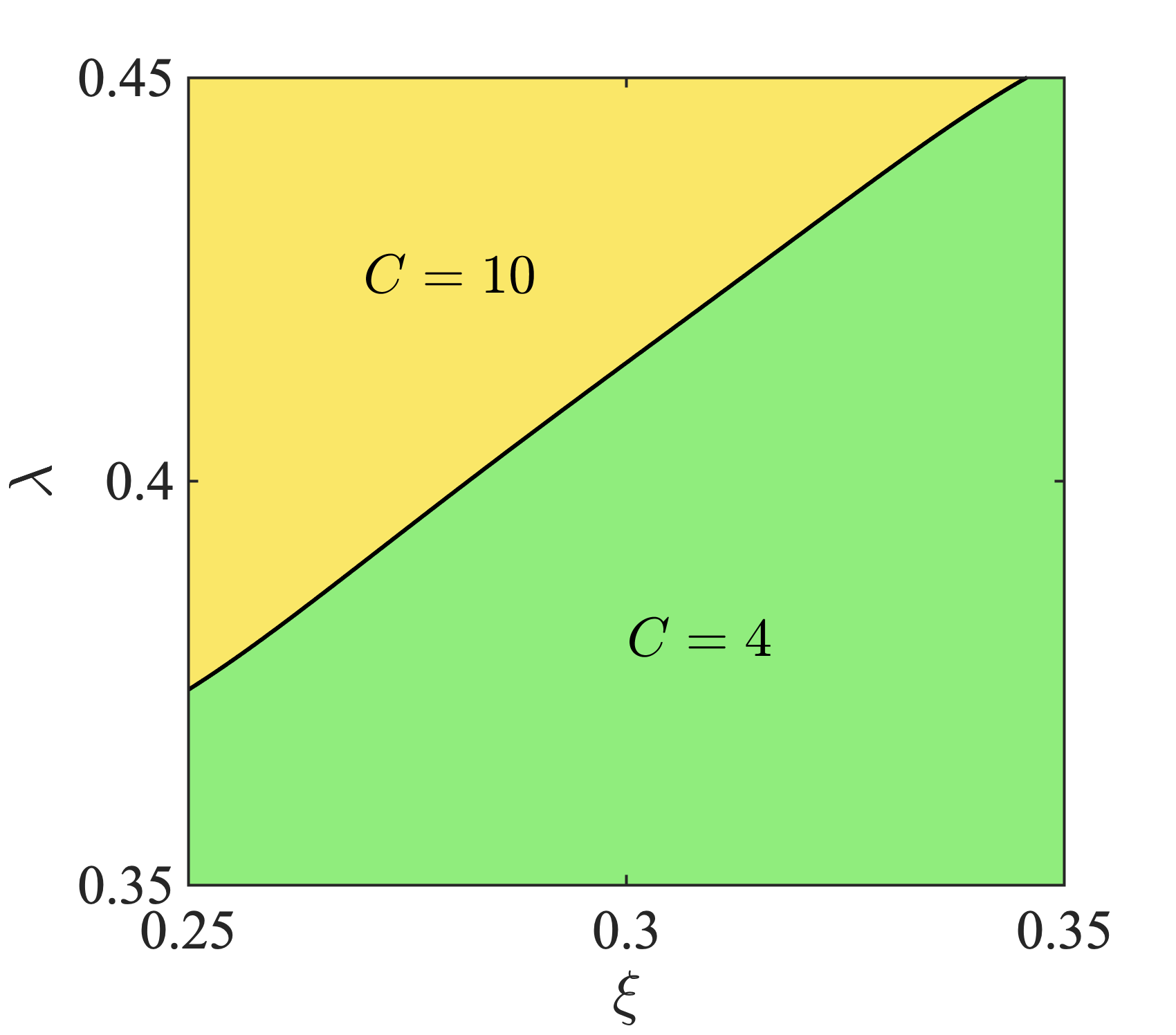}
			\label{fig.PhaseDiagramS_40_30_3_0}
		}\\
		\subfigure[]
		{
			\includegraphics[width=1.6in]{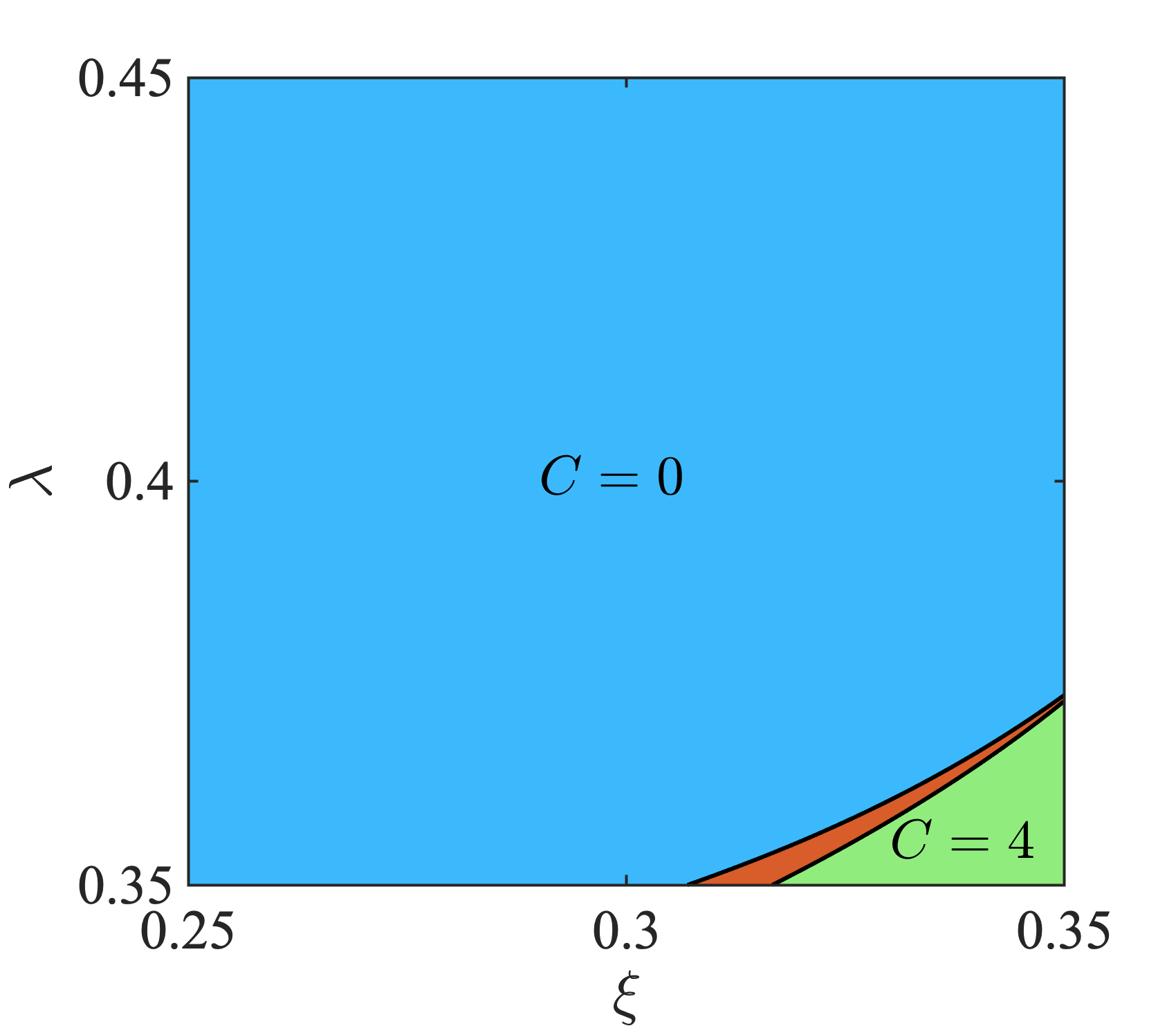}
			\label{fig.PhaseDiagramDp_40_30_3_0}
		}
		\subfigure[]
		{
			\includegraphics[width=1.6in]{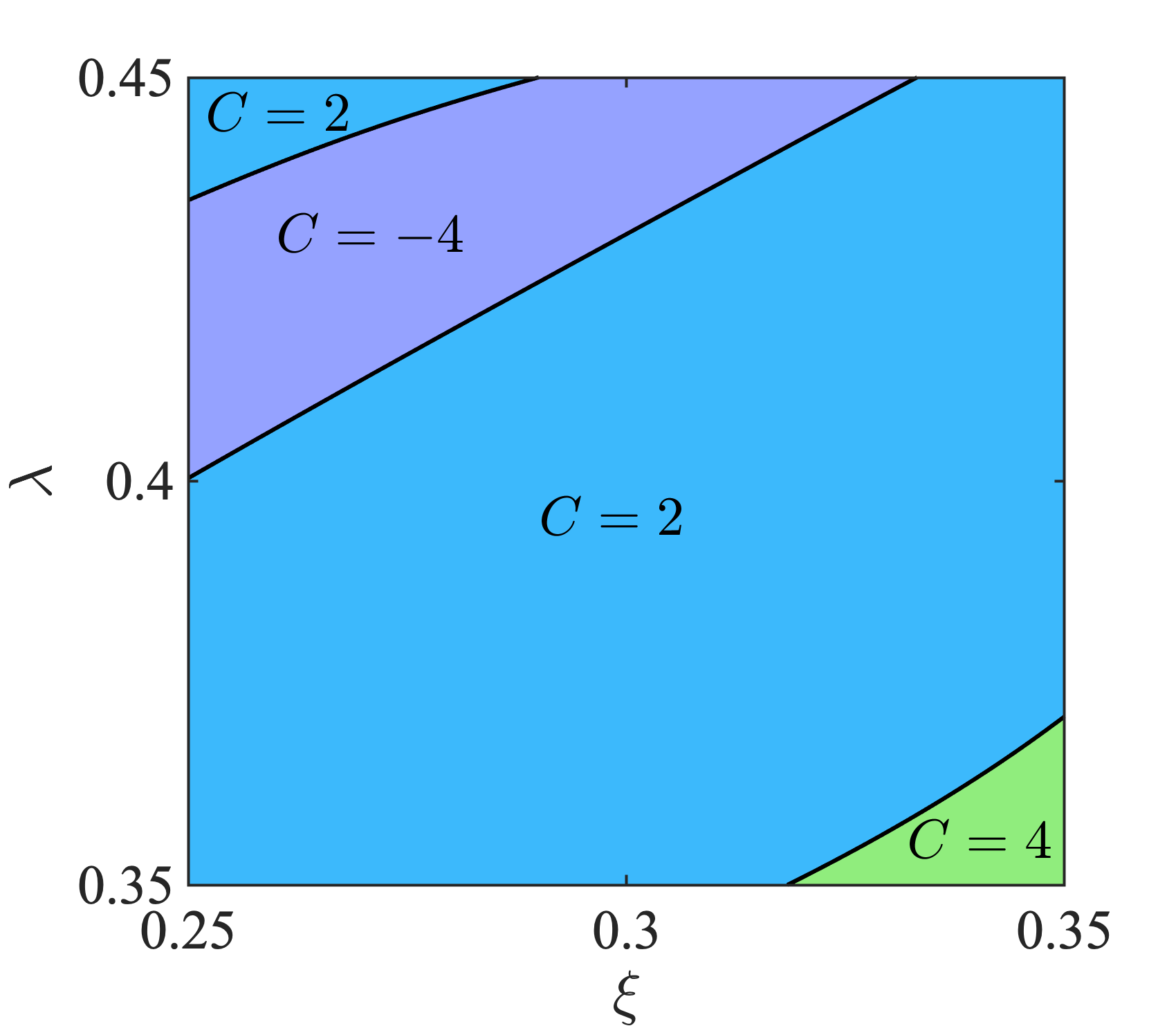}
			\label{fig.PhaseDiagramDm_40_30_3_0}
		}
		\caption{These are topological phase diagrams for (a) $ s $-wave, (b) $ d $+i$ d $-wave and (c) $ d $-i$ d $-wave pairing symmetry SC states. The orange zone in (b) represents Chern number of $ C=6 $. All of them are calculated at the parameter $ \left(\Delta,\mu\right)=\left(0.03,0\right) $.}
		\label{fig.PhaseDiagram_40_30_3_0}
	\end{figure}

	The phase diagrams in Fig. \ref{fig.PhaseDiagram_40_30_3_0} illustrate the $ s $-wave, $ d $+i$ d $-wave, and $ d $-i$ d $-wave pairing symmetry SC states. Our focus is on understanding how the Chern number changes with parameters $ \xi $ and $ \lambda $, and how different Chern numbers control the shapes of thermal Hall conductivity curves. Specifically, we are interested in the region with well-defined Chern number where there is a complete Fermi surface and well-defined quasiparticle transports.

	The first boundary is a diagonal line running from around $ (\xi,\lambda)=(0.32,0.35) $ to $ (0.35,0.37) $ for $ d $+i$ d $-wave and $ d $-i$ d $-wave pairings (Fig.\ref{fig.PhaseDiagramDp_40_30_3_0},\ref{fig.PhaseDiagramDm_40_30_3_0}). Under this boundary, the Chern number equals 4 for complex $ d $-wave pairing symmetry states, indicating that there is no superconducting state in this parameter zone. Note that we do not observe any points whose Chern number is not well-defined, and therefore we believe that there is no phase transition at this boundary for s-wave. The second boundary is a diagonal line running from about $ (\xi,\lambda)=(0.25,0.38) $ to $ (0.34,0.45) $ exclusively for $ s $-wave pairing (Fig. \ref{fig.PhaseDiagramS_40_30_3_0}). Below this boundary, the Chern number equals 4, and above it, the Chern number equals 10. The third boundary is a diagonal line running from  {$ (\xi,\lambda)=(0.25,0.40) $ to $ (0.33,0.45) $} exclusively for d-id-wave pairing (Fig. \ref{fig.PhaseDiagramDm_40_30_3_0}), and it does not exist in the d+id-wave pairing state. The upper left corner in Fig. \ref{fig.PhaseDiagramDm_40_30_3_0} falls in the region $C=2$, and the small region beside the first boundary in Fig. \ref{fig.PhaseDiagramDp_40_30_3_0} falls in the region $C=6$. We did not analyze them because the phase space areas they cover are too small. Our focus is on regions with well-defined Chern numbers and larger phase space areas. Such regions are more representative and ensure the universality of our conclusions. Moreover, it is worth mentioning that the physics near the topological phase transition boundaries is also highly interesting. For example, near the phase transition boundaries of the phase diagrams shown in Fig. \ref{fig.PhaseDiagram_40_30_3_0}, we have discovered the presence of Weyl superconducting states. This finding deserves further investi- gation in subsequent research.
	
	
	We have observed that the topological phase diagrams in the case of $ \mu=0.1 $ exhibit some notable differences when compared to those for $ \mu=0 $. Firstly, the positions of boundaries are dissimilar for $ \mu=0 $, whereas they are quite similar for $ \mu=0.1 $. Secondly, the variation of Chern number is not uniform for $ \mu=0 $, whereas it is uniform for $ \mu=0.1 $. To understand these differences, we can examine the Berry curvature and explore the nature of interactions that might account for the dissimilarities in both the topology and quasiparticle transport. In essence, we aim to investigate the topological origin of the differences in thermal Hall conductivity curves for the three pairing symmetry states and the reason behind the contrast in the phase diagrams for $ s $-wave and complex $ d  $-wave pairing symmetry states.
	
	
	\begin{figure*}[ht]
		\centering
		\subfigure[]
		{
			\includegraphics[width=1.64in]{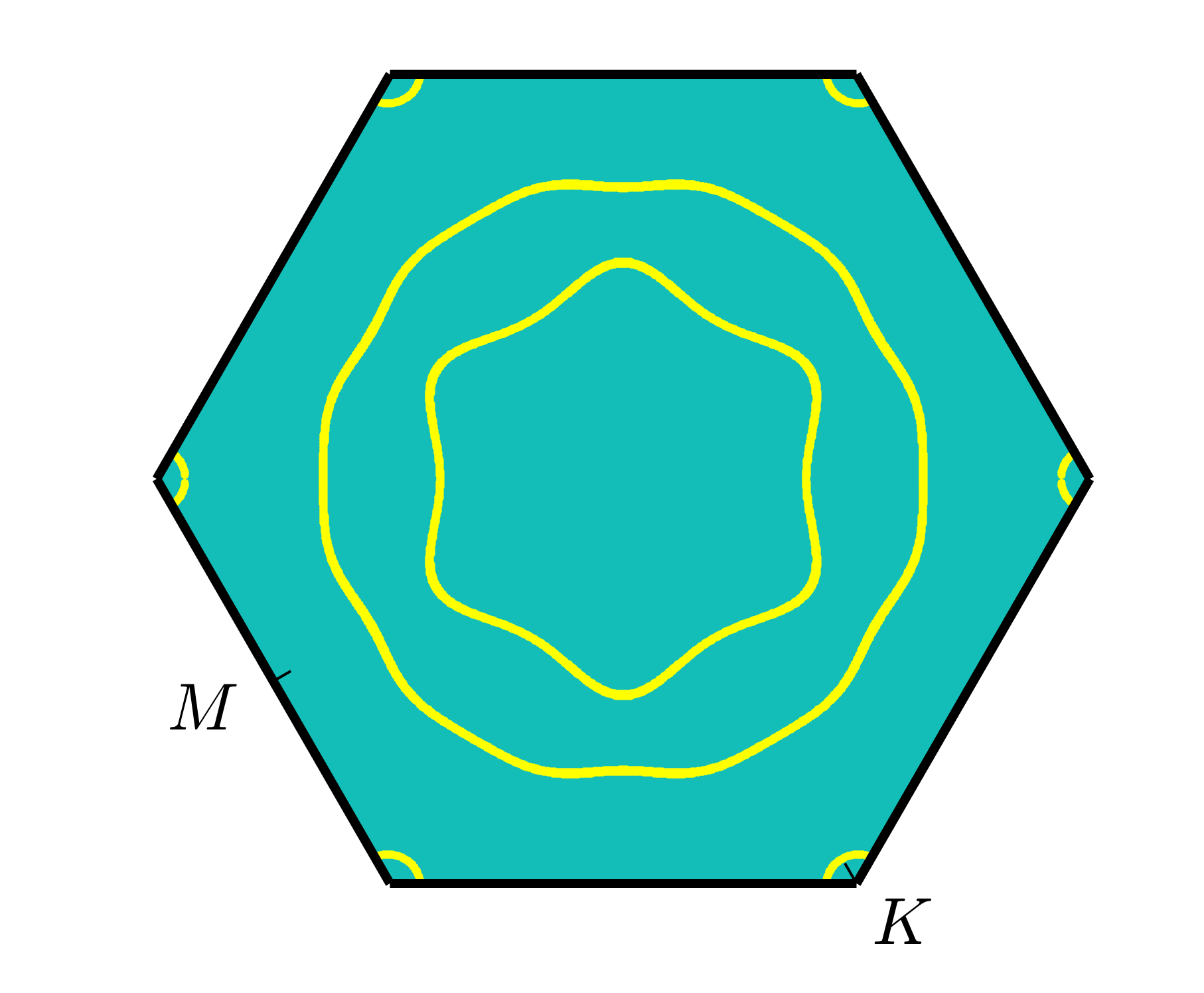}
			\label{fig.Ferimi_0.4_0.3_0}
		}
		\subfigure[]
		{
			\includegraphics[width=1.64in]{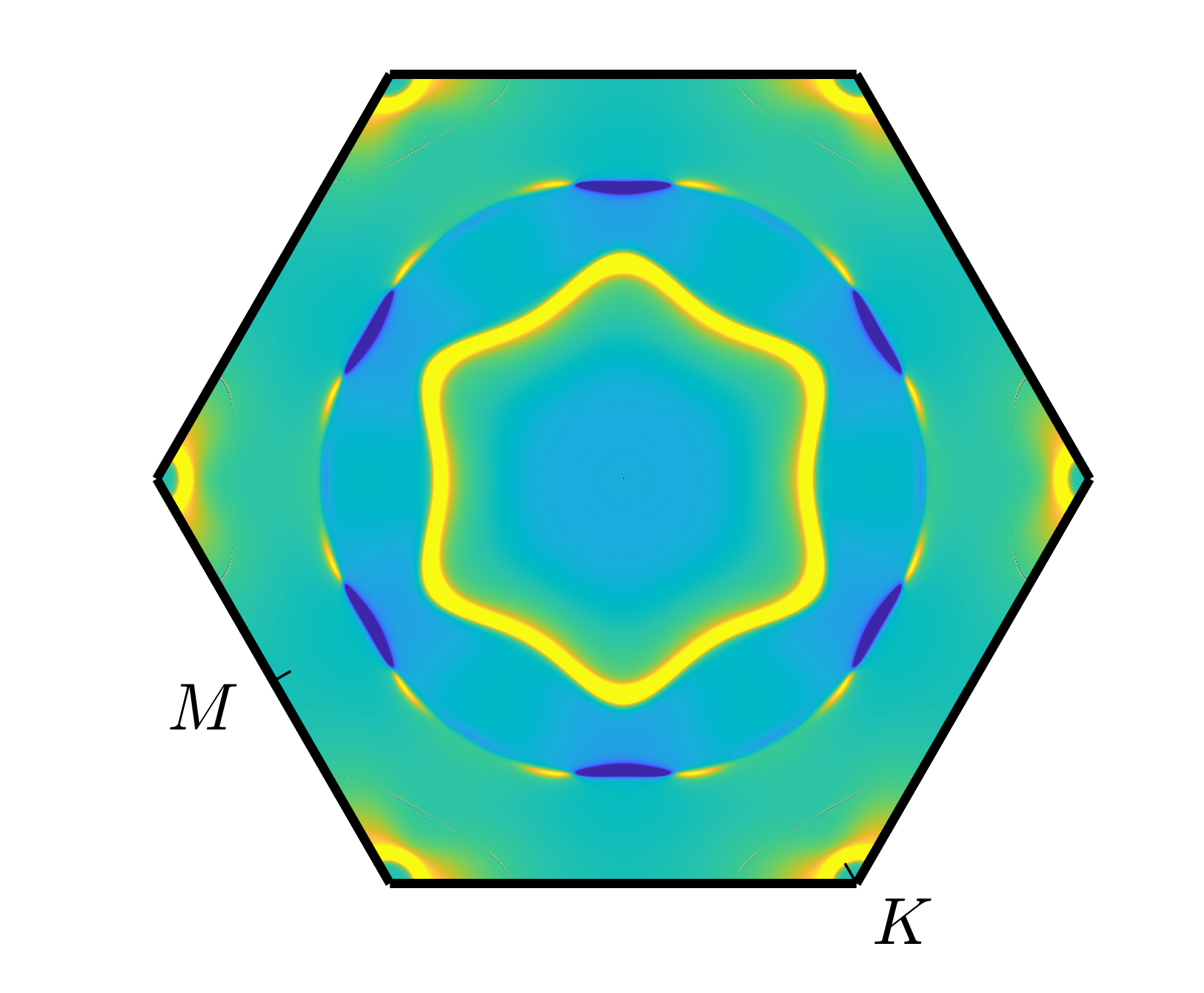}
			\label{fig.Berry_S_40_30_3_0}
		}
		\subfigure[]
		{
			\includegraphics[width=1.64in]{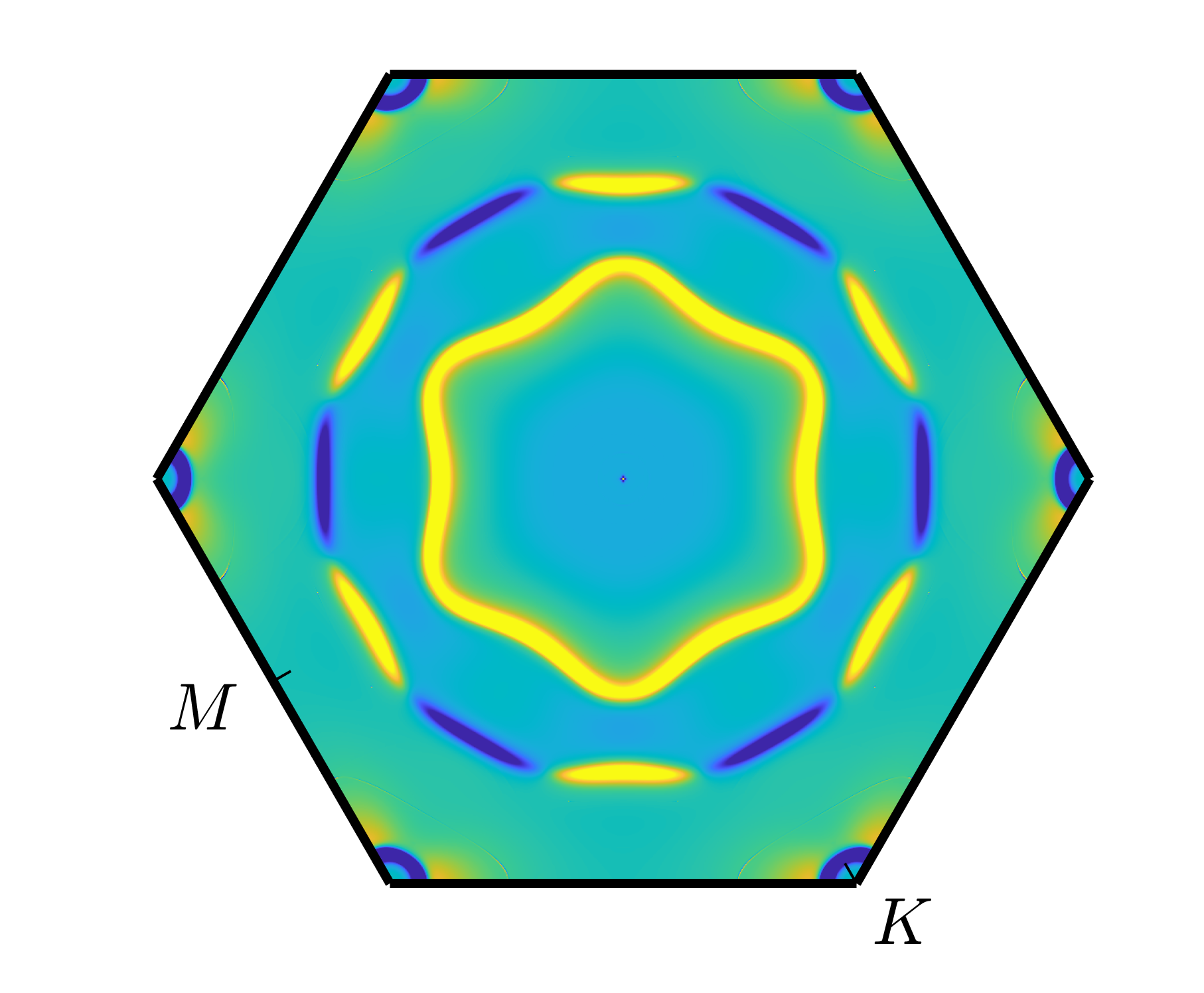}
			\label{fig.Berry_Dp_40_30_3_0}
		}
		\subfigure[]
		{
			\includegraphics[width=1.64in]{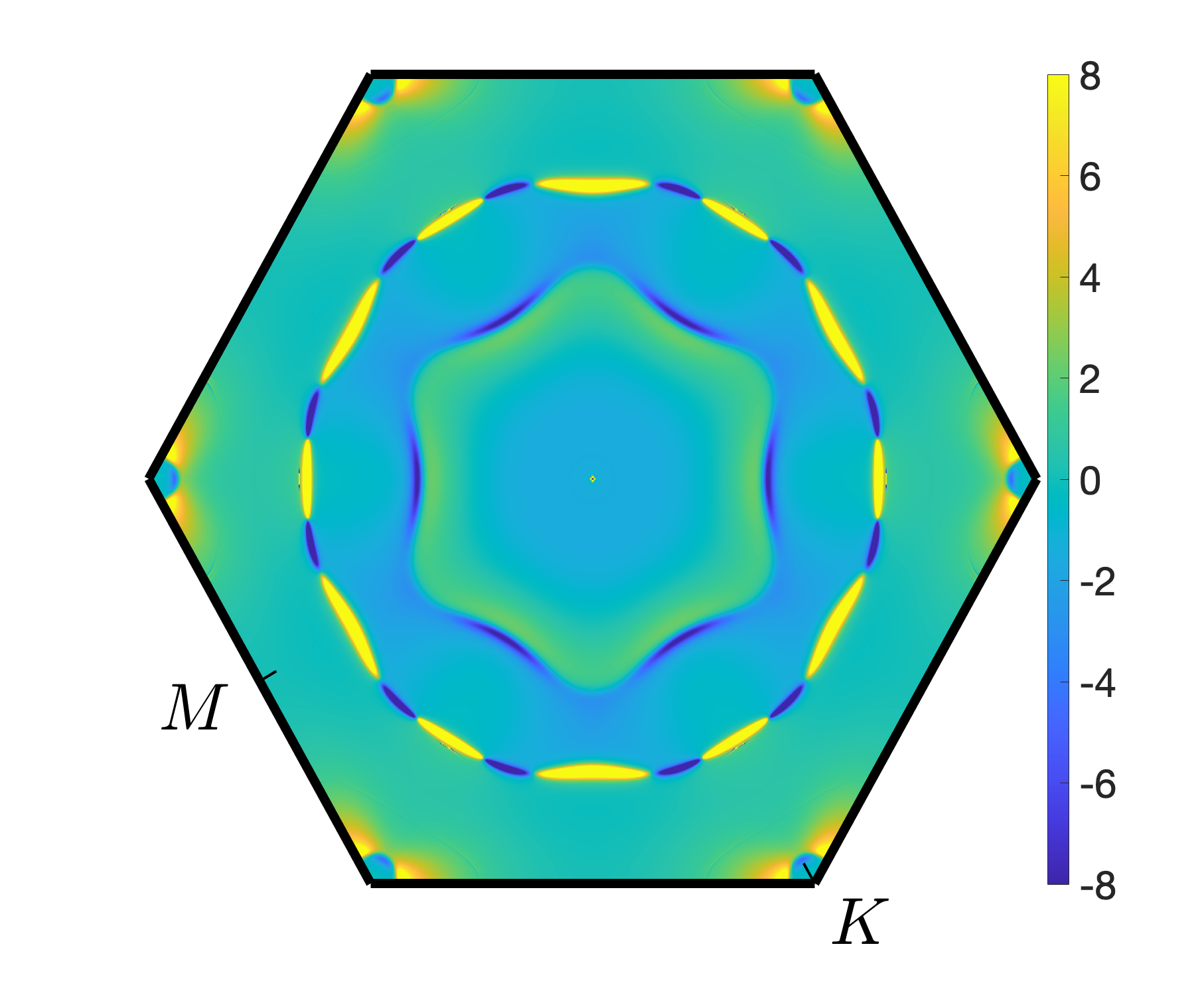}
			\label{fig.Berry_Dm_40_30_3_0}
		}
		\subfigure[]
		{
			\includegraphics[width=1.64in]{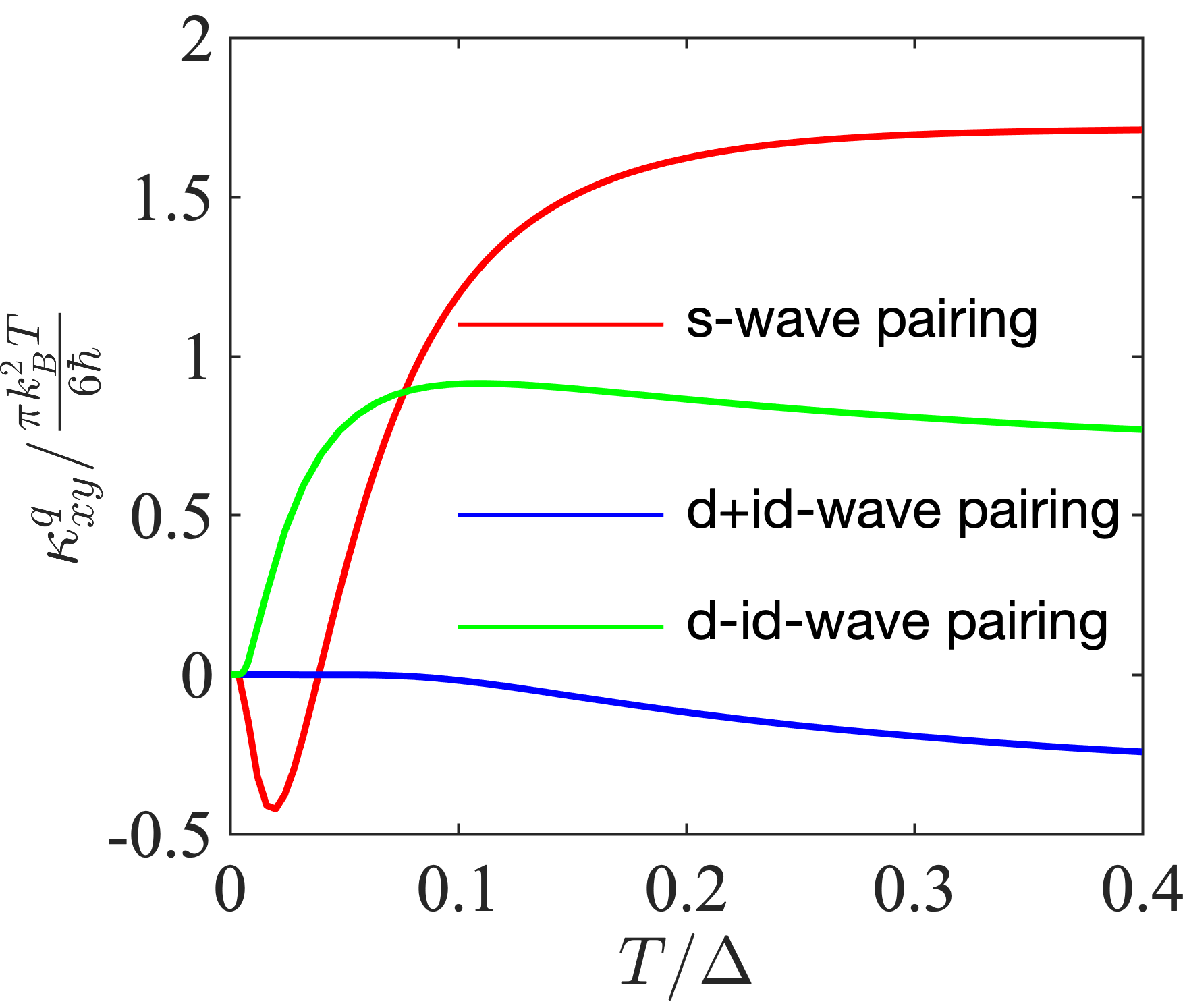}
			\label{fig.SC_40_30_3_0}
		}
		\subfigure[]
		{
			\includegraphics[width=1.64in]{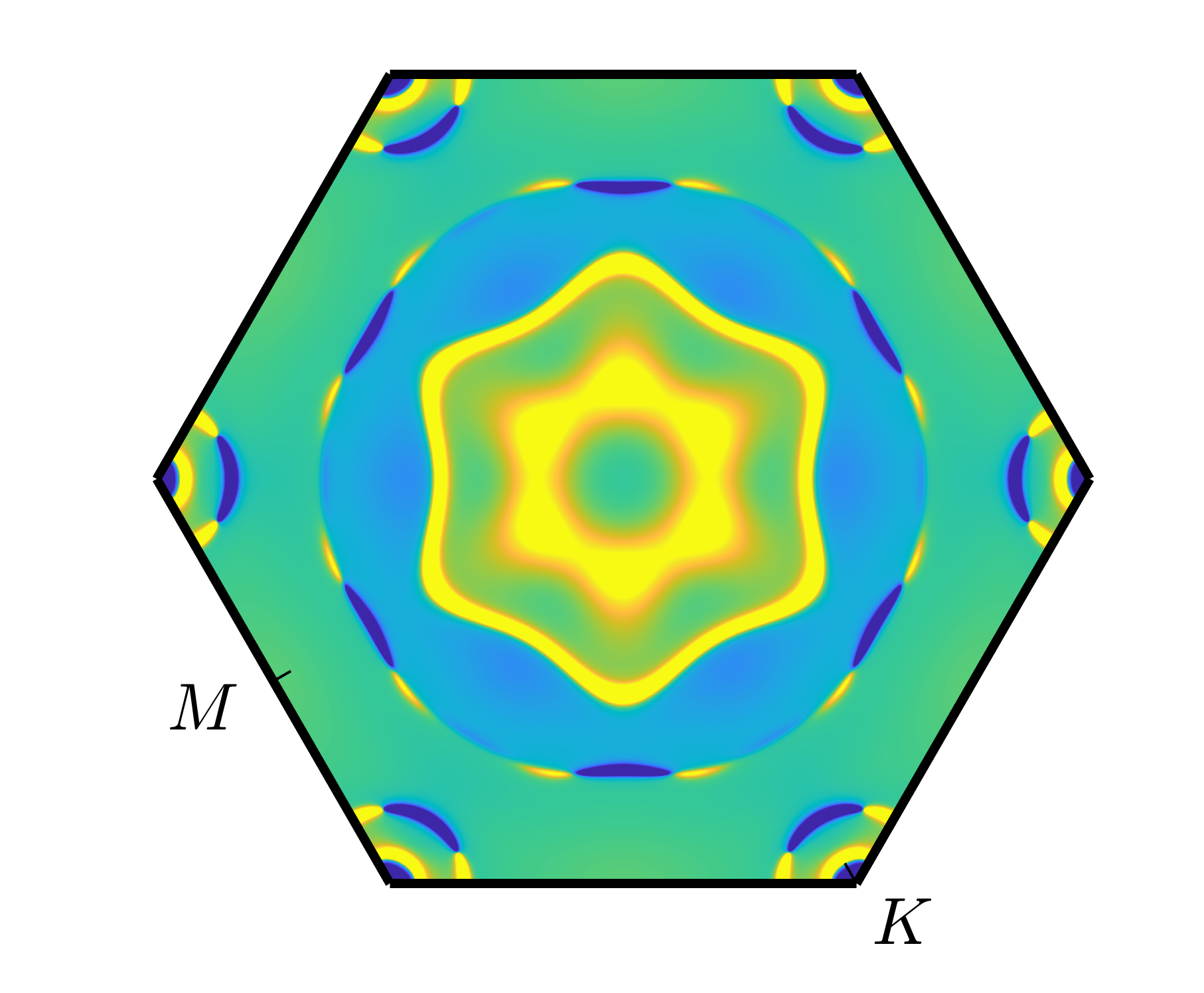}
			\label{fig.Berry25_S_40_30_3_0}
		}
		\subfigure[]
		{
			\includegraphics[width=1.64in]{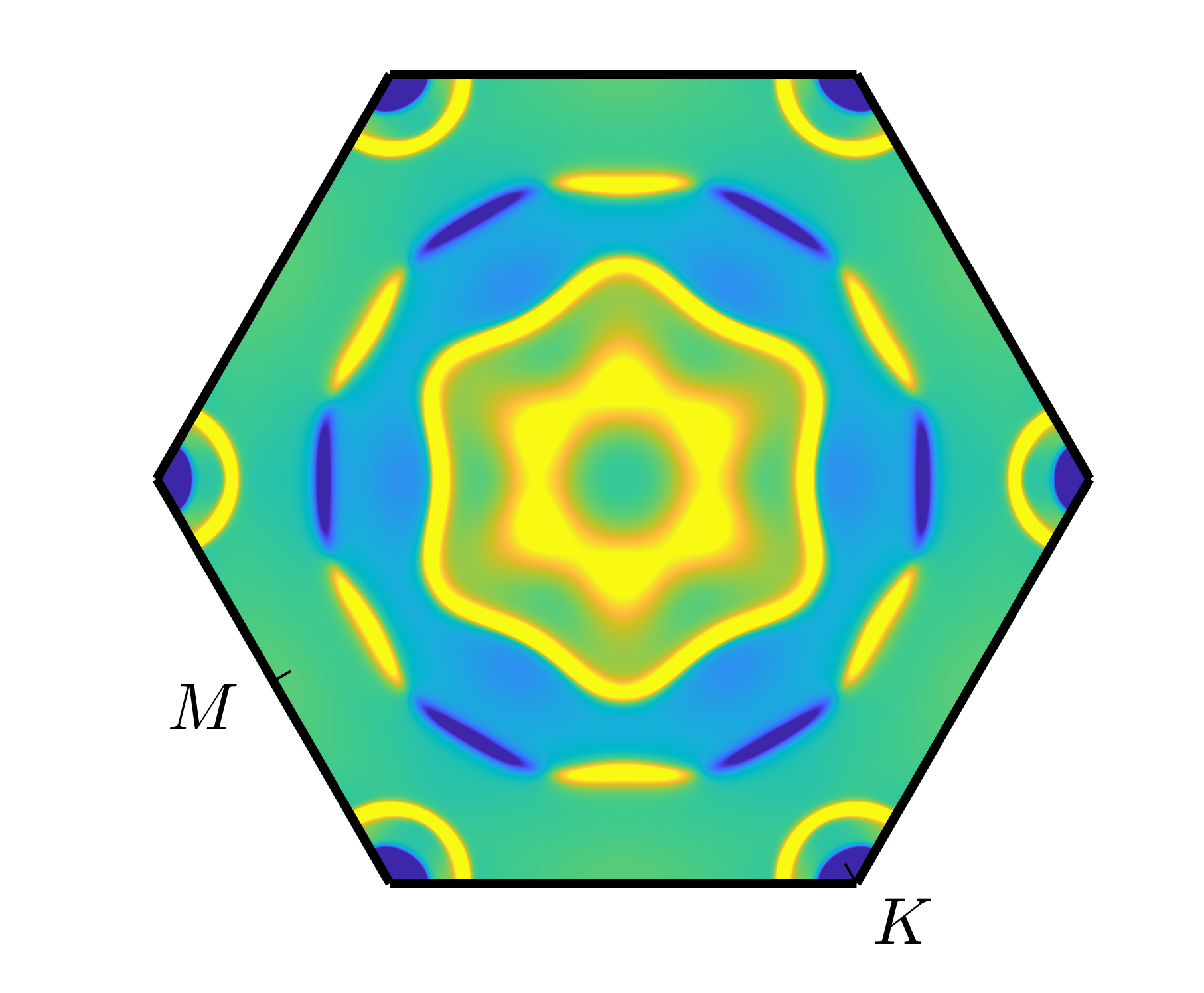}
			\label{fig.Berry25_Dp_40_30_3_0}
		}
		\subfigure[]
		{
			\includegraphics[width=1.64in]{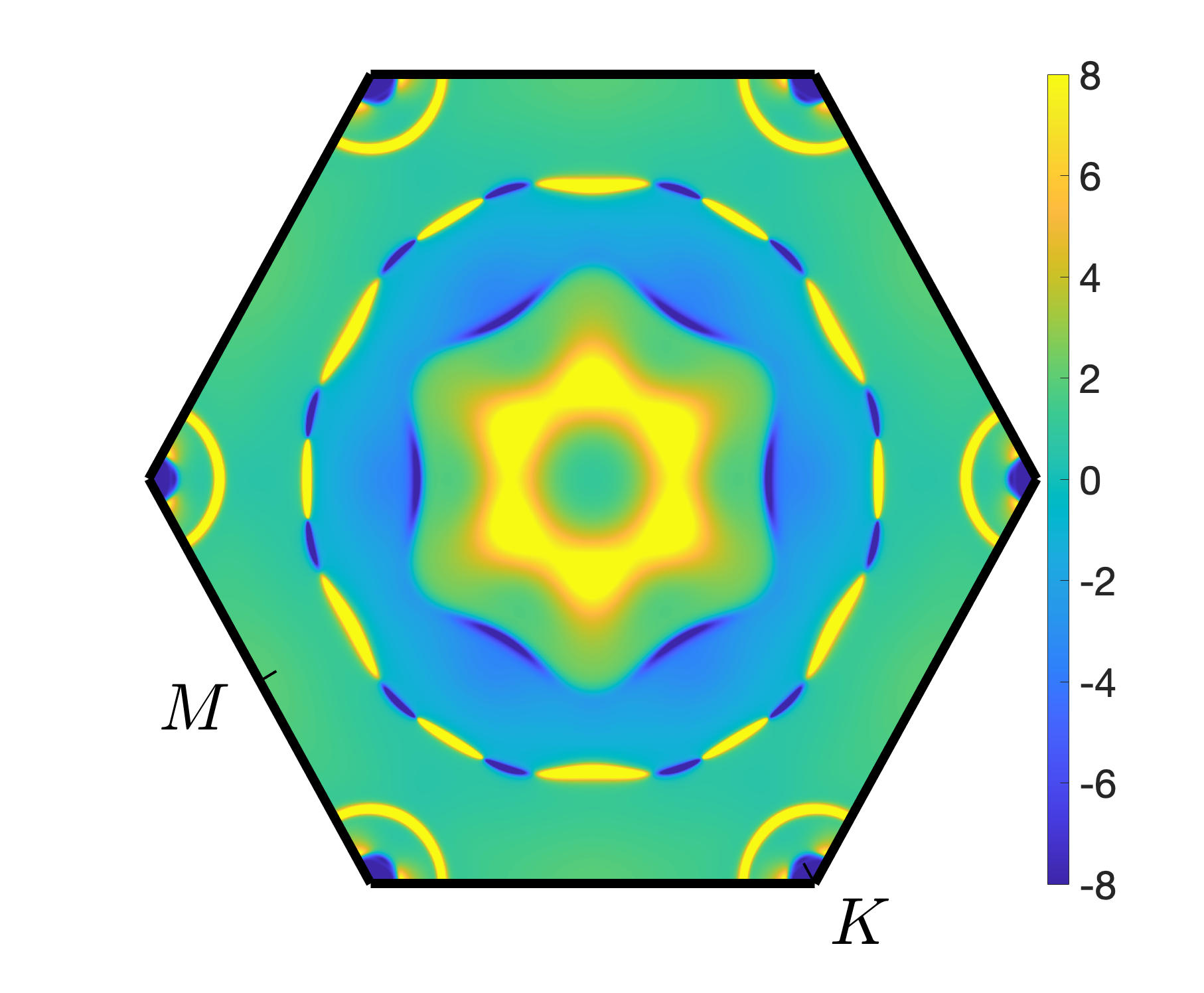}
			\label{fig.Berry25_Dm_40_30_3_0}
		}
		\caption{(a) Fermi surface. [(b-d)] Berry curvature for $ s $-wave, $ d $+i$ d $-wave and $ d $-i$ d $-wave pairings related to Chern number which is the summation of occupied bands. (e) Thermal Hall conductivity curves. [(f-h)] Berry curvature for $ s $-wave, $ d $+i$ d $-wave and $ d $-i$ d $-wave pairings related to the low-temperature quasiparticle transport which belongs to the  25th band. All the the data are calculated under the parameter $ (\xi,\lambda,\Delta,\mu)=(0.3,0.4,0.03,0) $,  {which is based on the experimental data.}}
	\end{figure*}
	
	
	We will focus on two regions: (i) the region in which the Chern number for $ s $-wave equals 4, $ d  $+i$ d  $-wave equals 0, and $ d  $-i$ d  $-wave equals 2, and (ii) the region where the Chern num- ber for $ s $-wave equals 10, $ d  $+i$ d  $-wave equals 0, and $ d  $-i$ d  $-wave equals -4. We do not consider the situation where the Chern number for all three pairing symmetry states is 4 because it implies they are all topologically trivial. Furthermore, we do not analyze the situation where the Chern number for $ s $-wave equals 10, $ d  $+i$ d  $-wave equals 0, and $ d  $-i$ d  $-wave equals 2 separately, as the differences can be inferred by examining the aforementioned regions.
	
	In the first case, we consider the parameters $ (\xi,\lambda,\Delta,\mu)=(0.3,0.4,0.03,0) $, resulting in Chern numbers of 4, 0, and 2 for the $ s $-wave, $ d $+i$ d $-wave, and $ d $-i$ d $-wave pairing SC states, respectively. As shown in Fig. \ref{fig.Ferimi_0.4_0.3_0}, the Fermi surface is notably different from that of the situation when $ \mu=0.1 $. Although the Fermi surface in the ring-like region also splits into two pieces, the Fermi surface around $ M $ points moves to $ K $ points, which is a distinguishing feature. 
	
	We cannot directly compare our current results with the previous findings; therefore, we further verified the symmetry of the system. It has been confirmed that the Fermi surface displays a sixfold rotational symmetry, consistent with the $D_{6h}^*$ symmetry of the CDW term \cite{Feng2021Chiral} and at least $C_{6h}$ symmetry of the SOC term in our model. Furthermore, we explored the topological superconductivity of spin singlet pairing at the atomic level, where the energy gap function of the s-wave pairing possesses the same $D_{6h}$ symmetry as the lattice, and the modulus of the energy gap function of the d-wave pairing has at least $C_{6h}$ symmetry. Hence, we can infer that the Berry curvature of all energy bands also exhibits at least a sixfold rotational symmetry. This confirms the precision of our calculations.
	
	Let us examine the Berry curvature, whose integral equals the Chern number, shown in Fig.\ref{fig.Berry_S_40_30_3_0}-\ref{fig.Berry_Dm_40_30_3_0}. Two main characteristics emerge: (i) the inner ring in the ringlike region is affected by the superconducting pairing symmetry. The inner ring of the $ s $-wave pairing state is the largest, followed by the $ d  $+i$ d  $-wave pairing state, while the $ d  $-i$ d  $-wave pairing state has a negative inner ring. (ii) The outer ring of the ringlike region is influenced by the interaction between SOC and SC term. Fragments on the ring facing M points of s-wave pairing states are negative, while those of $ d  $+i$ d  $-wave and $ d  $-i$ d  $-wave pairing states are positive. Furthermore, the number of fragments on outer ring for $ s $-wave and $ d  $+i$ d  $-wave pairing states is 12, while that for $ d $-i$ d $-wave is double.

	Moving on to the thermal Hall conductivity curves in Fig.\ref{fig.SC_40_30_3_0}, we observe significant differences from the situation when $ \mu=0.1 $. The curve for the $ s $-wave pairing state goes negative in the low-temperature region and quickly becomes positive again. The curve for the $ d  $+i$ d  $-wave pairing state remains flat, while that for the $ d  $-i$ d  $-wave pairing state goes positive in the low-temperature region. Combined with the above, we conclude that the low-temperature behavior of the thermal Hall conductivity curves depends on the outer ring of the ringlike region around $\Gamma$ point, which is determined by the topology contributed by the interaction of SOC and SC. Al- though there are other differences in the Berry curvature of the 25th bands (Fig.\ref{fig.Berry25_S_40_30_3_0}-\ref{fig.Berry25_Dm_40_30_3_0}), they are either too small to contribute to the curve shape or far away from the Fermi surface.

\begin{figure*}[ht]
		\centering
		\subfigure[]
		{
			\includegraphics[width=1.64in]{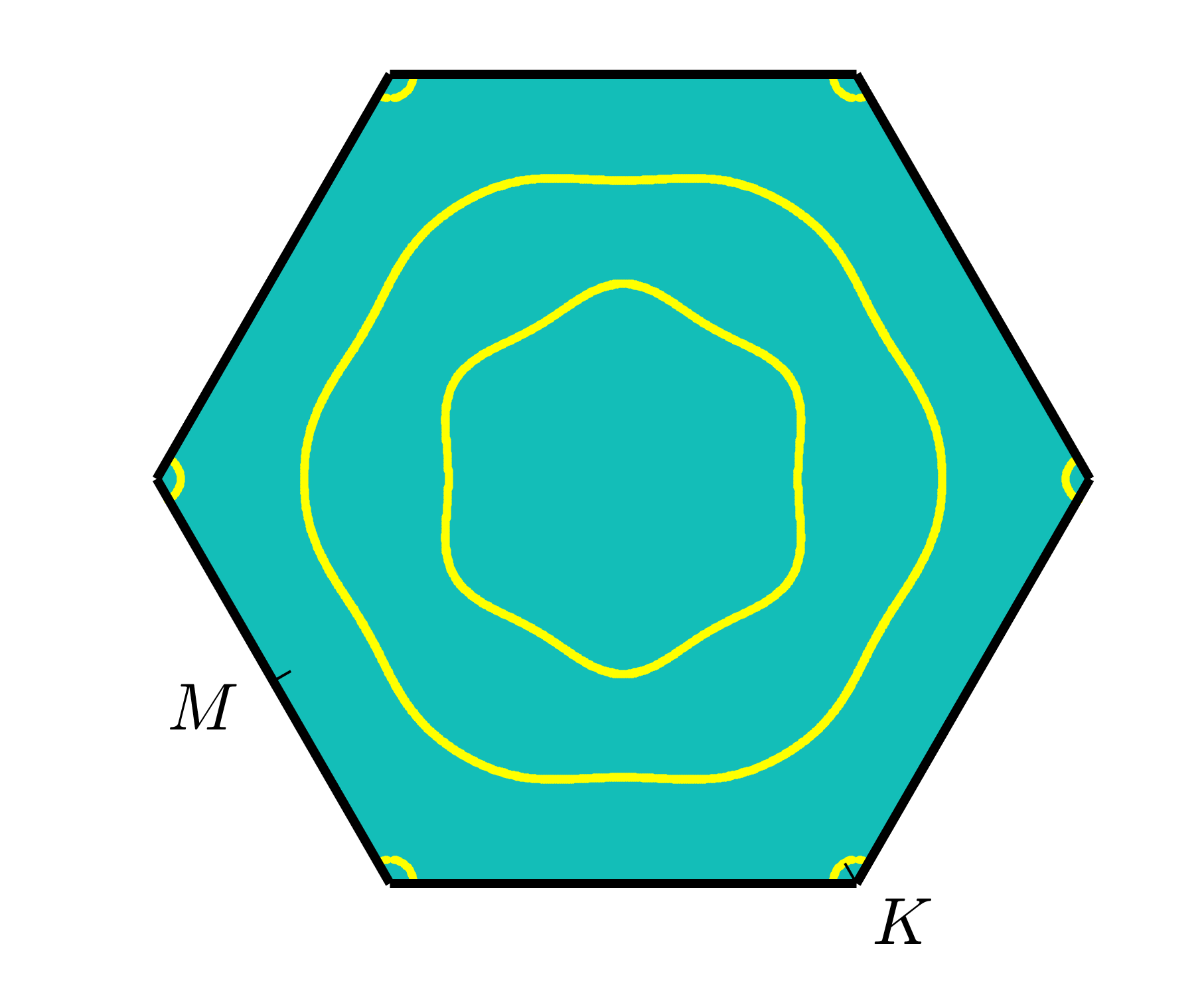}
			\label{fig.Ferimi_0.43_0.27_0}
		}
		\subfigure[]
		{
			\includegraphics[width=1.64in]{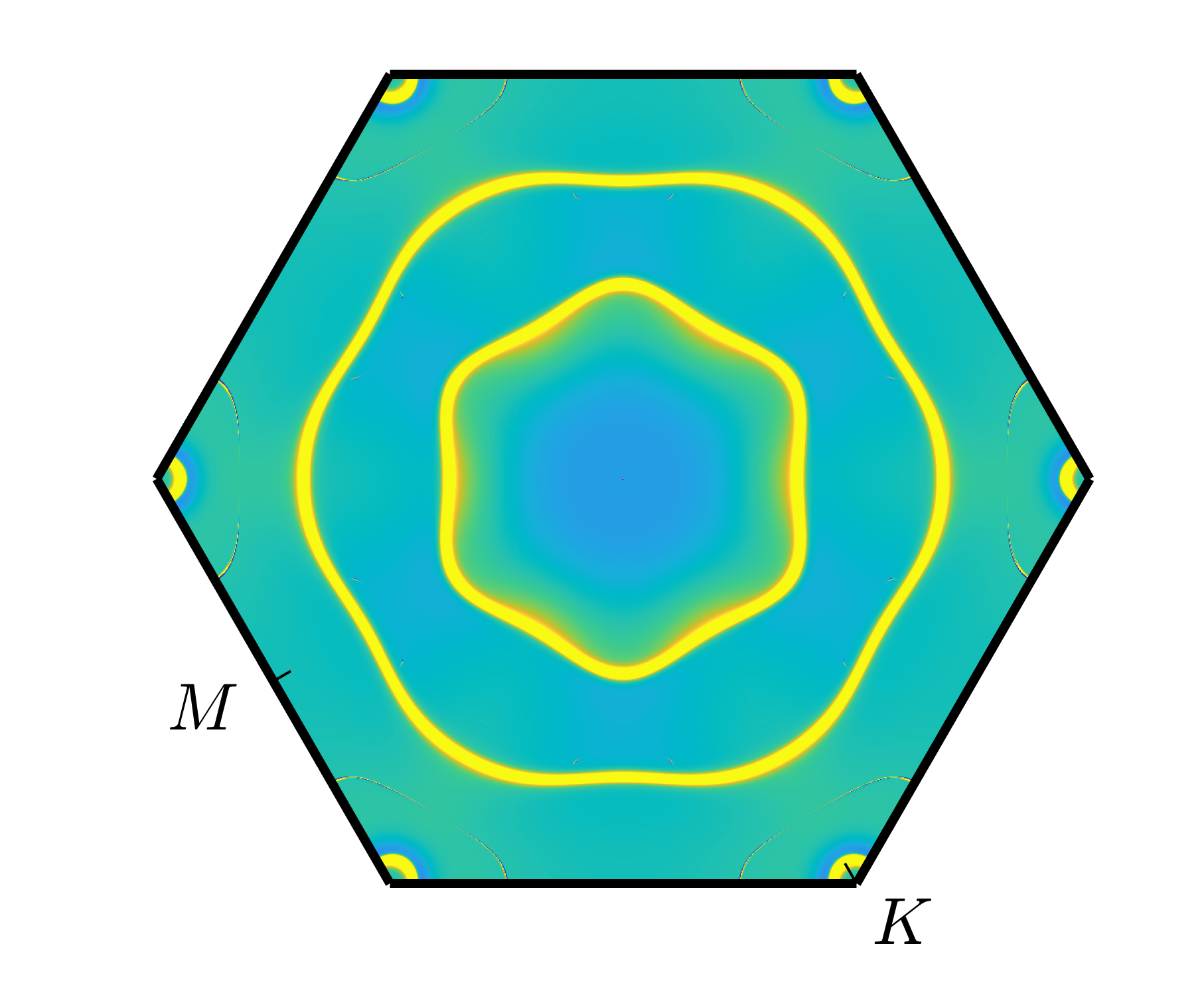}
			\label{fig.Berry_S_43_27_3_0}
		}
		\subfigure[]
		{
			\includegraphics[width=1.64in]{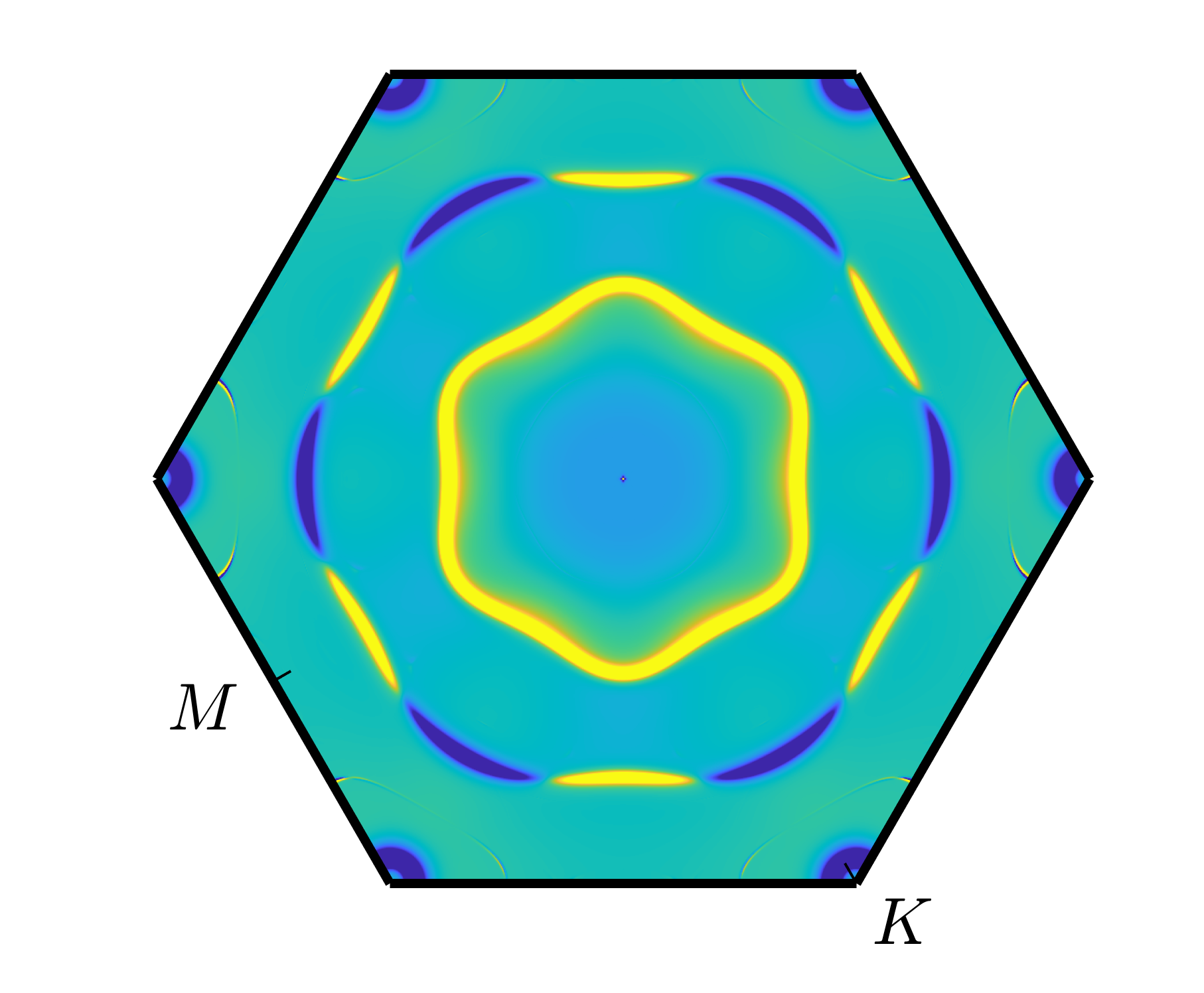}
			\label{fig.Berry_Dp_43_27_3_0}
		}
		\subfigure[]
		{
			\includegraphics[width=1.64in]{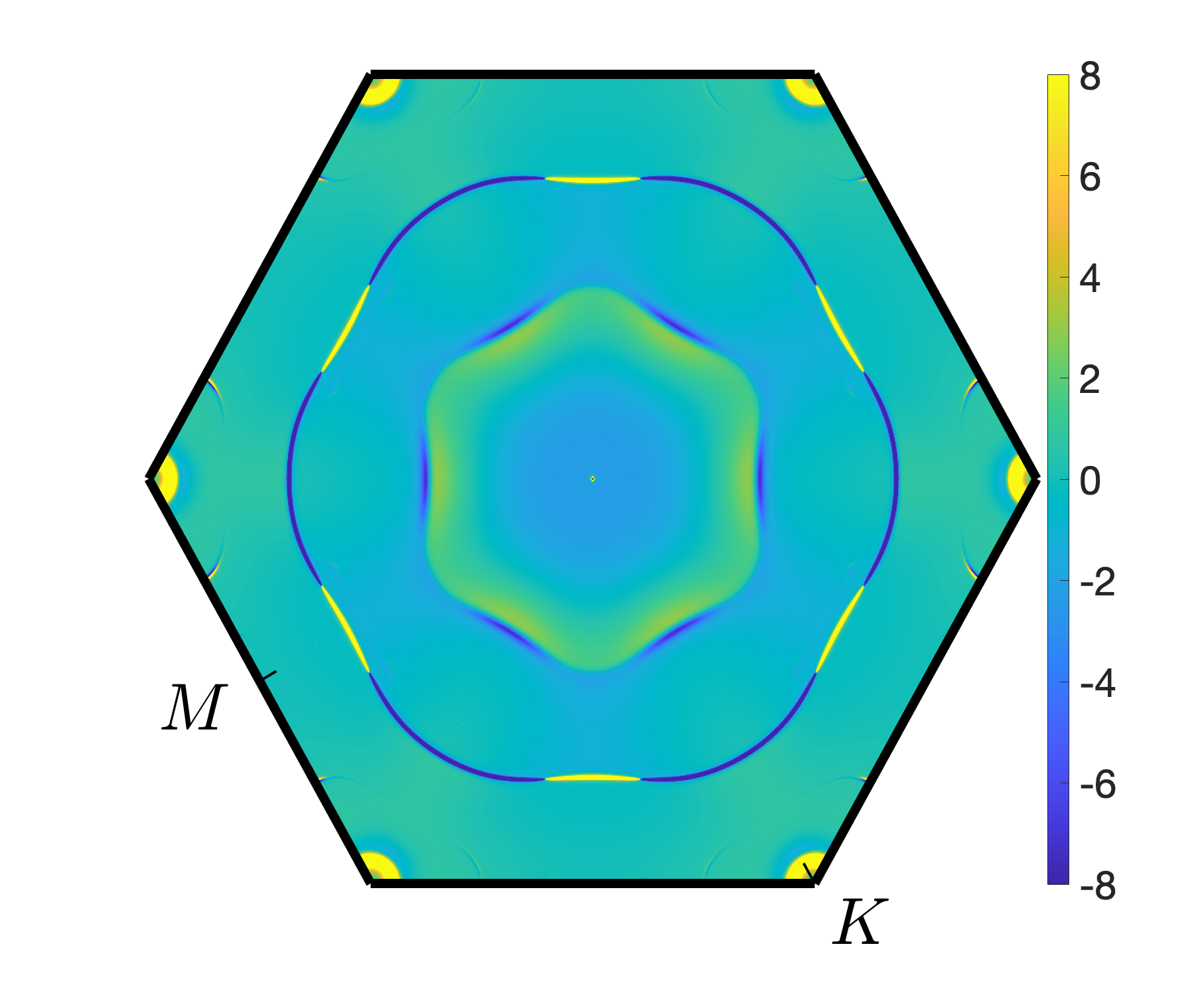}
			\label{fig.Berry_Dm_43_27_3_0}
		}
		\subfigure[]
		{
			\includegraphics[width=1.64in]{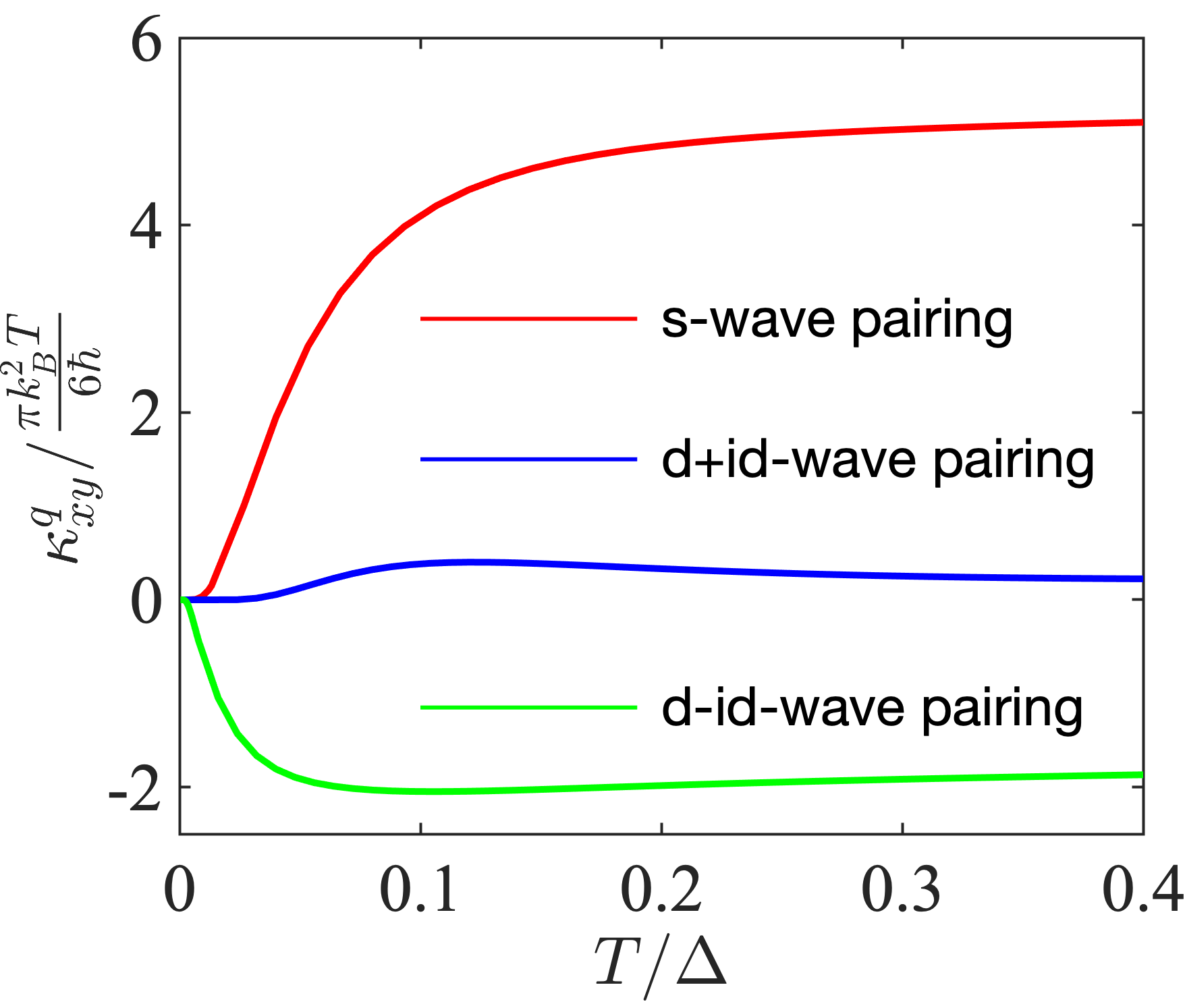}
			\label{fig.SC_43_27_3_0}
		}
		\subfigure[]
		{
			\includegraphics[width=1.64in]{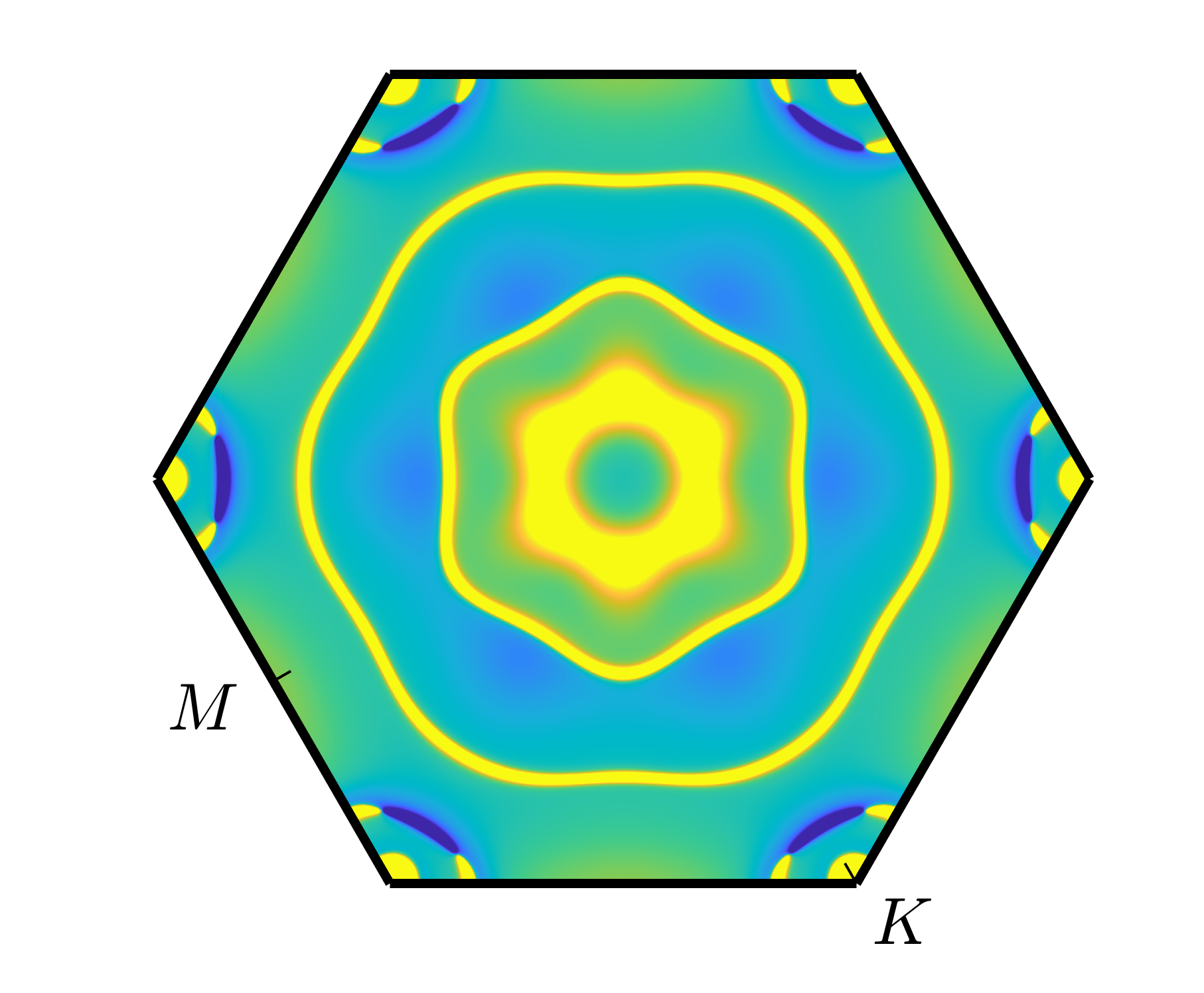}
			\label{fig.Berry25_S_43_27_3_0}
		}
		\subfigure[]
		{
			\includegraphics[width=1.64in]{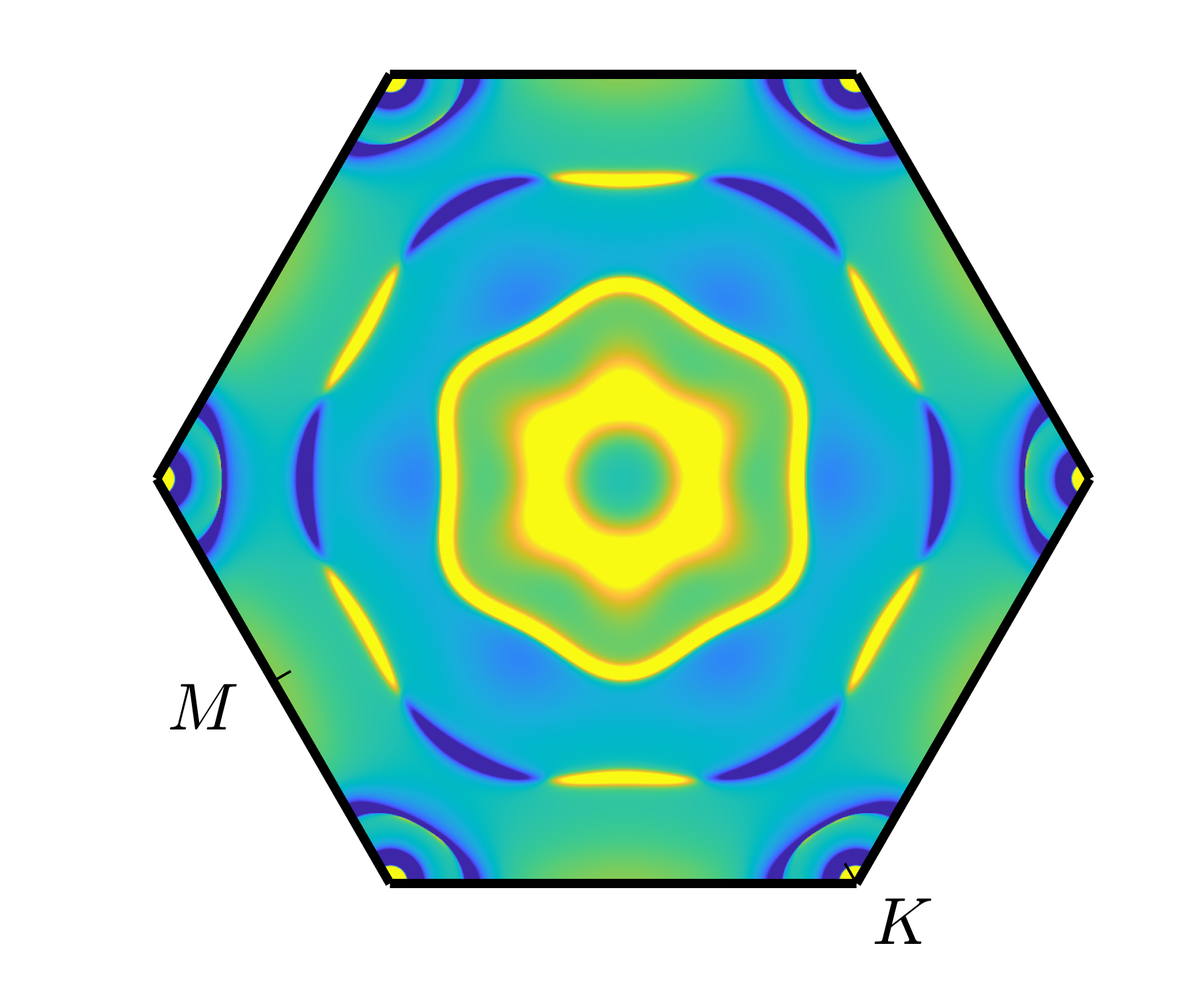}
			\label{fig.Berry25_Dp_43_27_3_0}
		}
		\subfigure[]
		{
			\includegraphics[width=1.64in]{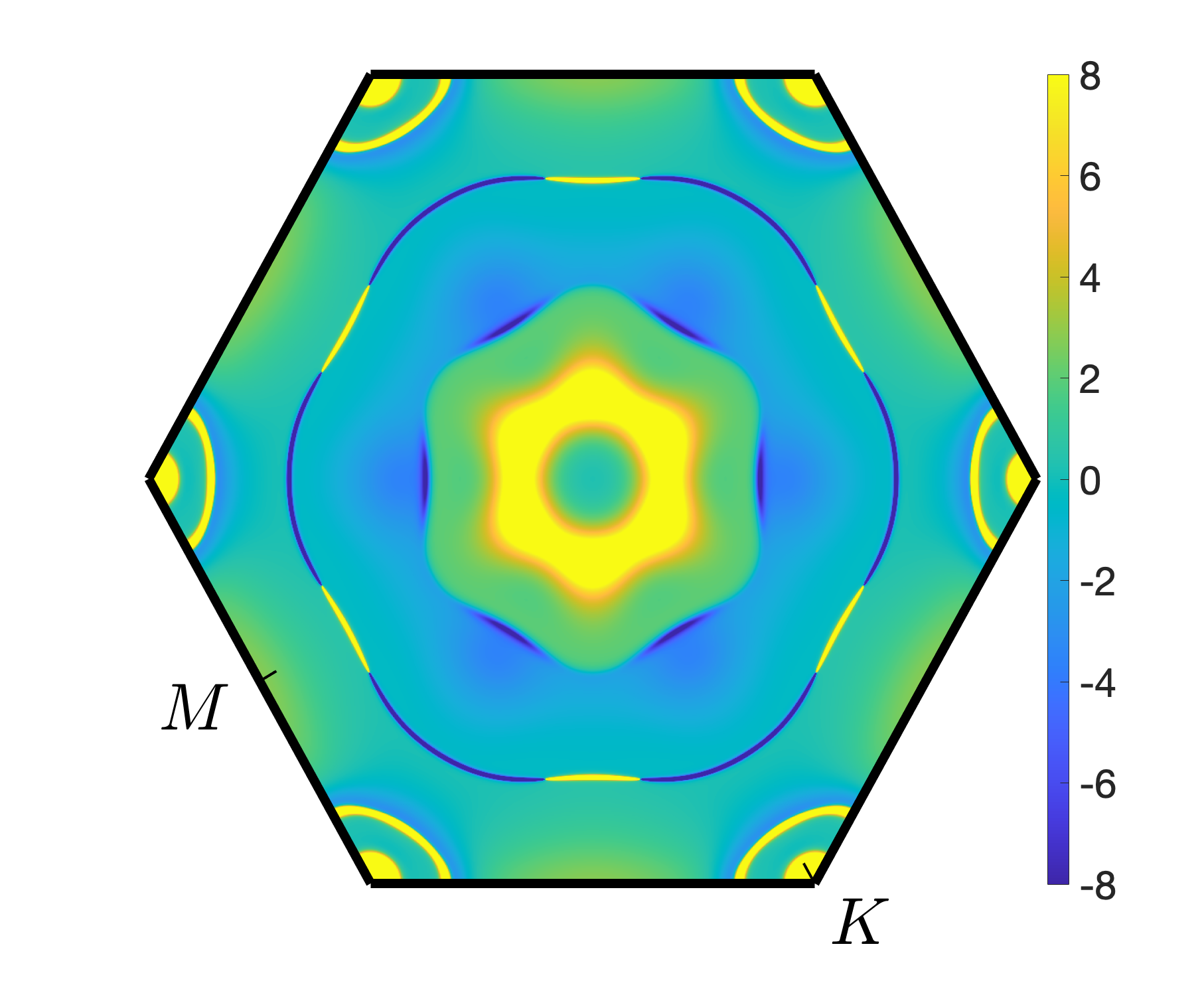}
			\label{fig.Berry25_Dm_43_27_3_0}
		}
		\caption{(a) Fermi surface. [(b-d)] Berry curvature for $ s $-wave, $ d $+i$ d $-wave and $ d $-i$ d $-wave pairing related to Chern number which is the summation of occupied bands. (e) Thermal Hall conductivity curves. [(f-h)] Berry curvature for $ s $-wave, $ d $+i$ d $-wave and $ d $-i$ d $-wave pairing related to the low-temperature quasiparticle transport which belongs to the  25th band. All the the data is calculated under the parameter $ (\xi,\lambda,\Delta,\mu)=(0.27,0.43,0.03,0) $.}
	\end{figure*}

	Here we consider the second case, where the parameter values of $(\xi, \lambda, \Delta, \mu) = (0.27, 0.43, 0.03, 0)$ are used as an example. In this case, the Chern numbers for the $ s $-wave, $ d  $+i$ d  $-wave, and $ d  $-i$ d  $-wave pairing SC states are 10, 0, and $-4$, respectively. The differences in the Fermi surfaces between $(\xi, \lambda, \Delta, \mu) = (0.3, 0.4, 0.03, 0)$ [Fig. \ref{fig.Ferimi_0.4_0.3_0}] and $(\xi, \lambda, \Delta, \mu) = (0.27, 0.43, 0.03, 0)$ [Fig. \ref{fig.Ferimi_0.43_0.27_0}] are minimal, with only a slight expansion of the outer ring and a slight shrinkage of the inner ring.
	
	The summation of the Berry curvature of the occupied bands is shown in Fig. \ref{fig.Berry_S_43_27_3_0}-\ref{fig.Berry_Dm_43_27_3_0}. The change in the Chern number of the $ s $-wave pairing symmetry SC state from 4 to 10, a large leap, is due to the change from negative to positive on the outer ring. The change in the Chern number of the$  d $+i$ d $-wave pairing from 2 to $- 4$ is due to some of the positive segments on the outer ring becoming negative. Consequently, the topological phase transitions arise from the outer ring of the ringlike region.
	
	The thermal Hall conductivity is shown in Fig. \ref{fig.SC_43_27_3_0}. Examining the Berry curvature shown in Fig. \ref{fig.Berry25_S_43_27_3_0}-\ref{fig.Berry25_Dm_43_27_3_0}, we observe that they are almost identical except for the region around the Fermi surface. The curve for $ s $-wave pairing becomes positive, the curve for $ d  $-i$ d  $-wave pairing becomes negative, and the curve for $ d  $+i$ d  $-wave pairing remains almost flat. All of these differences arise from the outer ring of the ringlike region, which is the origin of the topological properties at this parameter.
	
	Hence, it can be concluded that, for chemical potential $\mu=0$, the shapes of the thermal Hall conductivity curves are determined by the outer ring of the ringlike region of the Berry curvature. This outer ring is closely linked to the topology that arises from the interplay between the SOC and SC terms.
	
	
	\section{DISSUCUSION AND CONCLUSION}\label{Sec.Conclusion}

	Based on the $ A\mathrm{V_3Sb_5}$ system, where ${A}$ is either K, Rb, or Cs, we have developed a model on the kagome lattice to describe topological superconducting states characterized by chiral charge density waves. We have also predicted their quasiparticle transport. We have considered different superconducting pairing symmetry states, namely $ s $-wave, $ d $+i$ d $-wave, and $ d $-i$ d $-wave pairings, and we aim to differentiate these states based on a measurable value, which is the thermal Hall conductivity.
	
	In the absence of spin-orbit coupling, the phase diagrams of the superconducting states are divided into two regions. For $\mu \in [-0.1,0]$, the normal states are insulators, and thus, topological superconducting states do not exist. For $\mu\in \left(\left.0,0.1\right]\right.$, topological superconducting states exist, and the thermal Hall conductivity curves of different superconducting pairing symmetry states are qualitatively distinct. The Chern number contributed by the $ s $-wave pairing state’s SC term is 0, whereas that contributed by the $ d $$\pm$i$ d $-wave pairing states is $\pm2$. Notably, the primary difference in the Chern number arises from the ringlike region around the $\Gamma$ point of the Berry curvature, which also determines the qualitative difference among the three superconducting pairing symmetry states. Hence, the thermal Hall conductivity curves of the three superconducting pairing symmetry states are qualitatively distinct and topologically protected in the absence of SOC.
	
	In the presence of spin-orbit coupling, our analysis can be divided into two classical scenarios. First, we add an SOC term as a perturbation to a topological superconducting state ($\mu=0.1$). Second, we apply a large SOC term to drive the system into an insulator-metal phase transition, followed by adding an SC term ($\mu=0$).
	
	At $\mu=0.1$, as the strength of the SOC term increases, the system undergoes a topological phase transition, and the SOC contribution to the Chern number equals $- 3$, which primarily arises from the ringlike region of the Berry curvature. Thus the qualitative differences in the thermal Hall conductivity curves are topologically protected by the contribution of the SC term, since the SOC term contributes the same Chern number for all three superconducting pairing symmetry states.
	
	At $\mu=0$, varying the strength ratio of the SOC and CDW leads to topological phase transitions in our parameter range. Notably, the topological phase transition boundaries for the $ s $-wave, $ d $+i$ d $-wave, and $ d $-i$ d $-wave pairing symmetry SC states are different, owing to the complex interaction among the CDW, SOC, and SC terms. We have considered two parameters as examples and concluded that the differences in the thermal Hall conductivity curves arise from the outer ring of the ringlike region of the Berry curvature. Therefore the qualitative differences in the thermal Hall conductivity are also topologically protected by the interaction between SOC and SC terms of the system.
	
	 In summary, our study investigates topological supercon- ducting states and the thermal Hall effect on a kagome lattice. The interplay of superconductivity, CDW, and SOC on the kagome lattice results in the emergence of topological properties in the system. Different superconducting pairing symmetries give rise to distinct topological phases and dif- ferent Chern numbers. We emphasize that the qualitative differences in thermal Hall conductivity curves originate from the different topological regimes. This disparity is primarily driven by the superconducting term, which is topologically protected. Therefore employing the thermal Hall effect as a criterion for determining superconducting pairing symmetry is very helpful.

	\begin{acknowledgments}
		We would like to thank Fan Yang, Zhongbo Yan, Zheng-Yang Zhuang, and Shanbo Chow for the helpful discussions. This project is supported by NKRDPC-2018YFA0306001, NKRDPC-2022YFA1402802, NSFC-92165204, NSFC-12174453, NSFC-11974432, Leading Talent Program of Guangdong Special Projects (201626003), and Shenzhen International Quantum Academy (Grant No. SIQA202102).
	\end{acknowledgments}

	\appendix
	\section{Construction of Tight-Binding Model}\label{App.A}
	
	\subsection{ Nearstest Neighbor Tight-Binding Model}\label{App.1A}
	
	Fourier transformation of creation and annihilation operators can be written as {
		\begin{equation}
			\begin{aligned}
				c_{\mathbf{j}\alpha\sigma}^\dagger&=\frac{1}{\sqrt{N}}\sum_{\mathbf{k}} e^{+i\mathbf{k}\cdot\mathbf{r_j}}c_{\mathbf{k}\alpha\sigma}^\dagger\\
				c_{\mathbf{j}\alpha\sigma}&=\frac{1}{\sqrt{N}}\sum_{\mathbf{k}} e^{-i\mathbf{k}\cdot\mathbf{r_j}}c_{\mathbf{k}\alpha\sigma}\quad .
			\end{aligned}
		\end{equation}
	}
	
	So the tight-binding model can be written in $ \mathbf{k} $ space as
	\begin{equation}\label{key}
		H_{TB}=
		\begin{bmatrix}
			-\mu & -2t\cos\left(k_1/2\right) &-2t\cos\left({k_2}/{2}\right)  \\
			-2t\cos\left({k_1}/{2}\right) &\mu &-2t\cos\left({k_3}/{2}\right) \\
			-2t\cos\left({k_2}/{2}\right) &-2t\cos\left({k_3}/{2}\right) &-\mu
		\end{bmatrix}.
	\end{equation}
	
	\subsection{ Spin-Orbit Coupling}\label{App.1C}
	
	In this section, we are deriving the Rashba SOC Hamiltonian, which can be written as\cite{PhysRevB.64.121202} 
	\begin{equation}\label{key}
		H_{SOC}(\mathbf{r})=-\lambda\left( \boldsymbol{\sigma }\times \mathbf{p}\right) \cdot \mathbf{\hat{z}} = \lambda \left(\mathbf{p} \times \boldsymbol{\sigma }\right) \cdot \mathbf{\hat{z}}\quad,
	\end{equation}
	where we can see that the momentum of the electron is perpendicular to the Pauli matrix vector. 
	 {Note that in the language of second quantization, the electron's transition between lattice sites can be expressed in the form of hopping, with its momentum direction aligned with the lattice vector. Now, we are able to rewrite the Hamiltonian into the second quantization form in the basis of $ c_{i\alpha}^\dagger = \left(c_{i\alpha,\uparrow}^\dagger, c_{i\alpha,\downarrow}^\dagger\right) $. }
	\begin{equation}\label{fun.SOC2}
		\begin{aligned}
			H_{SOC}(\mathbf{r})&=\lambda\sum_{<i,j>}\sum_{<\alpha,\alpha'>}c_{i\alpha}^\dagger \sigma_{i\alpha,j\alpha'} c_{j\alpha'}\\
			&=\lambda\sum_{<i,j>}\sum_{<\alpha,\alpha'>}c_{i\alpha}^\dagger  \left(R_{\pi/2}\mathbf{e}_{i\alpha,j\alpha'}\cdot \boldsymbol{\sigma}\right) c_{j\alpha'}\quad ,
		\end{aligned}
	\end{equation}
	where $ \sigma_{i\alpha,j\alpha'} $ represents the Pauli matrix vector that is perpendicular to the bond direction. $ R_{\pi/2} $ is the 3-D in-plane rotation matrix with rotation angle $ \pi/2 $. $ \mathbf{e}_{i\alpha,j\alpha'}=\mathbf{r_{i\alpha}}-\mathbf{r_{j\alpha'}} $ is the vector connecting the nearest sites. $ \lambda  $ is the parameter of SOC strength. More clearly, when electrons hop or bond with the neighbor electron, there is a certain direction for the momentum, which is actually the bond direction represented by the creation and annihilation operators on certain sites. 
	
	Fourier transformation of the SOC Hamiltonian can be written as
	\begin{equation}\label{key}
		\hat{H}_{SOC}(\mathbf{k})=\lambda\sum_{\mathbf{k}}\sum_{<\alpha,\alpha'>} e^{i\mathbf{k}\cdot\mathbf{e}_{i\alpha,j\alpha'}}c_{\mathbf{k}\alpha}^\dagger \left(R_{\pi/2}\mathbf{e}_{i\alpha,j\alpha'}\cdot \boldsymbol{\sigma}\right)c_{\mathbf{k}\alpha'},
	\end{equation}
	where $ c_{\mathbf{k}\alpha}^\dagger=\left(c_{\mathbf{k}\alpha,\uparrow}^\dagger, c_{\mathbf{k}\alpha,\downarrow}^\dagger\right) $ is the Fourier transformation operator of $ c_{i}^\dagger $. The $ \hat{H}_{SOC}(\mathbf{k}) $is the Fourier transformation value of $ H_{SOC}(\mathbf{r}) $, marked as $ H_{SOC}(\mathbf{r})  {\stackrel {\mathcal {F}}{\longleftrightarrow }} \hat{H}_{SOC}(\mathbf{k}) $. For the   kagome lattice with CDW modulation, there are 12 atoms in the unit cell, so the Hamiltonian expands to a $ 24\times 24 $ matrix in the basis of $ c_{\mathbf{k}}^\dagger=\left(c_{\mathbf{k}1}^\dagger,c_{\mathbf{k}2}^\dagger,\cdots, c_{\mathbf{k}12}^\dagger\right) $. 
	
     {We calculated the energy spectrum of the system without and with SOC (as shown in Fig. \ref{fig.Appendix_E}(a, e)). It can be seen that after considering SOC, the original spin degenerate energy bands will split.}

     {It is worth emphasizing that the Rashba spin-orbit interaction we adopted is not the only choice. However, the qualitative differences in the thermal Hall conductivity curve arise from variations in the Chern numbers. Rashba interaction, as the simplest and the most possible SOC in the system, offers limited topological phases. If Rashba interaction alone is sufficient for us to distinguish between different parameter regimes in terms of superconducting pairing symmetry, more complex interactions will be even more effective.
	}
	
\section{Superconductivity}\label{App.2}
	
	$ \mathrm{AV_3Sb_5} $ is the first quasi-2D superconductor in kagome lattice. As the result, the superconductivity is widely interested. However, the SC pairing symmetry is not yet clear, and there are conflicting experimental results that indicate different pairing symmetries. We are considering both spin singlets and spin triplets.

    \renewcommand{\thefigure}{S1}
	\begin{figure*}[ht]
		\centering
		\includegraphics[width=7in]{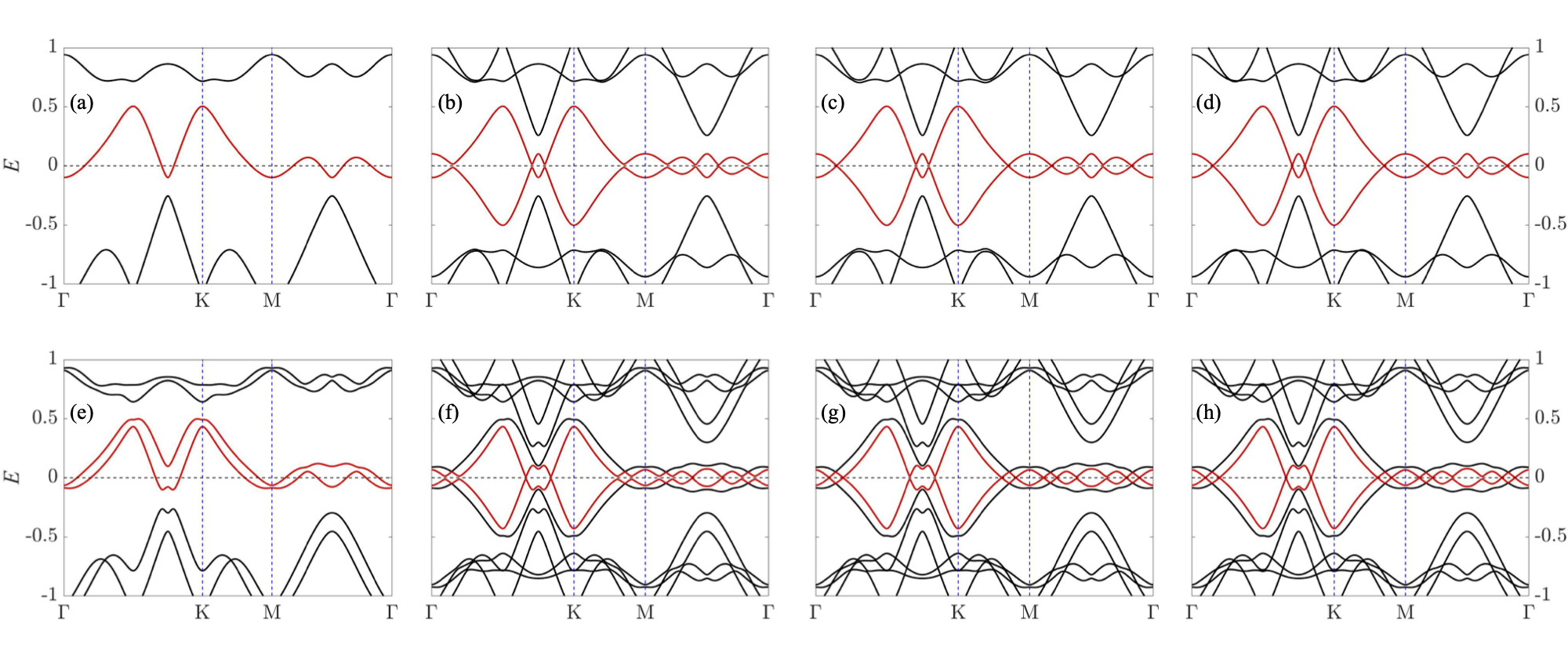}
		\caption{Calculated energy spectrum of model.
			(a)-(d) System without SOC for normal state, $ s $-wave, $ d $+i$ d $-wave and $ d $-i$ d $-wave pairings, and the parameter is $(\xi,\Delta,\mu)=(0.3,0.03,0.1)$. 
			(e)-(h) System with SOC for normal state, $ s $-wave, $ d $+i$ d $-wave and $ d $-i$ d $-wave pairings, and the parameter is $(\xi,\lambda,\Delta,\mu)=(0.3,0.1,0.03,0.1)$. The red line labeled the bands closest to the Fermi surface.
		}
		\label{fig.Appendix_E}
	\end{figure*}
	
	\subsection{$ s $-wave pairing}\label{App.2A}
	
	$ s $-wave SC is the so-called conventional SC which is explained by the BCS theory. Based on the self-consistent field approximation, the BCS Hamiltonian can be written as
	\begin{equation}\label{key}
		\hat{H}_{s-wave}=\frac{V}{2}\sum_{\mathbf{k}\mathbf{k}'}\left(c_{\mathbf{k}'\alpha,\uparrow}^\dagger c_{-\mathbf{k}'\alpha,\downarrow}^\dagger c_{-\mathbf{k}\alpha,\downarrow} c_{\mathbf{k}\alpha,\uparrow}+c_{\mathbf{k}'\alpha,\downarrow}^\dagger c_{-\mathbf{k}'\alpha,\uparrow}^\dagger c_{-\mathbf{k}\alpha,\uparrow} c_{\mathbf{k}\alpha,\downarrow}\right),
	\end{equation}
	where we ignore the interband coupling. This is the simplest pairing, and the self-consistent field approximation(SCFA) is the usual simplification of the Hamiltonian, which can be written as 
	\begin{equation}\label{key}
		\left\{
		\begin{aligned}
			&\left<c_{-\mathbf{k}\alpha,\downarrow} c_{\mathbf{k}\alpha,\uparrow}\right>=-\left<c_{-\mathbf{k}\alpha,\uparrow} c_{\mathbf{k}\alpha,\downarrow}\right>\neq 0\\
			&\left<c_{\mathbf{k}'\alpha,\uparrow}^\dagger c_{-\mathbf{k}'\alpha,\downarrow}^\dagger\right> = -\left<c_{\mathbf{k}'\alpha,\downarrow}^\dagger c_{-\mathbf{k}'\alpha,\uparrow}^\dagger\right>\neq 0\quad,
		\end{aligned}
		\right.
	\end{equation}
	where we ignore the difference of the index $ \mathbf{k} $ because the average energy is $ \mathbf{k} $-independent. Define that $ \Delta = \sum_{\mathbf{k}}\left<c_{-\mathbf{k}\downarrow}c_{\mathbf{k}\uparrow}\right> $, so the SCFA Hamiltonian can be written as 
	\begin{equation}\label{key}
		\begin{aligned}
			\hat{H}_{s-wave}=\frac{\Delta}{2}& \sum_{\mathbf{k},\alpha}\left(c_{\mathbf{k}\alpha,\uparrow}^\dagger c_{-\mathbf{k}\alpha,\downarrow}^\dagger - c_{\mathbf{k}\alpha,\downarrow}^\dagger c_{-\mathbf{k}\alpha,\uparrow}^\dagger\right)\\
			+\frac{\Delta^*}{2}&\sum_{\mathbf{k},\alpha}\left(c_{-\mathbf{k}\alpha,\downarrow} c_{\mathbf{k}\alpha,\uparrow}-c_{-\mathbf{k}\alpha,\uparrow} c_{\mathbf{k}\alpha,\downarrow}\right)\quad,
		\end{aligned}
	\end{equation}
	where we have used the equation  \begin{equation}\label{key}
		\begin{aligned}
			&\sum_{\mathbf{k},\alpha} c_{\mathbf{k}\alpha,\downarrow}^\dagger c_{-\mathbf{k}\alpha,\uparrow}^\dagger=\sum_{-\mathbf{k},\alpha} c_{-\mathbf{k}\alpha,\downarrow}^\dagger c_{\mathbf{k}\alpha,\uparrow}^\dagger\\
			=&\sum_{\mathbf{k},\alpha} c_{-\mathbf{k}\alpha,\downarrow}^\dagger c_{\mathbf{k}\alpha,\uparrow}^\dagger=\sum_{\mathbf{k}\neq0,\alpha} -c_{\mathbf{k}\alpha,\uparrow}^\dagger c_{-\mathbf{k}\alpha,\downarrow}^\dagger \quad,
		\end{aligned}
	\end{equation}
	which is the same as $ \sum_{\mathbf{k},\alpha}c_{-\mathbf{k}\alpha,\uparrow} c_{\mathbf{k}\alpha,\downarrow}$.
	
	We hope that we can easily expand the BCS Hamiltonian into a topological SC Hamiltonian, so we rewrite the s-wave SC in a tight-binding form which means the superconducting electrons have limited spatial mobility. The Hamiltonian can be written as 
	\begin{equation}\label{key}
		\begin{aligned}
			\hat H_{s-wave}=&\frac{\Delta}{2}\sum_{\mathbf{k}, \alpha}e^{i\mathbf{k}\cdot\mathbf{e}_{i\alpha,j\alpha}}c_{\mathbf{k}\alpha,\uparrow}^\dagger c_{-\mathbf{k}\alpha,\downarrow}^\dagger + h.c.\\
			-&\frac{\Delta}{2}\sum_{\mathbf{k}, \alpha}e^{i\mathbf{k}\cdot\mathbf{e}_{i\alpha,j\alpha}}c_{\mathbf{k}\alpha,\downarrow}^\dagger c_{-\mathbf{k}\alpha,\uparrow}^\dagger + h.c.\quad.
		\end{aligned}
	\end{equation}
	
	Bogoliubov Hamiltonian is written in the basis of Nambu representation
	\begin{equation}\label{key}
		\begin{aligned}
			\left(c_{\mathbf{k}}^\dagger, c_{-\mathbf{k}}\right)_r=
			\left(\right.&\left.c_{\mathbf{k}1,\uparrow}^\dagger,\cdots,c_{\mathbf{k}12,\uparrow}^\dagger,c_{\mathbf{k}1,\downarrow}^\dagger,\cdots,c_{\mathbf{k}12,\downarrow}^\dagger,\right.\\
			&\left.c_{-\mathbf{k}1,\uparrow},\cdots,c_{-\mathbf{k}12,\uparrow},c_{-\mathbf{k}1,\downarrow},\cdots,c_{-\mathbf{k}12,\downarrow}\right) ,
		\end{aligned}
	\end{equation}
	where the subscript $ r $ on the left of the equation represents the rearrangement of the creation and annihilation operators. Note that we are here considering the in-band coupling and ignore the interband coupling, so the site indexes $ i,j $ should be considered to be the identical sublattice of different unit cells. 
	
     {Under certain parameters, the energy spectrum of an s-wave superconducting states without and with SOC are shown in Fig. \ref{fig.Appendix_E}(b, f). It can be seen that s-wave superconductivity successfully gap out the energy bands that cross through the Fermi surface, and causes the particle hole symmetry.}
	
	\subsection{$ d $$ \pm $i$ d $-wave pairing}\label{App.2B}
	
	For another spin-singlet pairing SC, we are here consider a chiral $ d $+i$ d $-wave SC. This kind of SC has an angular-momentum two times faster than the rotation speed of $ \mathbf{k} $ vector. One of the simple methods to construct such a gap function $ \Delta_{d+id}(\mathbf{k}) $ on the lattice model is transforming a real-space effective Hamiltonian to $ \mathbf{k} $-space. The real-space Hamiltonian can be written as
	\begin{equation}\label{key}
		H_{d+id-wave} = \sum_{i,j}\sum_{\alpha}\Delta\left({i\alpha,j\alpha}\right)c_{i\alpha,\uparrow}^\dagger c_{j\alpha,\downarrow}^\dagger+h.c.\quad,
	\end{equation}
	where $ \Delta({i\alpha,j\alpha})=\Delta e^{i2\theta_{i\alpha,j\alpha}} $ is the SC gap function which depends on sublattice of the unit cell. $ \theta_{ij} $ is the angle between vector $ \mathbf{e}_x $ and $ \mathbf{e}_{i\alpha,j\alpha}=\mathbf{r}_{i\alpha}-\mathbf{r}_{j\alpha} $, and the double-angle phase represent the $ l=2 $ angular momentum of d-wave. 
	
	The Fourier transformation of the $ d $+i$ d $-wave pairing Hamiltonian can be written as
	\begin{equation}\label{key}
		\hat H_{d+id-wave}=\Delta\sum_{\mathbf{k}, \alpha}e^{i2\theta_{i\alpha,j\alpha}}e^{i\mathbf{k}\cdot\mathbf{e}_{i\alpha,j\alpha}}c_{\mathbf{k}\alpha,\uparrow}^\dagger c_{-\mathbf{k}\alpha,\downarrow}^\dagger + h.c.\quad.
	\end{equation}
	
	It might be a little bit weird to involve real-space indexes into a $ \mathbf{k} $-space Hamiltonian. However, $ \mathbf{e}_{ij} $ is independent of the position of cell $ i $ and $ j $,  but depend on the relative position between $ i $ and $ j $, which is actually some confirmed vector and we can tell them without knowing the site index. Here we can do a similar trick, rewrite the Hamiltonian as
	\begin{equation}\label{key}
		\begin{aligned}
			\hat H_{d+id-wave}=&\frac{\Delta}{2}\sum_{\mathbf{k}, \alpha}e^{i2\theta_{i\alpha,j\alpha}}e^{i\mathbf{k}\cdot\mathbf{e}_{i\alpha,j\alpha}}c_{\mathbf{k}\alpha,\uparrow}^\dagger c_{-\mathbf{k}\alpha,\downarrow}^\dagger + h.c.\\
			-&\frac{\Delta}{2}\sum_{\mathbf{k}, \alpha}e^{i2\theta_{i\alpha,j\alpha}}e^{-i\mathbf{k}\cdot\mathbf{e}_{i\alpha,j\alpha}}c_{\mathbf{k}\alpha,\downarrow}^\dagger c_{-\mathbf{k}\alpha,\uparrow}^\dagger + h.c.\quad.
		\end{aligned}
	\end{equation}
	 {
		As is well-known, the d-id-wave superconducting pairing is the chiral opposite of the d+id-wave complex d-wave pairing. Hence, the difference in their forms in real space lies solely in the sign of the phase, and we can directly provide its form as}
	 {
	\begin{equation}\label{key}
		\begin{aligned}
			\hat H_{d-id-wave}=&\frac{\Delta}{2}\sum_{\mathbf{k}, \alpha}e^{-i2\theta_{i\alpha,j\alpha}}e^{i\mathbf{k}\cdot\mathbf{e}_{i\alpha,j\alpha}}c_{\mathbf{k}\alpha,\uparrow}^\dagger c_{-\mathbf{k}\alpha,\downarrow}^\dagger + h.c.\\
			-&\frac{\Delta}{2}\sum_{\mathbf{k}, \alpha}e^{-i2\theta_{i\alpha,j\alpha}}e^{i\mathbf{k}\cdot\mathbf{e}_{i\alpha,j\alpha}}c_{\mathbf{k}\alpha,\downarrow}^\dagger c_{-\mathbf{k}\alpha,\uparrow}^\dagger + h.c.\quad ,
		\end{aligned}
	\end{equation}}
	 {	
		where the sign convention for the parameter $ \theta $ is the same as that for the d+id-wave case.}

         {We used the same parameters as for the s-wave superconducting states, and Fig. \ref{fig.Appendix_E}(c, d, g, h) displays the band structure of $ d $$ \pm $i$ d $-wave superconducting states. It is challenging to discern the differences between different superconducting pairing states directly from the band structure. Therefore, we rely more on thermal Hall conductivity curves as observable quantities to help distinguish between different superconducting pairing states, which is the core focus of this paper.}

	\section{Proof of Eq.\ref{formula.chern_number}}\label{App.C}
	
	Consider the situation that $ n,n'\in occ $. One of the terms of $ \Omega^n_{\mu\nu} $ can be written as
	\begin{equation}\label{key}
		i\frac{\bra{n}\partial H/\partial R^\mu \ket{n'}\bra{n'}\partial H/\partial R^\nu\ket{n}-(\mu\leftrightarrow \nu)}{\left(\epsilon_n-\epsilon_{n'}\right)^2}\quad.
	\end{equation}
	And one of the terms of $ \Omega^{n'}_{\mu\nu} $ can be written as
	\begin{equation}\label{key}
		\begin{aligned}
			&i\frac{\bra{n'}\partial H/\partial R^\mu \ket{n}\bra{n}\partial H/\partial R^\nu\ket{n'}-(\mu\leftrightarrow \nu)}{\left(\epsilon_{n'}-\epsilon_{n}\right)^2}\\
			=&-i\frac{\bra{n}\partial H/\partial R^\mu \ket{n'}\bra{n'}\partial H/\partial R^\nu\ket{n}-(\mu\leftrightarrow \nu)}{\left(\epsilon_n-\epsilon_{n'}\right)^2}\quad.
		\end{aligned}
	\end{equation}
	Therefore, the two terms that belongs to $ \Omega_{\mu\nu}^n $ and $ \Omega_{\mu\nu}^{n'} $ cancelled, which means when we add them together, the summation do not include the matrix elements labeled by $ \left(n,n'\right) $, even if there is energy degeneracy.  
	
	As a result, when we add up all the $ \Omega_{\mu\nu}^n,n\in occ $, the summation will not include the matrix elements labeled by $ \left(n,n'\right),\forall n,n'\in occ $. The result can be written as 
	\begin{equation}\label{key}
		\sum_{n\in occ}\Omega_{\mu\nu}^n =   \sum_{n\in occ}\sum_{n'\notin occ}i\frac{\bra{n}\partial H/\partial R^\mu \ket{n'}\bra{n'}\partial H/\partial R^\nu\ket{n}-(\mu\leftrightarrow \nu)}{\left(\epsilon_n-\epsilon_{n'}\right)^2},
	\end{equation}
	which can guarantee we are able to obtain a well-define Chern number.  

     {
    \section{Effect of SOC on Berry curvature}\label{App.E}}

    \renewcommand{\thefigure}{S2}
    \begin{figure}[t]
        \centering
        \includegraphics[width=3.5in]{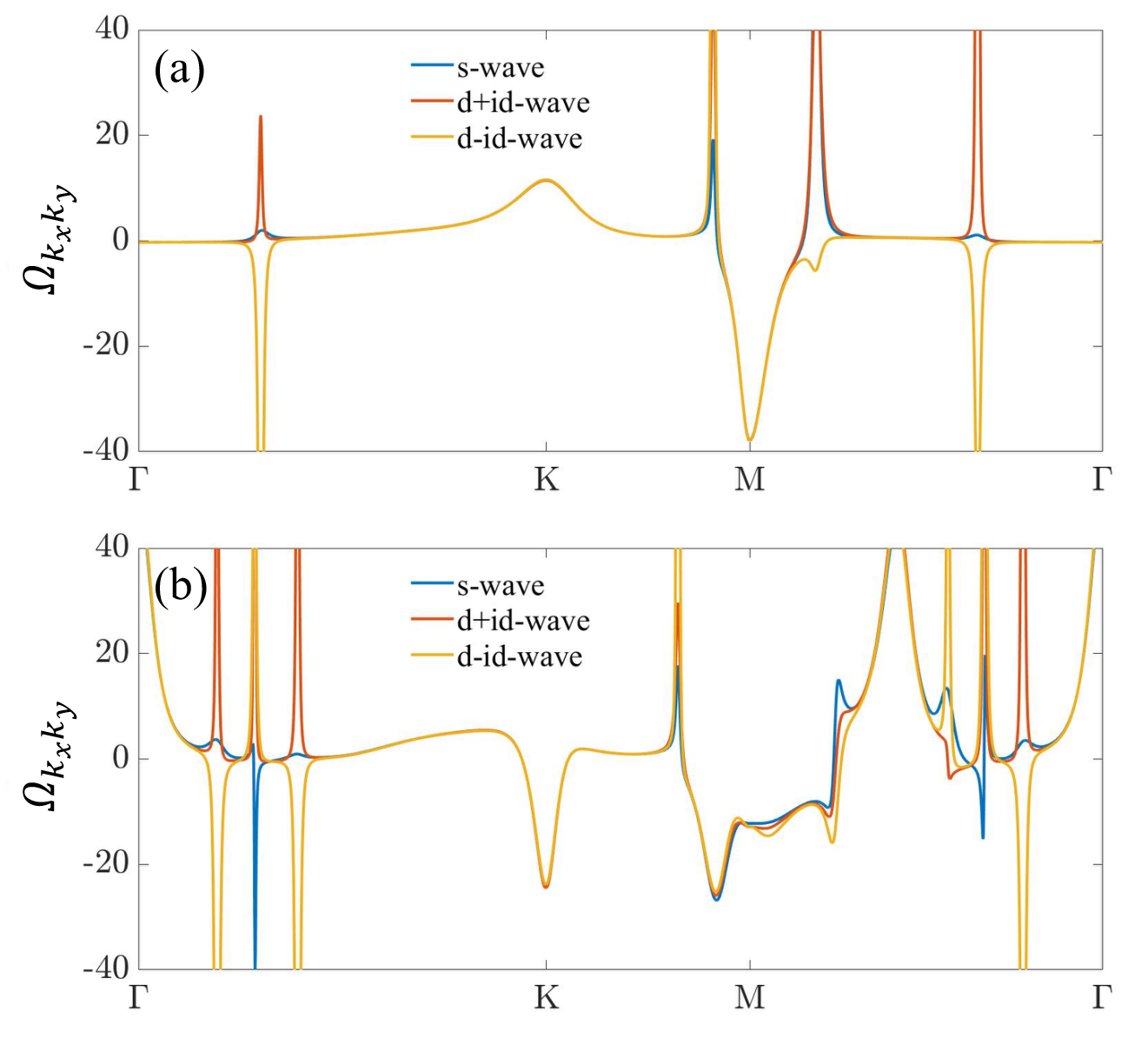}
        \caption{(a) Berry curvature of the system without SOC under the parameters $(\xi,\Delta,\mu)=(0.3,0.03,0.1)$. (b) Berry curvature of the system with SOC under the parameters $(\xi,\lambda,\Delta,\mu)=(0.3,0.1,0.03,0.1)$. It is clear that an SOC is to split the Berry curvature peaks, and the primary reason behind the splitting of Berry curvature peaks is the splitting of the Fermi surface. }
        \label{fig.S2}
    \end{figure}

   In this section, we are going to explain the similarity between Figs. \ref{fig.SC_30_3_10} and Fig.\ref{fig.SC_10_30_3_10}. We have to explain that the similarity is not the primary task of our study, while our primary aim is to distinguish different SC pairing symmetry states by calculating (or measuring) thermal Hall conductivity curves. Therefore, while we mentioned little change in thermal Hall conductivity when $ \lambda $ changes from 0 to 0.1, the actual significance lies in the fact that even as $ \lambda $ shifts from 0 to 0.1, we can still distinguish different superconducting pairing symmetry states qualitatively from the behavior of thermal Hall conductivity curves. Essentially, we seek to elucidate that the qualitative differences in thermal Hall con- ductivity curves solely originate from the superconducting terms.

    The first rationale is deduced from topological numbers. The Chern number of $ s $-wave shifts from 4 to 1, of $ d $+i$ d $-wave from 8 to 5, and of $ d $-i$ d $-wave from 0 to -3, all undergoing a change of -3. Notably, this uniform -3 shift across all Chern numbers corresponding to different superconducting terms signifies a contribution from the SOC. Thus, upon the affect of the SOC term into the system, the differences in Chern numbers for distinct superconducting pairing states continue to be solely shaped by the superconducting terms.

    The second rationale stems from Berry curvature: The shift in $\lambda$ from 0 to 0.1 causes a splitting of the Fermi surface near the $\Gamma$ point, transforming the ringlike region around the $\Gamma$ point in Figs. \ref{fig.3}(b)-\ref{fig.3}(d) into a double-ring pattern seen in Figs. \ref{fig.5}(b)-\ref{fig.5}(d). The Berry curvature distribution contributed by complex $ d $-wave pairing on a lattice with hexagonal symmetry (e.g., honeycomb, kagome) forms a circular ring around the $\Gamma$ point. Notably, the primary discrepancy between Figs. \ref{fig.5}(b)-\ref{fig.5}(d) and \ref{fig.5}(f)-\ref{fig.5}(h) lies in the vicinity of the $\Gamma$ point. Therefore we posit that the key factor behind the differences in thermal Hall conductivity curves at this juncture predominantly stems from variations in superconducting pairing symmetries, rather than from the SOC term.

    We have drawn Berry curvature distribution as Fig. \ref{fig.S2}. These graphs highlight that the main impact of the SOC on high-symmetry lines involves the splitting of significant Berry curvature peaks. However, substantial rearrangements of Berry curvature are not widespread, particularly among the smaller peaks around the $\Gamma$ point. Notably, significant shifts are observed near the K and M points, where the SOC primarily competes with the CDW term, resulting in changes in Chern numbers, while exhibiting less substantial interaction with the SC term. (In this context, the primary reason behind the splitting of Berry curvature peaks is the splitting of the Fermi surface.) These findings substantiate our viewpoint that the qualitative differences in the behavior of thermal Hall conductivity curves solely originate from the superconducting terms. It is worth noting that in regions with small band gaps, the absolute value of the Berry curvature tends to be very large, although this does not alter the Chern number. In order to provide a clearer visualization of the distribution of Berry curvature at different positions in $\mathbf{k}$ space, we have chosen the range [-40, 40] for plotting purposes.

\bibliography{Kagome}
	
\end{document}